\documentclass[12pt]{article}
\usepackage{booktabs} 
\usepackage{enumitem} 

\usepackage{natbib} 
\defcitealias{camsdata}{INERIS et.~al, 2022}
\defcitealias{nasadataset}{CIESIN et.~al, 2018}
\defcitealias{meteoseasons}{NCEI, 2023}
\defcitealias{cds}{CDS, 2020}

\usepackage[pdftex,colorlinks,citecolor=blue,urlcolor=blue,linkcolor=red]{hyperref} 

\usepackage{doi}

\usepackage{graphicx} 
\usepackage[small,bf,hang]{caption}	
\usepackage{subcaption}
\usepackage{nicefrac}
\usepackage{tabularx}
\usepackage[export]{adjustbox}
\usepackage{afterpage}

\renewcommand*{\thesubfigure}{\alph{subfigure}}
\newcommand\nsubcap[1]{\phantomcaption%
       \caption*{\textbf{(\thesubfigure}) #1}}

\renewcommand*{\thesubtable}{\alph{subtable}}
\newcommand\nsubcapp[1]{\phantomcaption%
       \caption*{\textbf{(\thesubtable}) #1}}
              
\usepackage{amssymb} 
\usepackage{amsmath} 
\usepackage{accents}
\usepackage{bbm} 
\usepackage{amsbsy} 
\usepackage{multirow} 
\usepackage[dvipsnames]{xcolor} 
\usepackage{mathtools}
\usepackage{eurosym}

\usepackage{floatrow}
\newfloatcommand{capbtabbox}{table}[][\FBwidth]

\usepackage[ruled]{algorithm2e} 
\SetAlCapFnt{\small} 
\SetKw{KwTo}{in}

\usepackage{authblk} 

\usepackage{footnote}

\bibpunct[, ]{(}{)}{;}{a}{,}{,}

\usepackage{tikz,pgfplots}
\usetikzlibrary{fit, positioning,chains,shapes.arrows,fit}
\usetikzlibrary{shapes,arrows, arrows.meta}
\usetikzlibrary{decorations.pathreplacing}
\usetikzlibrary{calc} 
\usetikzlibrary{matrix} 
\usepackage{pgf}
\usetikzlibrary{decorations.text}
\tikzset{cross/.style={cross out, draw=black, minimum size=2*(#1-\pgflinewidth), inner sep=0pt, outer sep=0pt},
cross/.default={1pt}}
\usetikzlibrary{shapes.misc}
\usepackage{environ}
\makeatletter
\newsavebox{\measure@tikzpicture}
\NewEnviron{scaletikzpicturetowidth}[1]{%
  \def\tikz@width{#1}%
  \begin{lrbox}{\measure@tikzpicture}%
  \BODY
  \end{lrbox}%
  \pgfmathparse{#1/\wd\measure@tikzpicture}%
  \BODY
}
\makeatother
\tikzdeclarecoordinatesystem{page}{
    \parsecomma#1\endparsecomma
    \pgfpointanchor{current page}{north east}
    \pgf@xc=\pgf@x%
    \pgf@yc=\pgf@y%
    \pgfpointanchor{current page}{south west}
    \pgf@xb=\pgf@x%
    \pgf@yb=\pgf@y%
    \pgfmathparse{(\pgf@xc-\pgf@xb)/2.*\page@x+(\pgf@xc+\pgf@xb)/2.}
    \expandafter\pgf@x\expandafter=\pgfmathresult pt
    \pgfmathparse{(\pgf@yc-\pgf@yb)/2.*\page@y+(\pgf@yc+\pgf@yb)/2.}
    \expandafter\pgf@y\expandafter=\pgfmathresult pt
}
\makeatother
\usetikzlibrary{arrows,automata}
\usepackage{ragged2e}

\tikzset{
    max width/.style args={#1}{
        execute at begin node={\begin{varwidth}{#1}},
        execute at end node={\end{varwidth}}
    }
}

\definecolor{newblue}{RGB}{86, 180, 233}
\definecolor{newred}{RGB}{248, 118, 109}
\definecolor{newgreen}{RGB}{163, 213, 150}

\colorlet{textcolor}{white}
\colorlet{bordercolor}{white}

\pgfdeclarelayer{background}
\pgfsetlayers{background,main}

\definecolor{airforceblue}{rgb}{0.36, 0.54, 0.66}
\definecolor{forestgreen}{rgb}{0.13, 0.55, 0.13}\definecolor{fulvous}{rgb}{0.86, 0.52, 0.0}
\definecolor{gray}{rgb}{0.5, 0.5, 0.5}
\definecolor{bistre}{rgb}{0.24, 0.17, 0.12}\definecolor{bostonuniversityred}{rgb}{0.8, 0.0, 0.0}
\definecolor{purpleheart}{rgb}{0.41, 0.21, 0.61}
\definecolor{lightsalmonpink}{rgb}{1.0, 0.6, 0.6}\definecolor{arrowcolor}{rgb}{0.92, 0.92, 0.92}
\tikzset{
inner/.style={
  on chain,
  circle,
  inner sep=4pt,
  fill=circlecolor,
  line width=1.5pt,
  draw=bordercolor,
  text width=1.2em,
  align=center,
  text height=1.25ex,
  text depth=0ex
},
on grid
}

\usepackage{regexpatch}

\usepackage{pbox}
\newcommand\drawarrow{

\node[on chain] (f) {};
\begin{pgfonlayer}{background}
\node[
  inner sep=10pt,
  single arrow,
  single arrow head extend=0.6cm,
  draw=none,
  fill=arrowcolor,
  fit= (c1) (f)
] (arrow) {};
\fill[white] 
  (arrow.before tail) -- (c1|-arrow.west) -- (arrow.after tail) -- cycle;
\end{pgfonlayer}
}

\numberwithin{equation}{section}

\usepackage{setspace}
\def\spacingset#1{\renewcommand{\baselinestretch}%
{#1}\small\normalsize} \spacingset{1}

\usepackage{xr}
\makeatletter
\let\ref\@refstar
\makeatother

\begin{document}

\def\spacingset#1{\renewcommand{\baselinestretch}%
{#1}\small\normalsize} \spacingset{1}

\title{\bf The short-term association between environmental variables and mortality: evidence from Europe}
\author[1,2,3,4,5,*]{Jens Robben}
\author[1,2,3,4,5]{Katrien Antonio}
\author[2,5]{Torsten Kleinow}
\affil[1]{Faculty of Economics and Business, KU Leuven, Belgium.}
\affil[2]{Faculty of Economics and Business, University of Amsterdam, The Netherlands.}
\affil[3]{LRisk, Leuven Research Center on Insurance and Financial Risk Analysis, KU Leuven, Belgium.}
\affil[4]{LStat, Leuven Statistics Research Center, KU Leuven, Belgium.}
\affil[5]{RCLR, Research Centre for Longevity Risk, University of Amsterdam, The Netherlands.}
\affil[*]{Corresponding author: \href{mailto:jens.robben@kuleuven.be}{j.robben@uva.nl}}
\date{\today} 
\maketitle
\thispagestyle{empty}

\begin{abstract} \noindent
Using fine-grained, publicly available data, this paper studies the short-term association between environmental factors, i.e., weather and air pollution characteristics, and weekly mortality rates in small geographical regions in Europe. Hereto, we develop a mortality modeling framework where a baseline model describes a region-specific, seasonal trend observed within the historical weekly mortality rates. Using a machine learning algorithm, we then explain deviations from this baseline using features constructed from environmental data that capture anomalies and extreme events. We illustrate our proposed modeling framework through a case study on more than 550 NUTS 3 regions (Nomenclature of Territorial Units for Statistics, level 3) in 20 European countries. Using interpretation tools, we unravel insights into which environmental features are most important when estimating excess or deficit mortality relative to the baseline and explore how these features interact. Moreover, we investigate harvesting effects through our constructed weekly mortality modeling framework. Our findings show that temperature-related features are most influential in explaining mortality deviations from the baseline over short time periods. Furthermore, we find that environmental features prove particularly beneficial in southern regions for explaining elevated levels of mortality, and we observe evidence of a harvesting effect related to heat waves.
\end{abstract}

\noindent%
\vfill

\newpage
\spacingset{1}
\section{Introduction}
In the epidemiological and medical literature, studies have consistently unveiled short-term associations between temperature and mortality statistics. For instance, both heat waves and cold spells can cause immediate increases in mortality, possibly with a delay of a few days or weeks \citep{basu2002relation}. Extreme weather events such as heavy rainfall, extreme drought, or wind storms also play a crucial role on mortality in the short term \citep{weilnhammer2021extreme}. Furthermore, epidemiological evidence underscores the adverse health effects of air pollution \citep{brunekreef2002air, ruckerl2011health}. Specifically, \cite{orellano2020short} find positive short-term associations between air pollution, including particulate matter, nitrogen dioxide, and ozone, and daily mortality outcomes. In this paper, we aim to identify and analyze the primary environmental factors associated with mortality deviations from a region-specific baseline mortality model. Our focus is on short-term associations at a weekly and regional level within Europe. 

The EuroMOMO project has emerged as an initiative for monitoring weekly excess mortality across 28 participating European countries or sub-national regions.\footnote{Website of the EuroMOMO project: \url{https://www.euromomo.eu/}.} Hereto, the Serfling model is used as mortality baseline \citep{serfling1963methods}. This baseline includes a log-linear long-term trend on the one hand, and Fourier terms to address the seasonality inherent in the weekly mortality pattern on the other hand. \cite{vestergaard2020excess} use this EuroMOMO modeling framework to estimate European-wide weekly mortality rates and excess all-cause mortality during the COVID-19 pandemic. Furthermore, \cite{nielsen2018influenza} introduce the FluMoMo model as an extension of the Serfling-type baseline mortality model by adding influenza activity and extreme temperatures as explanatory variables. 

Various methodologies have been explored in the epidemiological and medical literature to investigate the association between a particular environmental factor and daily (or weekly) mortality statistics. These methodologies range from observational studies \citep{keatinge2000heat} to the more advanced distributed lag (non-linear) models that capture non-linear effects from both predictor space and lag dimensions on mortality outcomes \citep{gasparrini2010distributed}. For a more detailed and technical overview of these methodologies, we refer to Suppl.~Mat.~\ref{appendix:literaryreview}. In the actuarial and economic literature, the impact of environmental factors on mortality remains largely unexplored. However, its importance is underscored by the European Insurance and Occupational Pensions Authority (EIOPA) which highlights the need to integrate climate change scenarios into insurers' own risk and solvency assessments \citep{eiopa2021opinion}. Among the scant economic literature, \cite{carleton2022valuing} find that extreme temperatures, both hot and cold, significantly increase mortality rates, particularly among the elderly, with higher incomes and adaptation reducing some of these effects. Within the actuarial field, there is a substantial body of literature dedicated to modeling mortality rates, particularly focusing on one-year mortality rates and their evolution across time and age. This literature encompasses a variety of approaches tailored to single and multiple populations \citep{booth2008mortality,enchev2017multi}, potentially integrating socio-economic characteristics \citep{Cairns2019modelling, wen2023modelling,villegas2014modeling}, but studies regarding the impact of environmental factors on mortality remain rather limited.  One such example comes from the American Academy of Actuaries, the Casualty Actuarial Society, the Canadian Institute of Actuaries, and the Society of Actuaries who developed the Actuaries Climate Index, a quarterly measure derived from changes in extreme weather events and sea levels, aiming at offering a practical monitoring tool for tracking climate trends \citep{aci}.

Our results contribute to the literature in three ways. First, following \cite{serfling1963methods}, we construct a weekly mortality baseline model that estimates a seasonal mortality pattern observed in the historical mortality rates of a particular region. We enhance the calibration process of this baseline model with a quadratic penalty matrix to obtain parameter estimates for the baseline that exhibit smooth variations across adjacent regions. Second, we leverage insights from fine-grained open data acquired via the Copernicus Climate Data Store (CDS) for weather factors and the Copernicus Atmospheric Monitoring Service (CAMS) for air pollution factors. Particularly, we use the E-OBS daily gridded meteorological data for Europe from the CDS \citepalias{nasadataset} and the European air quality reanalysis from the CAMS \citepalias{camsdata}. These datasets offer high temporal and spatial resolutions, providing detailed information on environmental factors across Europe. We demonstrate how this raw, fine-grained data can be used to create indices for environmental anomalies and extreme environmental conditions for small geographic regions. Third, we use this large set of pre-engineered environmental anomalies and extreme environmental indices as inputs in a machine learning model to analyze mortality deviations from the baseline level. As such, we extend the FluMOMO mortality modeling methodology \citep{nielsen2018influenza}. Using a machine learning approach allows us to simultaneously examine the short-term impact of multiple environmental factors on mortality, model complex interactions between these factors, and avoid making assumptions about the functional form of how each factor affects mortality.

This paper is organized as follows: Section~\ref{sec:notationsanddata} provides an overview of the notations and data. In Section~\ref{sec:modelspecification}, we detail our mortality modeling framework. We outline the calibration strategy in Section~\ref{sec:calibrationstrategy}. To illustrate the practical application of our methodology, we conduct a case study in Section~\ref{sec:casestudy} on the age group 65+, applying our proposed framework to the NUTS 3 regions of 20 European countries. We provide details regarding the feature engineering process, discuss the calibration results, and provide further insights and applications. Section~\ref{sec:conclusions} concludes and summarizes the key findings of our study.
\vspace{-0.25cm}

\section{Notations and data} \label{sec:notationsanddata}
\subsection{Notations} \label{subsec:notations}
Let $d_{x,t,w}^{(r)}$ be the observed death count in region $r$ at age group $x$ during ISO week $w$ of ISO year $t$.\footnote{We follow the ISO 8601 standard maintained, see \url{https://www.iso.org/standard/70907.html}.} We denote the set of regions by $\mathcal{R}$, the age groups by $\mathcal{X}$, the ISO years by $\mathcal{T}$ and the ISO weeks in ISO year $t$ by $\mathcal{W}_t$. We omit explicit reference to gender, although the methodologies in this paper apply to male, female, and unisex data. Moreover, we denote $E_{x,t,w}^{(r)}$ for the exposure-to-risk in region $r$ at age group $x$ during ISO week $w$ of ISO year $t$. 

The region-specific weekly force of mortality, $\mu_{x,t,w}^{(r)}$, represents the instantaneous rate of mortality. We define the observed weekly death rate, i.e., $m_{x,t,w}^{(r)}$, as the death counts, $d_{x,t,w}^{(r)}$, divided by the exposure, $E_{x,t,w}^{(r)}$. Furthermore, $\smash{q_{x,t,w}^{(r)}}$ refers to the weekly mortality rate and represents the probability that an individual from region $r$ and age group $x$ and who is alive at the start of ISO week $w$ in ISO year $t$, will die within the next week. Under the assumption that the force of mortality $\smash{\mu_{x,t,w}^{(r)}}$ is constant within a week, we can estimate the weekly mortality rate, for each $x\in\mathcal{X}$, $t\in\mathcal{T}$, $w \in \mathcal{W}_t$ and $r \in \mathcal{R}$, as \citep{pitacco2009modelling}:
\begin{align} \label{eq:calcmortrates}
q_{x,t,w}^{(r)} \approx 1 - \exp\left(-\mu_{x,t,w}^{(r)}\right).
\end{align}

\subsection{Data sources: mortality and environmental data} \label{subsec:data}
\begin{figure}[!ht]
\centering
\begin{subfigure}[b]{0.4\textwidth}
\includegraphics[width = 0.75\textwidth]{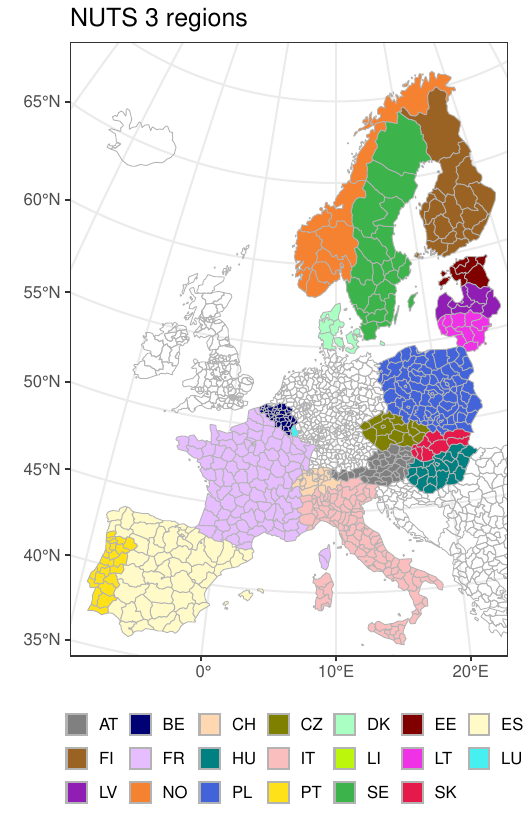}
\begin{minipage}{\textwidth}
\vspace{-0.5cm}
\nsubcap{\label{fig:nuts3}}
\end{minipage}
\end{subfigure}
\hspace{0.25cm}
\begin{subfigure}[b]{0.55\textwidth}
\centering
\includegraphics[width = \textwidth]{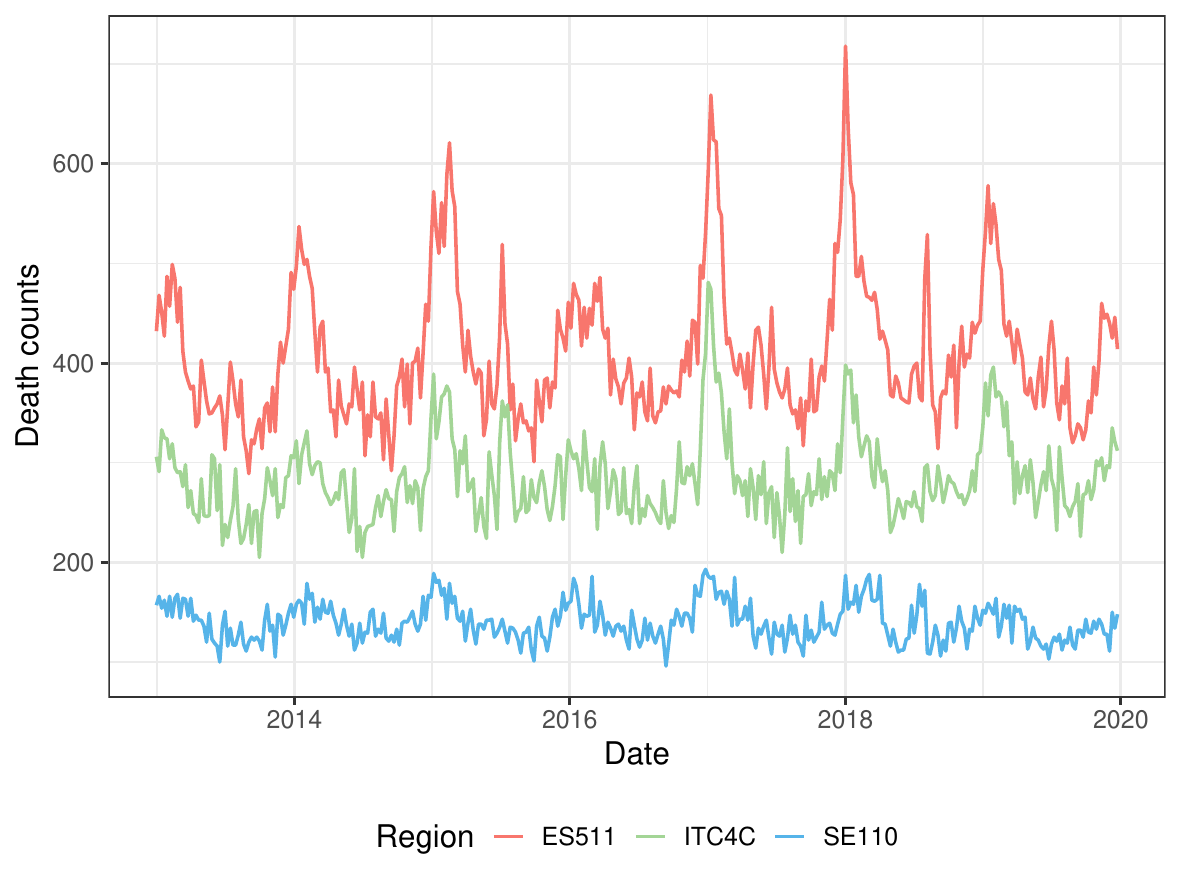}
\begin{minipage}{1.05\textwidth}
\vspace{-0.5cm}
\nsubcap{\label{fig:deathstime}}
\end{minipage}
\end{subfigure}
\caption{Panel (a) shows the NUTS 3 regions in 20 European countries, excluding overseas regions. Panel (b) shows the weekly female death counts throughout the years 2013-2019 in the NUTS 3 regions Barcelona (red), Milano (green), and Stockholm (blue).\label{fig:nuts3deathstime}}
\end{figure}

\paragraph{Mortality data.} We extract the deaths by ISO-week, sex, 5-year age group and NUTS 3 region from Eurostat throughout the years 2013-2019 for the 20 European countries highlighted in Figure~\ref{fig:nuts3}.\footnote{The Eurostat database can be consulted on \url{https://ec.europa.eu/eurostat/databrowser/view/demo_r_mweek3/default/table?lang=en}.} Countries such as Germany, the Netherlands, and the United Kingdom are excluded due to the unavailability of weekly death counts at the NUTS 3 level. The Nomenclature of Territorial Units for Statistics (NUTS) serves as a geocode standard used to designate the administrative divisions of countries for statistical purposes and corresponds for example to the departments in France (101 regions) and provinces and metropolitan cities in Italy (107 regions). We do not consider overseas regions in our analysis. Figure~\ref{fig:nuts3} visualizes the NUTS 3 European regions used in this study. 

In this paper, we specifically concentrate on older ages due to their increased vulnerability to the impacts of environmental factors. This approach aligns with previous studies, such as \cite{keatinge2000heat} and \cite{sunyer1996air}. Hereto, we aggregate the region-specific death counts $\smash{d_{x,t,w}^{(r)}}$ across the ages 65+, denoted as $\smash{d_{t,w}^{(r)}}$ in the sequel of this paper. Figure~\ref{fig:deathstime} illustrates the weekly death counts spanning the years 2013-2019 across the NUTS 3 regions of Barcelona (ES511), Milano (ITC4C), and Stockholm (SE110). The figure reveals a seasonal trend, with more deaths during winter weeks and relatively lower death counts during summer weeks. In order to calculate weekly mortality rates, it is necessary to construct a weekly exposure measure, $\smash{E_{t,w}^{(r)}}$, for the age group 65+ in each NUTS 3 region throughout the years 2013-2019. Suppl.~Mat.~\ref{app:weekexpo} details this construction.

\paragraph{Weather and air pollution data.} To retrieve weather and air pollution factors, we consult the Copernicus Climate Data Store (CDS) and the Atmosphere Data Store (ADS). Particularly, we use the E-OBS land-only, gridded meteorological data for Europe \citepalias{cds}, and the CAMS European air quality reanalyses data \citepalias{camsdata}. These datasets are based on observations gathered by meteorological and air pollution stations across Europe. The CAMS reanalysis enhances these observations with the inclusion of satellite data. The datasets are defined on a daily (E-OBS) or hourly (CAMS) high-resolution gridded basis, where weather and air pollution factors are measured on grids with a spatial resolution of $0.10^{\circ}$ ($\approx 10$ km) in both longitude and latitude direction.

Table~\ref{tab:climatevar} lists the weather factors that we will use in this paper. These factors have shown their relevance in epidemiological and medical research concerning their impact on mortality. For instance, \cite{barnett2010measure} finds that the daily minimum, average, and maximum temperatures all have a comparable level of predictive power to explain daily death counts, while \cite{braga2002effect} and \cite{alberdi1998daily} study the impact of humidity and wind speed, respectively, on daily mortality. Table~\ref{tab:envvar} outlines the air pollutants under consideration, which are selected based on their demonstrated significant short-term association with mortality in the literature. For example, \cite{pascal2014short} identify short-term impacts of PM$_{10}$ and PM$_{2.5}$ levels on daily mortality in nine French cities. In addition, \cite{orellano2020short} find short-term associations between ozone and nitrogen dioxide and daily mortality based on a systematic review including 196 studies. 

\begin{table}[htb!]
  \small
  \centering
  \begin{subtable}[t]{0.48\textwidth}
  \centering
  \begin{tabular}[t]{p{0.2\linewidth}p{0.73\linewidth}}
    \toprule
    \textbf{Weather factor} & \textbf{Explanation} \\ \midrule
    \texttt{Tmax} & {\small Daily maximum temperature, measured in degrees Celsius ($^{\circ}$C), 2 meters above surface.} \\
    \texttt{Tavg} & {\small Daily average temperature in $^{\circ}$C at 2 meters above surface.} \\
    \texttt{Tmin} & {\small Daily minimum temperature in $^{\circ}$C at 2 meters above surface.} \\
    \texttt{Hum} & {\small Daily average relative humidity at 2 meters above surface.} \\
    \texttt{Rain} & {\small Daily total precipitation in $mm$.} \\
    \texttt{Wind} & {\small Daily average wind speed in $\nicefrac{m}{s}$ at 10 metres above surface. } \\
    \bottomrule
  \end{tabular}
  \begin{minipage}{1.1\textwidth}
\nsubcapp{\label{tab:climatevar}}
\end{minipage}
\end{subtable}
\hspace{0.2cm}
\begin{subtable}[t]{0.48\textwidth}
\centering
\begin{tabular}[t]{p{0.22\linewidth}p{0.73\linewidth}}
    \toprule
    \textbf{Air pollutant} & \textbf{Explanation} \\ \midrule
    \texttt{O3} & {\small Hourly ozone levels, measured in micrograms per cubic meter ($\mu g/m^3$).} \\
    \texttt{NO2} & {\small Hourly nitrogen dioxide levels ($\mu g/m^3$). } \\
    \texttt{PM10} & {\small Hourly particular matter levels of particles with a diameter of 10 microns or less ($\mu g/m^3$).} \\
    \texttt{PM2.5} & {\small Hourly fine particular matter levels of particles with a diameter of 2.5 microns or less ($\mu g/m^3$). \:\:\:\:\:\:\:\:\:\:\:\:} \\
    \bottomrule
  \end{tabular}
    \begin{minipage}{1.1\textwidth}
\nsubcapp{\label{tab:envvar}}
\end{minipage}
\end{subtable}
\caption{The weather variables retrieved from the E-OBS gridded meteorological dataset (a) and the air pollutants from the CAMS European air quality dataset (b). \label{tab:climateenvvar}}
\end{table}

\begin{figure}[ht!]
\centering
\begin{subfigure}[b]{0.4\textwidth}
\centering
\includegraphics[width=0.8\textwidth]{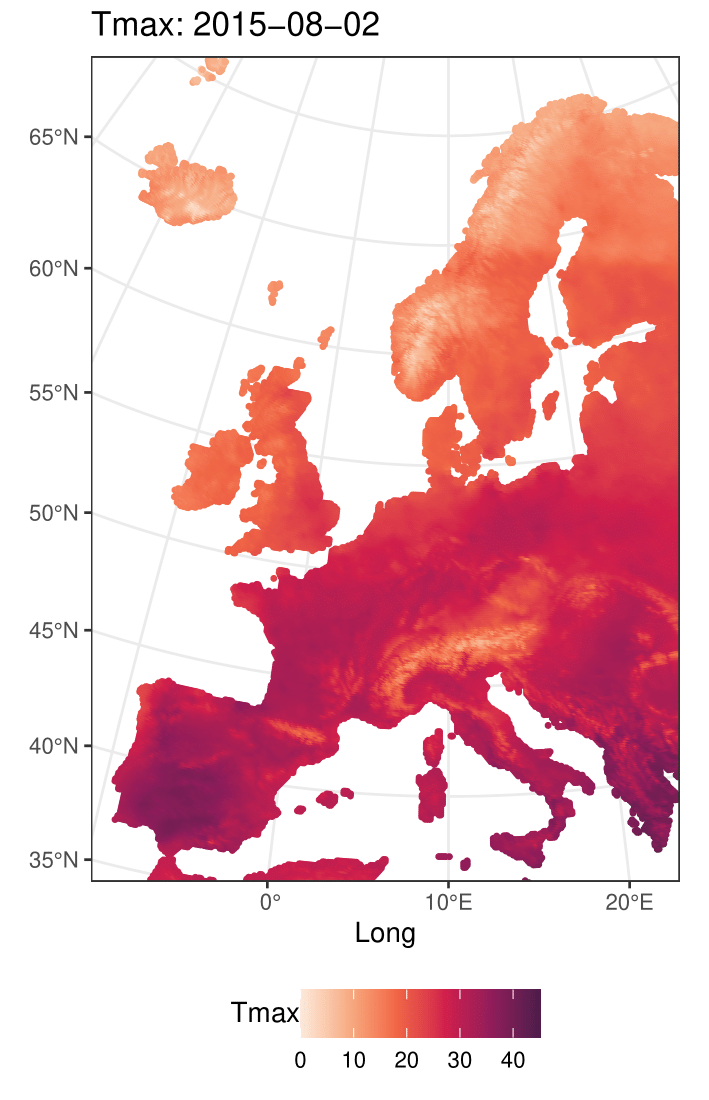}
\begin{minipage}{1.1\textwidth}
\vspace{-0.5cm}
\nsubcap{\label{fig:climvar1}}
\end{minipage}
\end{subfigure}
\hspace{0.5cm}
\begin{subfigure}[b]{0.4\textwidth}
\centering
\includegraphics[width=0.8\textwidth]{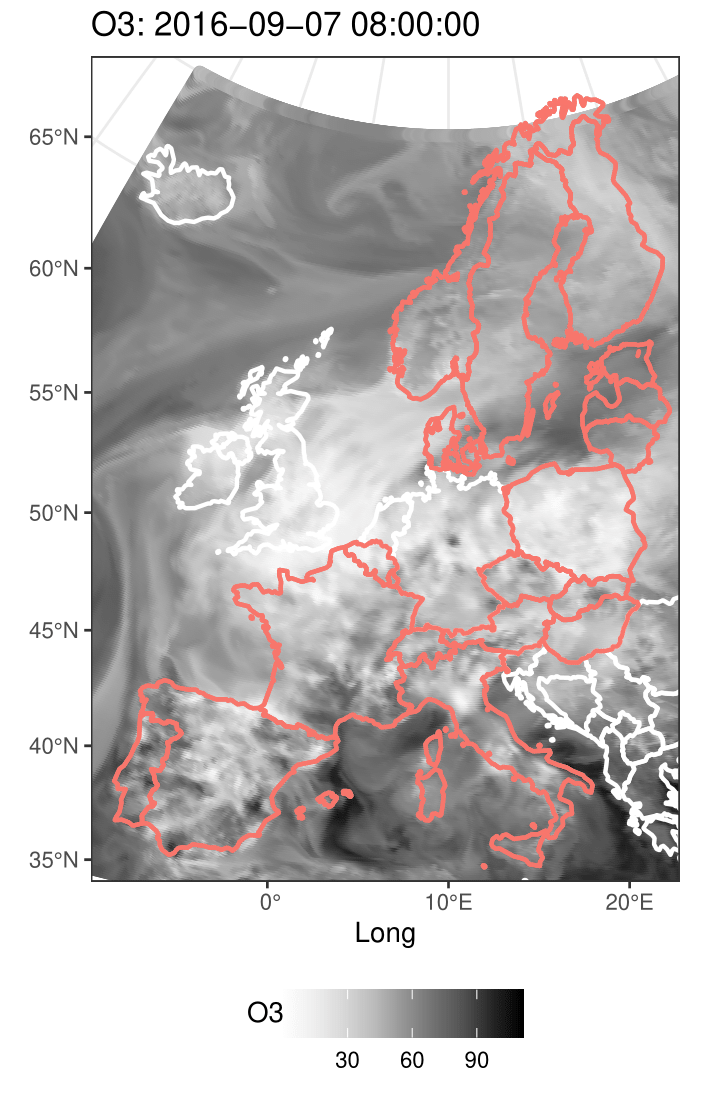}
\begin{minipage}{1.1\textwidth}
\vspace{-0.5cm}
\nsubcap{\label{fig:envvar1}}
\end{minipage}
\end{subfigure}
\caption{Panel (a) shows the maximum temperature on August 2, 2015. Panel (b) visualizes the ozone concentration on September 7, 2016 at 8:00 AM, where we highlight the boundaries of the 20 European countries under consideration.\label{fig:climenvvar1}}
\end{figure}

We extract the E-OBS daily weather factors and hourly CAMS air pollution measurements in Europe across a grid that covers the 20 European countries, and spans the years 2013-2019. As an example, Figures~\ref{fig:climvar1} and~\ref{fig:envvar1} display, respectively, the daily maximum temperature and the hourly ozone concentration at two randomly selected dates within the considered time range. Unlike the E-OBS dataset, the CAMS dataset covers both land and sea. Figure~\ref{fig:envvar1} therefore highlights the boundaries of the European countries under consideration in red to provide a clear visualization of the land component.

\section{Model specification} \label{sec:modelspecification}
Using environmental data, we aim to explain an excess or deficit of mortality, relative to a baseline mortality model. The baseline model estimates a weekly seasonal mortality pattern observed in the weekly mortality rates of each considered NUTS 3 region, resulting in an estimate for the expected number of deaths per week for each region, see Section~\ref{subsec:thebaselineweeklymortalitymodel}.

\begin{figure}[ht!]
\centering
\begin{adjustbox}{width=0.9\textwidth}
\begin{tikzpicture}[node distance=2cm,>=latex,auto,every edge/.append style={thick}]

\node[align=center] at (11.75,-0.75) {$\mbox{{\Huge \: Feature engineering}}\quad \vcenter{\hbox{\includegraphics[width=.1\textwidth]{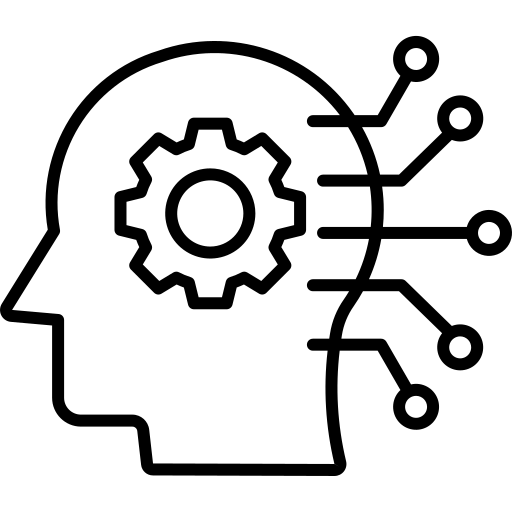}}}$};

\node[inner sep=0pt] (baseline) at (-3,3)
    {\includegraphics[width=.8\textwidth]{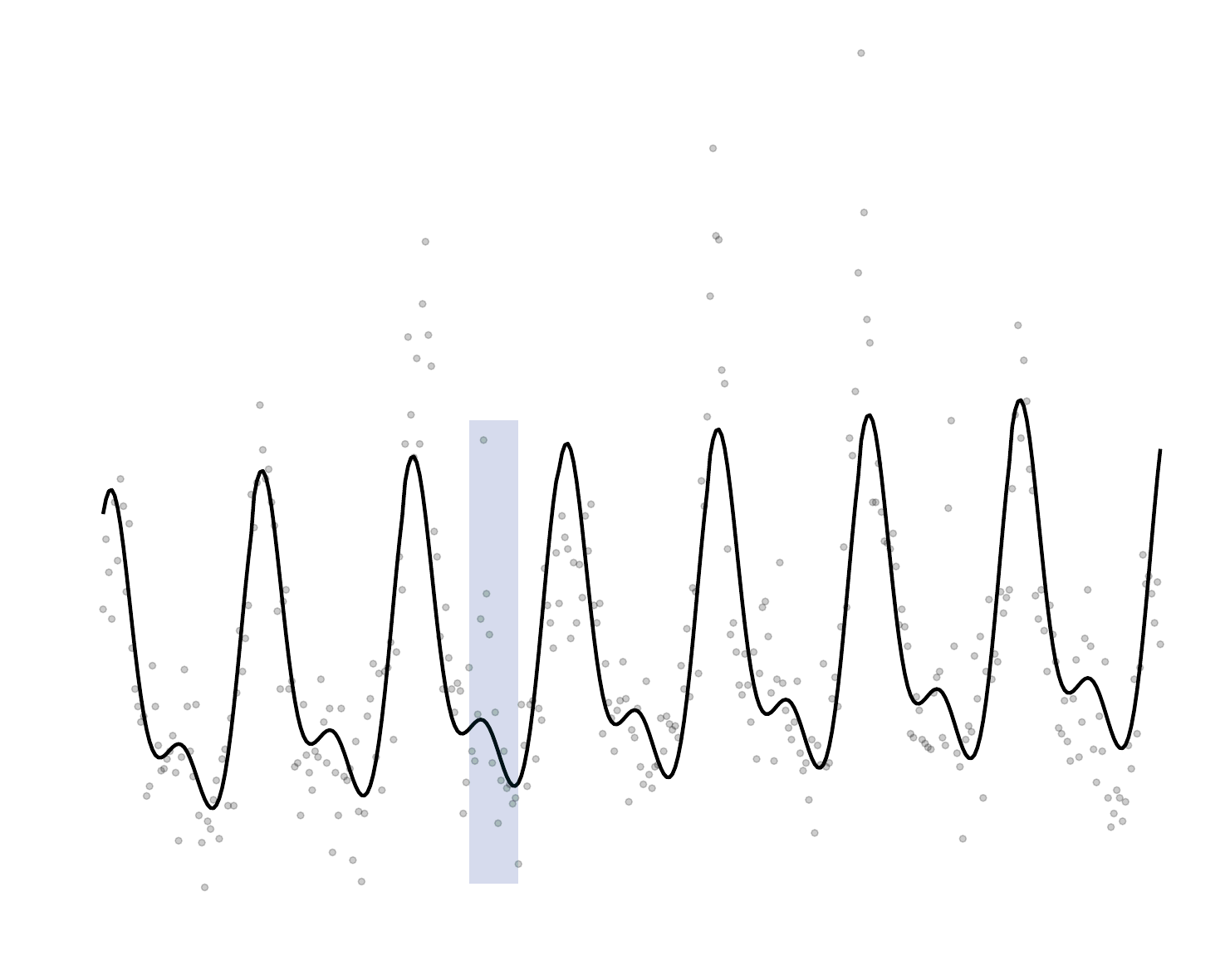}};
\node[draw=white] at (3.75,-1.5) {{\LARGE $t$}};
\draw [-{Stealth[length=3mm, width=2mm]},black] (-9,-2) to (3.75,-2);

\node[inner sep=0pt] (ml) at (11,5.5)
    {\includegraphics[width=.60\textwidth]{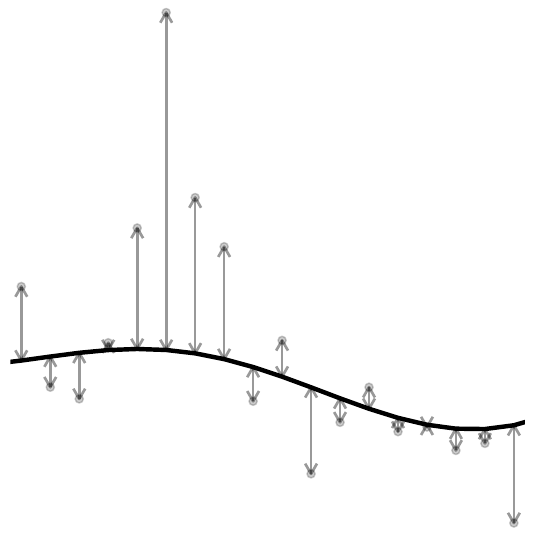}};
\node[inner sep=0pt] (ml) at (7.5,-3.25)
    {\includegraphics[width=.1\textwidth]{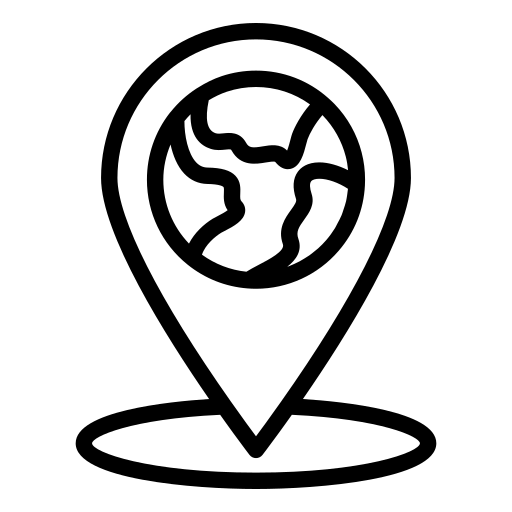}};
\node[inner sep=0pt] (ml) at (11.5,-3.25)
    {\includegraphics[width=.1\textwidth]{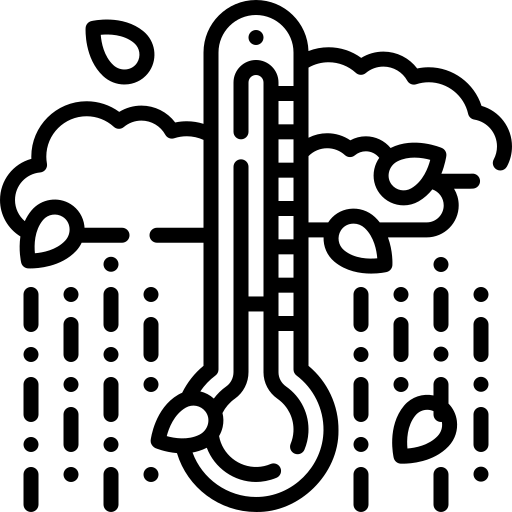}};
\node[inner sep=0pt] (ml) at (15.5,-3.25)
    {\includegraphics[width=.1\textwidth]{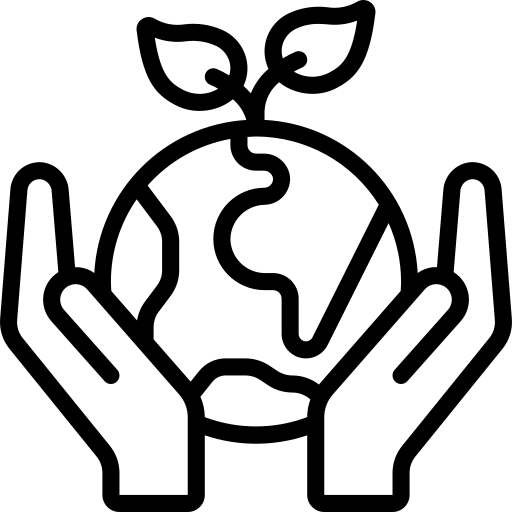}};

\draw [-{Stealth[length=3mm, width=2mm]},gray!70,dashed] (-4.5,3) to[out=40,in=140] (9.25,4.5);
\draw [{Stealth[length=3mm, width=2mm]}-,gray!70,dashed] (11,1.65) to[out=-90,in=90] (7.5,-2.3);
\draw [{Stealth[length=3mm, width=2mm]}-,gray!70,dashed] (11.4,1.65) to[out=-90,in=90] (11.5,-2.3);
\draw [{Stealth[length=3mm, width=2mm]}-,gray!70,dashed] (11.8,1.65) to[out=-90,in=90] (15.5,-2.3);
\end{tikzpicture}
\end{adjustbox}
\caption{Visual representation of the weekly mortality model, consisting of a baseline (left) and machine learning model (right). The baseline model is region-specific and captures the overall seasonal pattern in the observed mortality rates. Next, we explain the mortality deviations from this region-specific baseline model with a machine learning algorithm using a set of engineered region-specific environmental features. \label{tikz:model}}
\end{figure}

Mortality excess may arise from extreme environmental conditions. 
Therefore, we engineer features that quantify deviations from typical environmental conditions for each week of the year, referred to as anomalies. Furthermore, to allow for, e.g., heat waves or cold spells, we create extreme environmental indices that indicate how many days within a week a particular environmental feature exceeds or falls below certain thresholds. We then use these environmental anomalies and extreme indices, along with lagged versions of them, to explain excess or deficit mortality relative to the mortality baseline, see Section~\ref{subsec:themachinelearningmodel}.\footnote{Additional feature engineering techniques associate excess mortality with weekly average minimum and maximum temperatures \citep{nielsen2018influenza}, or create a heat wave indicator based on whether average temperatures exceed a certain threshold \citep{gasparrini2011impact}.} Section~\ref{subsec:featureengineering.casestudy} provides a detailed discussion of the feature engineering step outlined above. To capture any complex interaction effects between the environmental features and their possibly non-linear impact on mortality, we employ a machine learning model. This offers the additional advantage of automating the feature selection process in a dataset with numerous features. Figure~\ref{tikz:model} illustrates the proposed methodology. 

\subsection{The baseline weekly mortality model}\label{subsec:thebaselineweeklymortalitymodel}
The weekly, region-specific baseline mortality model estimates the overall, seasonal trend observed in the region's weekly death counts, see Figure~\ref{fig:deathstime} for an example of this seasonality. We adhere to the framework proposed by \cite{serfling1963methods}, incorporating seasonality through Fourier terms. We exclude the dependence on the age group $x$ since our focus will be on the age group 65+, as motivated in Section \ref{subsec:notations}. In this paper, we include population-exposures as an offset into a Poisson regression model for the weekly death counts, $D_{t,w}^{(r)}$, in region $r$:  
\begin{align} \label{eq:baselinemodelspec}
D_{t,w}^{(r)} \sim \text{Poisson}\left(E_{t,w}^{(r)}\: \mu_{t,w}^{(r)}\right), 
\end{align}
and consider the following structure for $\mu_{t,w}^{(r)}$ for each region $r\in \mathcal{R}$ \citep{serfling1963methods}:
\begin{equation} \label{eq:baselinedeathsstructure}
\begin{aligned}
\log \mu_{t,w}^{(r)} = \: &\beta_0^{(r)} + \beta_1^{(r)} t + \beta_2^{(r)} \sin\left(\frac{2\pi w}{52.18}\right) + \beta_3^{(r)} \cos\left(\frac{2\pi w}{52.18}\right) + \\ &\:\:\:\beta_4^{(r)} \sin\left(\frac{2\pi w}{26.09}\right) + \beta_5^{(r)} \cos\left(\frac{2\pi w}{26.09}\right),
\end{aligned}
\end{equation}
with $52.18$ the average number of weeks per year.\footnote{Alternatively, the exact number of ISO weeks in the year of consideration can be used.} Note that this baseline structure is region-, week-, and year-specific. Further, the region-specific parameters $\smash{\beta_{p}^{(r)}}$, for $p=0,1,..,5$, control for the level, trend, and seasonal variation. In Section~\ref{subsec:calibratingblm}, we show how to estimate this baseline model. We denote the estimated expected death counts from the baseline model in~\eqref{eq:baselinemodelspec} as:
\begin{align} \label{eq:btwr}
\hat{b}^{(r)}_{t,w} := \hat{E}\left[D_{t,w}^{(r)}\right] =  E_{t,w}^{(r)} \: \hat{\mu}_{t,w}^{(r)},
\end{align}
for each region $r\in\mathcal{R}$, with $\smash{\hat{\mu}_{t,w}^{(r)}}$ denoting the fitted baseline structure in~\eqref{eq:baselinedeathsstructure}. The black solid line in the left panel of Figure~\ref{tikz:model} shows an example.

\subsection{Modeling mortality deviations from the baseline model} \label{subsec:themachinelearningmodel}
We aim to explain the deviations from the baseline expected death counts using short-term effects of a set of features engineered from the available region-specific environmental factors, as visualized in the right panel of Figure~\ref{tikz:model}. We treat the fitted baseline death counts, $\smash{\hat{b}^{(r)}_{t,w}}$ in (\ref{eq:btwr}), as known and use them as an offset in a Poisson model:
\begin{equation} \label{eq:machinemodelspec}
\begin{aligned} 
 D_{t,w}^{(r)} &\sim \text{Poisson}\left(\hat{b}^{(r)}_{t,w} \: \phi_{t,w}^{(r)} \right), \\
\phi_{t,w}^{(r)} &= f\Big(\delta^{(r)},\:\gamma^{(r)},\:S_{t,w}, \:\boldsymbol{e}_{t,w}^{(r)},\:  L^1\left(\boldsymbol{e}_{t,w}^{(r)}\right),\: \ldots, L^s\left(\boldsymbol{e}_{t,w}^{(r)}\right) \Big).
\end{aligned}
\end{equation}
Here, $f(\cdot)$ denotes the outcome of a selected predictive model, such as a regression tree, random forest, gradient boosting, or neural network. Our preference for a machine learning model stems from its ability to highlight the important features from the high-dimensional set of environmental features, and to identify non-linear relationships and potential interaction effects. $\smash{\boldsymbol{e}_{t,w}^{(r)}}$ refers to the vector of the environmental anomalies and extreme indices constructed for week $w \in \mathcal{W}_t$ of year $t \in \mathcal{T}$ in region $r \in \mathcal{R}$. These anomalies quantify deviations from region-specific baseline conditions, on which we further elaborate in Section~\ref{subsec:featureengineering.casestudy}. We explicitly incorporate the longitude, i.e., $\smash{\delta^{(r)}}$, and latitude coordinate, i.e., $\smash{\gamma^{(r)}}$, of the centroid of each region $r$, as well as the season, $S_{t,w} \in \{\texttt{Spring}, \texttt{Summer}, \texttt{Autumn}, \texttt{Winter}\}$ to allow for potential interaction effects between regions, seasons, and the environmental features.\footnote{We define the seasons as: spring (March 15 - June 15), summer (June 15 - September 15), autumn (September 15 - December 15), and winter (December 15 - March 15). As our analysis focuses on weekly death counts, we define the season as the one in which the first day of the considered week falls.} Lastly, we include lagged values of the environmental features in the predictive model $f(\cdot)$. The functions $L^u(\cdot)$, for $u = 1,..,s$, are lag operators that shift the features $u$ weeks back in time. Lagged features play a crucial role in capturing various temporal phenomena and dependencies in the data, such as the impact of consecutive hot or cold weeks and potential harvesting effects \citep{schwartz2001there}. Harvesting effects occur when, for example, temperature-related excess mortality in the previous week leads to a mortality deficit in the current week.

\section{Calibration strategy and interpretation tools} \label{sec:calibrationstrategy}

\subsection{Calibrating the baseline model} \label{subsec:calibratingblm}
We calibrate the baseline weekly mortality model that captures the region-specific, seasonal pattern in the observed weekly death counts. Under the Poisson assumption in~\eqref{eq:baselinemodelspec} \citep{nelder1972generalized}, we consider all regions and add a penalty term to obtain smooth variations in the estimated parameters $\smash{\hat{\beta}_p^{(r)}}$ across neighbouring regions, for $p = 0,1,..,5$. Hereto, for notational convenience, we denote:
\begin{align*}
\boldsymbol{\beta}^{(r)} &:= \left(\beta_0^{(r)}, \beta_1^{(r)}, \beta_2^{(r)}, \beta_3^{(r)}, \beta_4^{(r)}, \beta_5^{(r)}\right) \in \mathbb{R}^{6} \\
\boldsymbol{z}_{t,w} &:= \left(1,t, \sin\left(\frac{2\pi w}{52.18}\right), \cos\left(\frac{2\pi w}{52.18}\right), \sin\left(\frac{2\pi w}{26.09}\right), \cos\left(\frac{2\pi w}{26.09}\right)\right) \in \mathbb{R}^{6},
\end{align*}
where $\boldsymbol{\beta}^{(r)}$ represents the region-specific parameter vector, and $\boldsymbol{z}_{t,w}$ the covariate vector for week $w$ in year $t$ in the Poisson GLM, as detailed in~\eqref{eq:baselinemodelspec} and \eqref{eq:baselinedeathsstructure}. Moreover, for $p=0,1,..,5$, we write $\boldsymbol{\beta}_p$ to denote the parameter vector across the considered regions, i.e., $\smash{\boldsymbol{\beta}_p := (\beta_p^{(r)})_{r\in\mathcal{R}}}$. We then calibrate the baseline weekly mortality model by minimizing the following objective function:
\begin{align} \label{eq:penpoisloglik}
\hat{\boldsymbol{\beta}} =  \operatorname*{argmin}_{\boldsymbol{\beta}} \: -l_P(\boldsymbol{\beta}) + \displaystyle \sum_{p=0}^5 \lambda_p \boldsymbol{\beta}_p^T \boldsymbol{S} \boldsymbol{\beta}_p,
\end{align}
where $\boldsymbol{\beta} := (\boldsymbol{\beta}^{(r)})_{r\in\mathcal{R}}$ and $l_P(\boldsymbol{\beta})$ is the Poisson log-likelihood. Additionally, the objective function in~\eqref{eq:penpoisloglik} contains penalty terms $\smash{\lambda_p \boldsymbol{\beta}_p^T \boldsymbol{S} \boldsymbol{\beta}_p}$, for $p= 0,1,..,5$, to impose smooth variations in the parameter vector $\smash{\boldsymbol{\beta}_p}$ across neighboring regions. Hereto, we use the following entries in the penalty matrix $\smash{\boldsymbol{S} := (s_{ij})_{i,j}}$, with $i,j \in \mathcal{R}$: 
\begin{equation}\label{eq:penmat}
s_{ij} = \begin{cases}
|\mathcal{N}_i| & \:\:\:\text{if} \:\: i=j \\
-1 & \:\:\: \text{if $i\neq j$ are neighboring regions} \\
0 & \:\:\: \text{elsewhere},
\end{cases}
\end{equation}
with $\mathcal{N}_i$ the set of neighbors of region $i$. The smoothness penalty in~\eqref{eq:penmat} penalizes based on the sum of the squared differences between the parameters of neighboring regions, see Suppl.~Mat.~\ref{app:choicepenaltymatrix} for a derivation. The parameter $\lambda_p$ in~\eqref{eq:penpoisloglik} controls the degree of smoothness in $\boldsymbol{\beta}_p$ across neighboring regions. A large $\smash{\lambda_p}$ results in the same $\smash{\beta_p^{(r)}}$ across all regions, while a $\lambda_p$ close to $0$ results in no penalization, yielding parameter estimates from a traditional, unregularized Poisson GLM. A similar smoothness technique has also been applied by \cite{li2024boosting} to shrink mortality forecasts in adjacent age groups and neighboring regions.

We estimate the parameters $\smash{\beta_p^{(r)}}$ in~\eqref{eq:penpoisloglik} using the penalized iteratively re-weighted least squares algorithm, where we select optimal smoothing parameters $\lambda_p$ using Un-Biased Risk Estimation scores \citep{wood2017generalized}.\footnote{We use the argument \texttt{paraPen} from the \texttt{gam} function in the \texttt{R}-package \texttt{mgcv} to implement the proposed spatially smoothed Poisson GLM, see \cite{wood2015package}.} We denote the optimal smoothing parameters as $\hat{\lambda}_p$, for $p = 0,1,..,5$, and the optimal parameter vector as $\smash{\hat{\boldsymbol{\beta}}}$. We then estimate the expected baseline death counts for each region $r \in \mathcal{R}$, year $t \in \mathcal{T}$, and week $w\in\mathcal{W}_t$ as:
\begin{align} \label{eq:baselinedeaths}
\smash{\hat{b}^{(r)}_{t,w}} = E_{t,w}^{(r)} \cdot \exp\left( \left(\hat{\boldsymbol{\beta}}^{(r)}\right)^T \boldsymbol{z}_{t,w} \right).
\end{align}

\subsection{Calibrating the mortality deviations model}\label{subsec:calibratingmlm}
Fixing the expected weekly baseline death counts, as obtained in Section~\ref{subsec:calibratingblm}, we now calibrate a predictive machine learning model to explain deviations from this baseline using environmental features, see Section~\ref{subsec:themachinelearningmodel}. Machine learning algorithms typically rely on a set of parameters of which some are carefully selected through a tuning process (tuning parameters), while others are set to predetermined values (hyper-parameters). In this paper, we opt for an extensive grid search that involves exploring a predefined grid of parameter values to identify optimal tuning parameter configurations. 

\paragraph{XGBoost algorithm.} While any predictive modeling technique can be deployed in our framework, we will specifically focus on the extreme gradient boosting machine \citep{chen2016xgboost} in the case study of Section~\ref{sec:casestudy}. Suppl.~Mat.~\ref{appendix:xgboostalgo} explains the XGBoost algorithm in case of a Poisson distributed outcome, see~\eqref{eq:machinemodelspec}, and details the considered tuning parameters. We choose the negative Poisson log-likelihood as loss function. 

\paragraph{Parameter tuning with cross-validation.} We use $K$-fold cross-validation to select optimal values for the tuning parameters, inspired by \citet{hastie2001elements}.\footnote{See \cite{bergmeir2018note} for a discussion on why traditional K-fold cross-validation can be used in a time series setting, conditionally on the lag structure in the model being adequately specified. Moreover, by incorporating the baseline number of deaths as an offset in our machine learning model, we effectively eliminate the trend and seasonal components from the time series of death counts, and, as such, enhance the applicability of traditional cross-validation techniques.} Suppl.~Mat.~\ref{appendix:xgboosttuning} describes the tuning strategy in further detail, with a visualisation in Figure~\ref{tikz:cv}. The optimal tuning parameters correspond to the parameter combination that yields the smallest average Poisson negative log-likelihood on the hold-out folds.

\paragraph{Training} Using the optimally chosen values for the tuning parameters, we calibrate the XGBoost model on the available historical data covering the entire duration of $T$ years. We denote the outcomes estimated from the XGBoost model as $\smash{\hat{f}_{\text{XGBoost}}\left(\cdot\right)}$ for $r \in \mathcal{R}$, $t\in \mathcal{T}$ and $w \in \mathcal{W}_t$. Consequently, the estimated death counts equal:
\begin{align*}
\hat{d}_{t,w}^{(r)} = \hat{b}_{t,w}^{(r)} \cdot \hat{f}_{\text{XGBoost}}\left(\boldsymbol{x}_{t,w}^{(r)}\right), \hspace{0.5cm} r\in\mathcal{R}, \: t\in\mathcal{T}, \: w \in \mathcal{W}_t,
\end{align*}
with $\smash{\boldsymbol{x}_{t,w}^{(r)}}$ the input vector of the XGBoost model consisting of the (lagged) environmental anomalies and extreme indices $\smash{\boldsymbol{e}_{t,w}^{(r)}}$, the season $S_{t,w}$, and the longitude $\smash{\delta^{(r)}}$ and latitude $\smash{\gamma^{(r)}}$ coordinates of region $r$, see Section~\ref{subsec:themachinelearningmodel}. As such, we can interpret the XGBoost's outcomes as a multiplier that can either augment or diminish the baseline number of death counts.

\subsection{Interpretation tools} \label{sec:interpretationtools}
Since machine learning models, like the XGBoost model, are often seen as black boxes, the use of interpretation tools is crucial to unravel insights in these models \citep{molnar2020interpretable}. 

\paragraph{Feature importance} In this interpretation tool, our objective is to unveil the features that significantly contribute to the predictions. We follow the approach proposed by \cite{breiman1984nonlinear} and measure the importance of a particular feature $X_l$ within a regression tree by aggregating the reductions in the considered loss function across all splits associated to this feature. Given that the XGBoost model consists of multiple regression trees, see Suppl.~Mat.~\ref{appendix:xgboostalgo}, we extend this analysis to measure the reduction in loss caused by feature $X_l$ across all trees in the ensemble. The higher the feature importance, the more important the feature is to the prediction process. Suppl.~Mat.~\ref{app:varimp} outlines the mathematical formulation of the feature importance.

\paragraph{Accumulated Local Effects} \cite{apley2020visualizing} propose Accumulated Local Effects (ALE) plots to visualize the impact of a particular feature on the predictions generated by a machine learning model. In contrast to the partial dependence plots introduced by \citet{friedman2001greedy}, ALE plots prove to be more effective in illustrating feature effects when dealing with correlated features \citep{molnar2020interpretable}. Given the strong correlation present in the environmental features, this paper places specific emphasis on the relevance of ALE plots for visualizing and interpreting the impact of a feature on the predictions. Suppl.~Mat.~\ref{app:ale} outlines the mathematical formulation and technical details of ALE effects.

\section{Case study on the NUTS 3 regions in 20 European countries} \label{sec:casestudy}

\subsection{Feature engineering} \label{subsec:featureengineering.casestudy}
To examine the short-term association between environmental factors and weekly mortality rates, we use data on weekly death counts for the age group 65+ within the NUTS 3 regions for the years 2013-2019, detailed in Section~\ref{subsec:data}. Hereto, we first aggregate across the temporal and spatial dimensions of the gridded, environmental data to construct daily features at NUTS 3 level. Suppl.~Mat.~\ref{subsubsec:hourlytodaily} and~\ref{subsubsec:populationweighted} outline this temporal and spatial aggregation step. To capture the effects of extreme environmental conditions, such as heat waves and cold spells, we engineer extreme environmental indices that indicate how many days within a week the environmental factors exceed a certain high quantile or are lower than a certain low quantile (see Suppl.~Mat.~\ref{subsubsec:extremeenvironmentalindices}). Furthermore, since the focus is on explaining deviations from a baseline model for weekly death counts, we engineer environmental anomalies that quantify deviations from normal, baseline conditions for each week in the year (see Suppl.~Mat.~\ref{subsubsec:baselinemodels}). Lastly, we aggregate the environmental features on a weekly basis and create lagged versions of it in Suppl.~Mat.~\ref{subsubsec:weeklyaggregation}. Figure~\ref{tikz:fengineering} illustrates the feature engineering process for the weather and air pollution factors in Table~\ref{tab:climateenvvar}. The flow chart also refers to the section in the Supplementary Material that details each step at the upper right corner of the box. 

\begin{figure}[ht!]
\centering
\adjustbox{max width=0.9\textwidth}{\begin{tikzpicture}[
node distance = 2.5cm and 3.5cm,
   arr/.style = {-Triangle,thick},
   box/.style = {rectangle, draw, semithick,
                 minimum height=1.25cm, minimum width=3cm,
                 max width = 3cm, fill=white, align = center},
   disc/.style = {shape=cylinder, draw, shape aspect=0.3,
                shape border rotate=90,
                text width=17mm, align=center,
                font=\linespread{0.8}\selectfont}
                        ]

\node (n12) [disc,label={\scriptsize \hspace{0cm} \hyperlink{subsection.2.2}{S2.2}}] {Air pollution \\ factor};

\node (n21) [box, right=of n12, label={\scriptsize \hspace{2cm} 
E.1}] {Daily \\ aggregation};
\node (n11) [disc, above=of n21, minimum height = 2cm,label={\scriptsize \hspace{0cm} \hyperlink{subsection.2.2}{S2.2}}] {Weather \\ factor};

\node (n31) [box, above right= 1.25cm and 3.5cm of n21, label={\scriptsize \hspace{2cm} E.2}] {Spatial \\ aggregation};

\node (n41) [box, above right = 1.25cm and 3.5cm of n31, label={\scriptsize \hspace{2cm} E.4}] {Environmental \\ anomalies};
\node (n42) [box, below right = 1.25cm and 3.5cm of n31, label={\scriptsize \hspace{2cm} E.3}] {Extreme \\ environmental \\ indices};

\node (n51) [box, above right= 1.25cm and 3.5cm of n42, label={\scriptsize \hspace{2cm} E.5}] {Weekly \\ aggregation \\ + \\ lagged weekly \\ features};

\node (a0) [above=of n12, yshift=2.75cm, xshift=-1.1cm] {};
\node (a1) [above=of n21, yshift=2.75cm, xshift=-2cm] {};
\node (a2) [above=of n21, yshift=2.75cm, xshift=-1.9cm] {};
\node (a3) [above=of n42, yshift=2.75cm, xshift=2cm] {};
\node (a4) [above right=2.5cm and 3.5cm of n42, yshift=2.75cm,xshift=1.75cm] {};

\node (b0) [above=of n12, yshift=1.85cm, xshift=-1.1cm] {};
\node (b1) [above=of n42, yshift=1.85cm, xshift=-5.2cm] {};
\node (b2) [above=of n42, yshift=1.85cm, xshift=-5cm] {};
\node (b3) [above right=2.5cm and 3.5cm of n42, yshift=1.85cm,xshift=1.75cm] {};

\draw[arr]   (n12) -> (n21);
\draw[arr]   (n11) -| (n31);
\draw[arr]   (n21) -| (n31);
\draw[arr]   ([yshift = 0.25cm]n31) -| (n41);
\draw[arr]   ([yshift = -0.25cm]n31) -| (n42);
\draw[arr]   ([yshift = 0.75cm]n41) -| (n51);
\draw[arr]   ([yshift = -0.75cm]n42) -| (n51);

\draw (a0)--(a1) node[midway,above, fill = gray!20]{Hourly};
\draw[<->]  (a0)--(a1);
\draw[<->]  (a2)--(a3) node[midway,above, fill = gray!20]{Daily};
\draw[<->]  (a2)--(a3);
\draw[<->]  (a3)--(a4) node[midway,above, fill = gray!20]{Weekly};
\draw[<->]  (a3)--(a4);

\draw[<->]  (b0)--(b1) node[midway,above, fill = gray!20]{(long,lat)};
\draw[<->]  (b0)--(b1);
\draw[<->]  (b2)--(b3) node[midway,above, fill = gray!20]{NUTS 3};
\draw[<->]  (b2)--(b3);
\end{tikzpicture}}
\caption{Flow chart explaining the feature engineering process for the weather-related factors and the air pollution factors, with the relevant section that details each step at the upper right corner of the box.\label{tikz:fengineering}}
\end{figure}

\begin{figure}[!ht]
\centering
\begin{subfigure}[b]{0.4\textwidth}
\centering
\includegraphics[width = 0.8\textwidth]{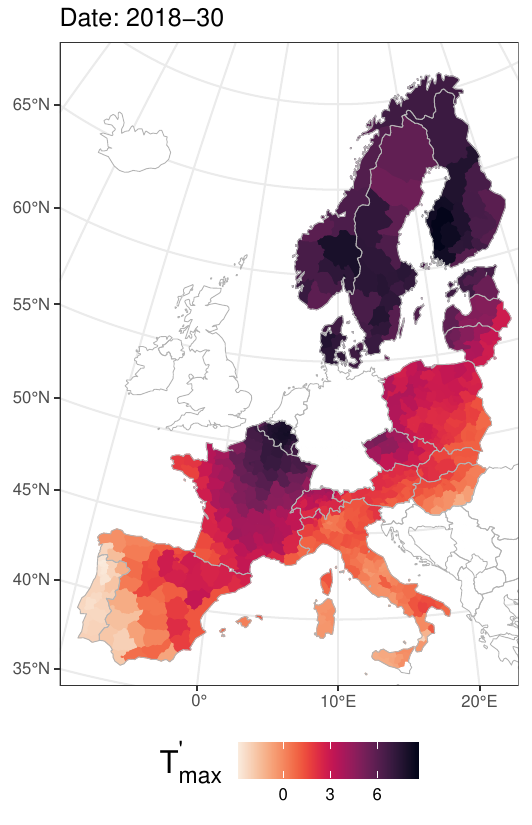}
\begin{minipage}{1\textwidth}
\vspace{-0.5cm}
\nsubcap{\label{fig:ff1}}
\end{minipage}
\end{subfigure}
\hspace{0.5cm}
\begin{subfigure}[b]{0.4\textwidth}
\centering
\includegraphics[width = 0.8\textwidth]{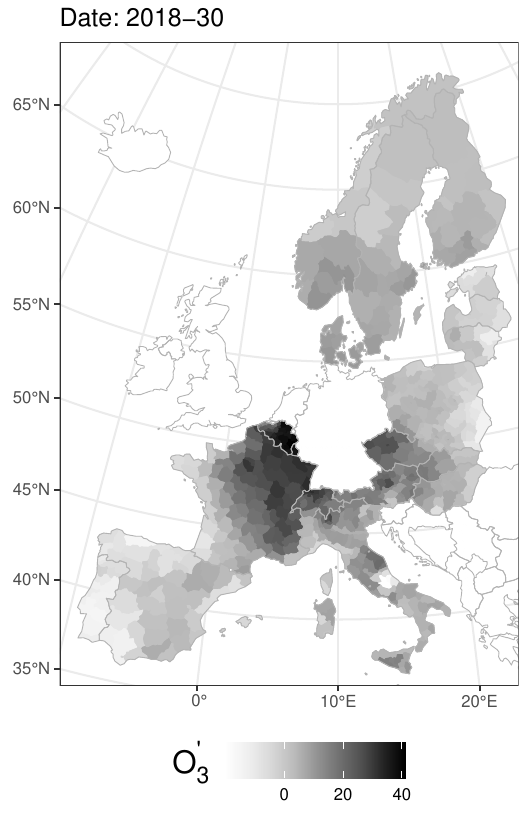}
\begin{minipage}{1\textwidth}
\vspace{-0.5cm}
\nsubcap{\label{fig:ff4}}
\end{minipage}
\end{subfigure}
\caption{Panel (a) displays the weekly average of the daily maximum temperature anomalies ($T_{\max}'$), while panel (b) presents the weekly average of the daily average ozone anomalies ($O_3'$), both for the 30th ISO week of 2018 across the NUTS 3 regions in Europe.\label{fig:finalfeatures}}
\end{figure}

Table~\ref{tab:features} lists the final features which we use as inputs in the machine learning model. Alongside these features, we also introduce the lagged feature values using the lag operator $L^u(\cdot)$, e.g., $\smash{L^1(T_{\max}')}$ denotes last week's weekly average of the daily maximum temperature anomalies. Figure~\ref{fig:finalfeatures} presents two examples of weekly aggregated features at the NUTS 3 level during the 30th ISO week of 2018. Figure~\ref{fig:ff1} illustrates the weekly average of the daily maximum temperature anomalies ($T_{\max}'$), while Figure~\ref{fig:ff4} displays the weekly average of the daily average ozone anomalies ($O_3'$). We conclude from Figure~\ref{fig:ff1} that the 30th ISO week of 2018 exhibited significantly elevated temperatures in the Nordic countries and the northern regions of France, surpassing region-specific baseline maximum temperature levels by more than six degrees. Furthermore, the northern French and Belgian regions exhibited higher ozone levels compared to baseline concentrations, as Figure~\ref{fig:ff4} suggests.
\begin{table}[!htb]
  \centering
\adjustbox{max width=0.48\textwidth}{  \begin{tabular}{ll}
    \toprule
    \textbf{Feature} & \pbox{5.5cm}{\textbf{Weekly average of the daily}}\\
    \midrule
        $T_{\max}'$ & maximum temperature anomalies \\
        $T_{\min}'$ & minimum temperature anomalies \\
        $I_{\text{hot}}$ & hot-day indicator, see~(\ref{eq:Tind}) \\
        $I_{\text{cold}}$ & cold-day indicator \\
        $H'$  & average relative humidity anomalies \\
        $I_{\text{highH}}$  & high humidity indicator \\
        $I_{\text{lowH}}$  & low humidity indicator \\
        $P'$ & total precipitation levels \\
        $I_{\text{highP}}$  & high precipitation indicator \\
        $I_{\text{lowP}}$  & low precipitation indicator \\
        $W'$ & average wind speed anomalies \\ 
        $I_{\text{highW}}$  & high wind speed indicator \\
        $I_{\text{lowW}}$  & low wind speed indicator \\       
    \bottomrule
  \end{tabular}}
  \quad
\adjustbox{max width=0.48\textwidth}{\begin{tabular}{ll}
    \toprule
    \textbf{Feature} & \pbox{5.5cm}{\textbf{Weekly average of the daily}}\\
    \midrule
        $\text{O}_3'$  & average ozone anomalies \\
        $I_{\text{high}\text{O}_3}$  & high ozone indicator \\
        $I_{\text{low}\text{O}_3}$  & low ozone indicator \\
        $\text{NO}_{2}'$  & average nitrogen dioxide anomalies \\
        $I_{\text{high}\text{NO}_2}$  & high nitrogen dioxide indicator \\
        $I_{\text{low}\text{NO}_2}$  & low nitrogen dioxide indicator \\
        $\text{PM}_{10}'$  & average PM$_{10}$ anomalies \\
        $I_{\text{high}\text{PM}_{10}}$  & high PM$_{10}$ indicator \\
        $I_{\text{low}\text{PM}_{10}}$  & low PM$_{10}$ indicator \\
        $\text{PM}_{2.5}'$  & average PM$_{2.5}$ anomalies \\
        $I_{\text{high}\text{PM}_{2.5}}$  & high PM$_{2.5}$ indicator \\
        $I_{\text{low}\text{PM}_{2.5}}$  & low PM$_{2.5}$ indicator \\      
        \: & \: \vspace{-0.25cm}\\ 
    \bottomrule
  \end{tabular}}
  \caption{Weekly region-specific, environmental features at a NUTS 3 geographical level. The prime $'$ refers to anomalies, and $I$ refers to the indicator features.  \label{tab:features}
}
\end{table}

\subsection{Calibration} \label{subsec:calibration.casestudy}
We calibrate the baseline model for weekly death counts, as outlined in~\eqref{eq:baselinemodelspec} and~\eqref{eq:baselinedeathsstructure}, by minimizing the penalized Poisson negative log-likelihood, see Section~\ref{subsec:calibratingblm}. This penalization ensures a smooth variation of the fitted parameters $\smash{\beta_p^{(r)}}$ in the baseline model across neighboring regions, which is particularly useful for regions with a low number of weekly death counts. We consider the calibration period $\mathcal{T} = \{2013,2014,...,2019\}$. Suppl.~Mat.~\ref{subsubsec:baselinemodel.casestudy} provides further details and visualizations of the fitted parameters in the baseline model.

After the calibration of the baseline model for the weekly death counts, we focus on calibrating the XGBoost model as explained in Section~\ref{subsec:calibratingmlm}. This machine learning model incorporates the environmental features engineered in Section~\ref{subsec:featureengineering.casestudy} and listed in Table~\ref{tab:features}. Additionally, we incorporate the one-week lagged values of the features in Table~\ref{tab:features} and the seasonal indicator variables. We furthermore allow for spatial variations in the impact of specific features on the weekly deviations from the baseline death counts by incorporating the longitude ($\smash{\delta^{(r)}}$) and latitude ($\smash{\gamma^{(r)}}$) coordinates of the centre of each NUTS 3 region. This leads to a total of 56 features used as inputs in the machine learning model. Suppl.~Mat.~\ref{subsubsec:xgboostmodel.casestudy} details the tuning strategy and provides the results.

\subsection{Model fits, findings and discussion} \label{subsec:insights}
\subsubsection{In-sample fit and model performance} \label{subsubsec:insamplefit}
\begin{figure}[htb!]
\centering
\includegraphics[width=0.9\textwidth]{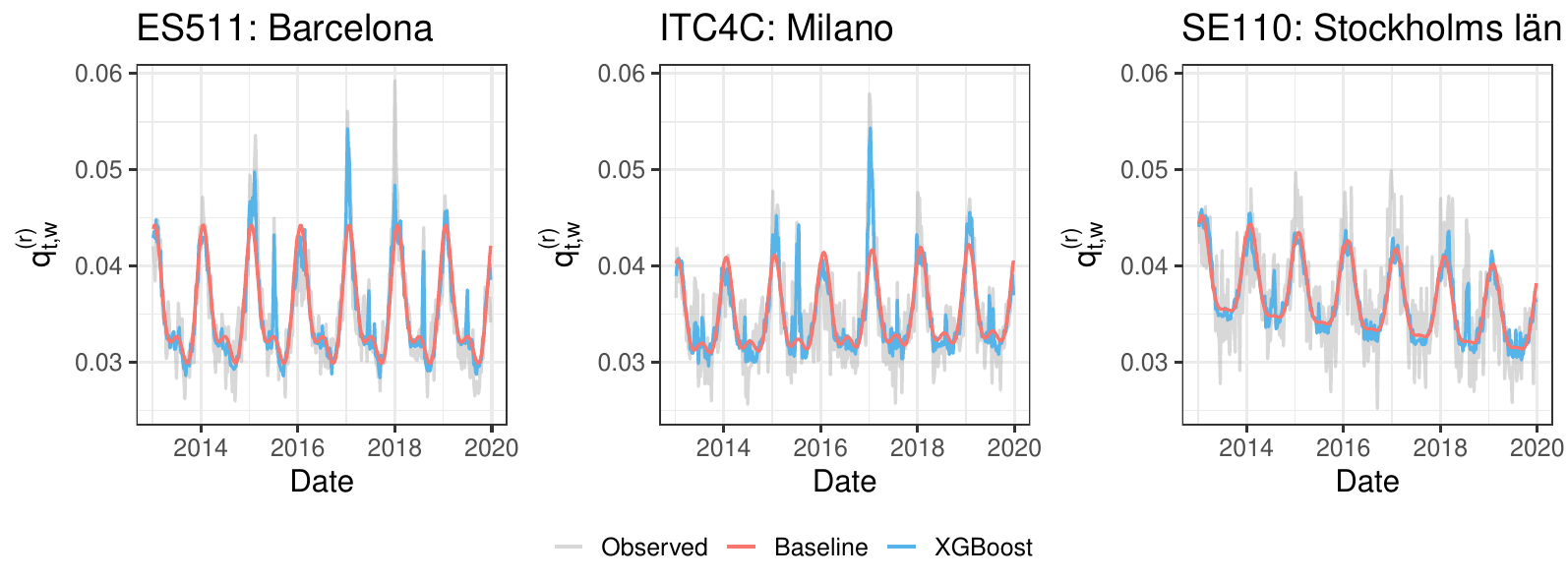}
\caption{The observed and estimated weekly mortality rates during the period 2013-2019 in the NUTS 3 regions of Barcelona (left), Milano (middle), and Stockholm (right). The observed weekly mortality rates are shown in gray and the mortality rates as obtained from the baseline model and the XGBoost model in red and blue, respectively. \label{fig:insampledtr}}
\end{figure}

Figure~\ref{fig:insampledtr} displays the observed mortality rates in gray, alongside the estimated mortality rates from both the baseline model (red) and the machine learning model integrating environmental features (blue). The comparison reveals that the machine learning model performs better in terms of an in-sample fit compared to the baseline model. In Barcelona and Milano, the proposed model effectively captures the excess mortality during the winters of 2015/2016 and 2017/2018, and the excess mortality observed during the summer of 2015. In Stockholm, the model identifies the excess mortality observed during the summer of 2014 and 2018. We show the residuals in Suppl.~Mat~\ref{app:residualsfit} and perform a statistical in-sample comparison in Suppl.~Mat.~\ref{appendix:in-sample-test}

\subsubsection{Interpretation tools}

\begin{figure}[ht!]
\centering
\includegraphics[width = 0.8\textwidth]{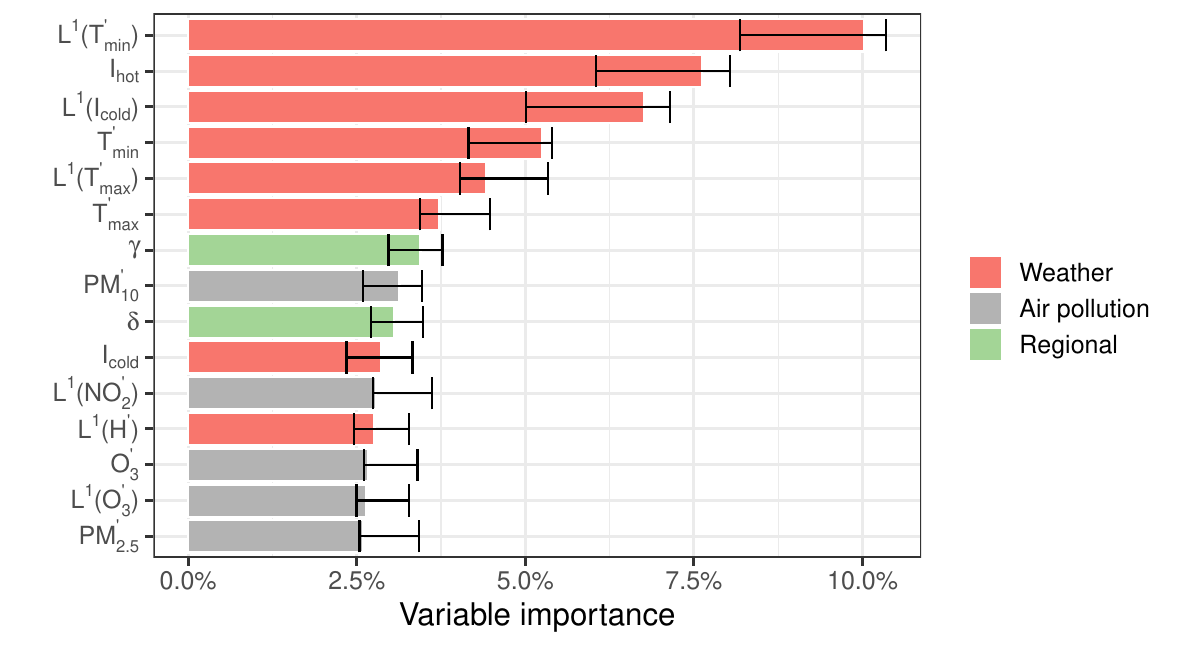}
\caption{The top 15 features with the highest feature importance. The feature importance of the features related to weather are visualized in red, air pollution in gray, and the centroid of the NUTS 3 region in green. Furthermore, the black error bars denote the 95$\%$ non-parametric bootstrap confidence interval for each feature based on $1\ 000$ bootstrap samples of the training data.  \label{fig:varimp}}
\end{figure}

\paragraph{Feature importance.} We analyze which features contribute most to the machine learning model that captures deviations in mortality from the baseline model. Hereto, we use the feature importance measure outlined in Section~\ref{sec:interpretationtools}. Figure~\ref{fig:varimp} presents the top 15 most important features, for which the description is given in Table~\ref{tab:features}. To illustrate the variability in these feature importance measures, we create 1\ 000 bootstrap copies of the same size as the training data.\footnote{To create a non-parametric bootstrap copy of the training data, we randomly sample observations with replacement from the training data.} We then recalibrate the machine learning model on each bootstrap copy and calculate the feature importance measure for each input feature. The black error bars in Figure~\ref{fig:varimp} show the resulting 95$\%$ bootstrap confidence intervals \citep{diciccio1996bootstrap}. For computational efficiency, we fit the machine learning model on each bootstrap sample using the tuning parameter configuration obtained in Section~\ref{subsec:calibration.casestudy}. The $x$-axis represents the feature importance in percentages, indicating the relative contribution of each feature to the predictive model. The six features with the highest feature importance are all associated with temperature. Additionally, based on the bootstrap confidence intervals, these features consistently exhibit the highest feature importance across the bootstrapped datasets. Among the top three features, we find the lagged weekly average of the daily minimum temperature anomalies ($L^1(T_{\min}')$), the weekly average of the hot-day indicator ($I_{\text{hot}}$), and the lagged weekly average of the cold-day indicator ($L^1(I_{\text{cold}})$). In contrast, features related to the wind speed, humidity, and precipitation demonstrate limited importance in the model. This is in line with earlier studies such as \cite{braga2002effect} and \cite{alberdi1998daily}.

\begin{figure}[!htb]
\centering
\begin{subfigure}{0.32\textwidth}
\centering
\includegraphics[width = \textwidth]{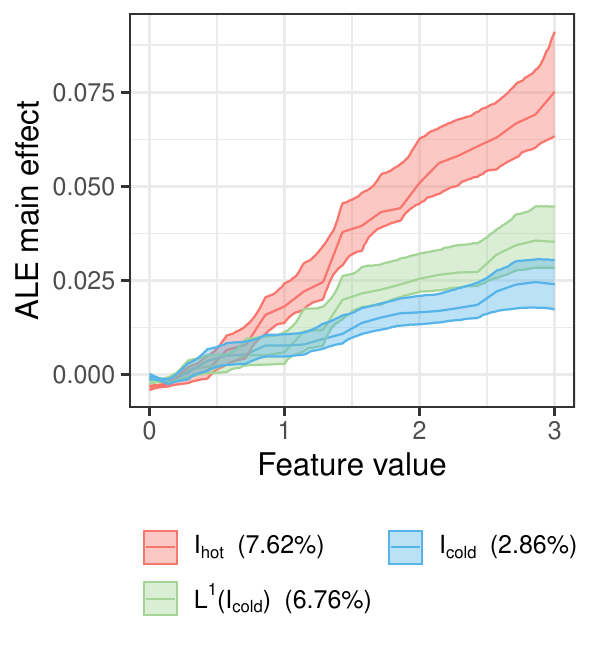}
\begin{minipage}{1.15\textwidth}
\vspace{0cm}
\nsubcap{\label{fig:alemain1}}
\end{minipage}
\end{subfigure}
\begin{subfigure}{0.32\textwidth}
\centering
\includegraphics[width = \textwidth]{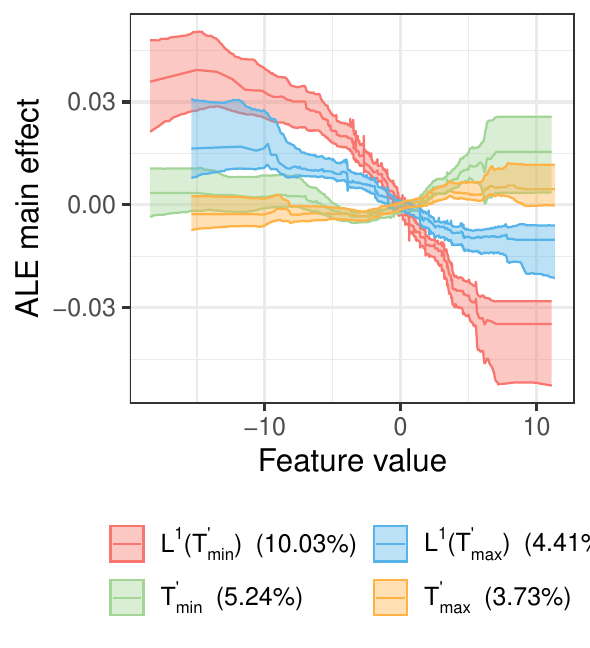}
\begin{minipage}{1.15\textwidth}
\vspace{0cm}
\nsubcap{\label{fig:alemain2}}
\end{minipage}
\end{subfigure}
\begin{subfigure}{0.32\textwidth}
\centering
\includegraphics[width = \textwidth]{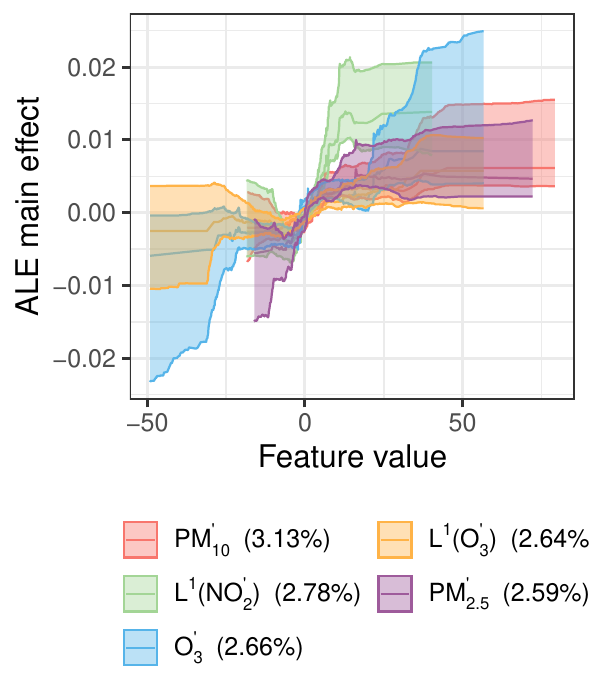}
\begin{minipage}{1.15\textwidth}
\vspace{-0.5cm}
\nsubcap{\label{fig:alemain4}}
\end{minipage}
\end{subfigure}
\caption{The ALE main effects for a selected subset of the top 15 most important features, where the feature importance is given within parentheses in each panel's legend. We present the 95$\%$ point-wise confidence interval based on 1000 bootstrap samples of the training data. The different panels display the ALE main effects of the extreme temperature indices (a), temperature anomalies (b), and air pollution anomalies (c).\label{fig:alemain}}
\end{figure}

\paragraph{ALE main effects.} Figure~\ref{fig:alemain} shows the ALE main effects for a few selected features, with the $95\%$ point-wise bootstrap confidence interval of the ALE main effect based on the fits constructed on each of the 1 000 bootstrap copies of the training data. Table~\ref{tab:features} gives a description of the considered features.

In Figure~\ref{fig:alemain1}, we observe an increasing trend in the ALE main effects corresponding to the three extreme temperature indices. We estimate that higher weekly averages of the daily hot or cold day indices, potentially lagged by one week, correspond to substantially higher excess deaths relative to the baseline. In accordance with the bootstrap-generated confidence intervals, these features have a statistically significant impact on the model's predictions. In Figure~\ref{fig:alemain2}, we show the ALE main effect for the four temperature anomalies. The effect is the strongest and most significant for the lagged weekly average of the daily minimum temperature anomalies ($\smash{L^1(T_{\min}')}$). When this value is higher, the ALE main effect is lower, leading to a mortality deficit relative to the baseline on average. This effect might vary substantially from season to season. However, such an interaction effect can not be captured through ALE main effects. For the (lagged) weekly averages of the daily air pollution anomalies in Figure~\ref{fig:alemain4}, we see, to a lesser extent, an increasing pattern in the ALE main effect. While the effect size is relatively small compared to the temperature effects, most of these ALE main effects are statistically significant, as their confidence intervals do not include the value zero in most cases. We conclude that the (lagged) hot- and cold-week index features have the most substantial impact on excess mortality.\footnote{The `hot-week index' is short for the weekly average of the daily hot-day indicator (feature $I_{\text{hot}}$). We use a similar definition for the cold-week index (feature $I_{\text{cold}}$).}

\begin{figure}[ht!]
\centering
\begin{subfigure}{0.3\textwidth}
\centering
\includegraphics[width = 0.85\textwidth]{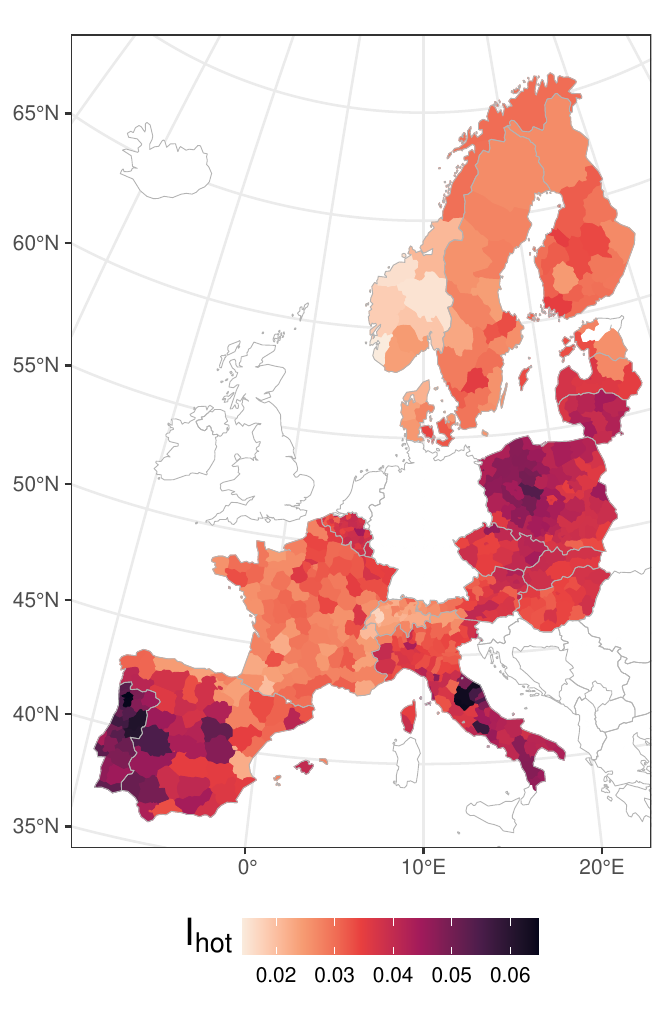}
\begin{minipage}{1.1\textwidth}
\vspace{-0.5cm}
\nsubcap{\label{fig:alereg1}}
\end{minipage}
\end{subfigure}
\hspace{0.1cm}
\begin{subfigure}{0.3\textwidth}
\centering
\includegraphics[width = 0.85\textwidth]{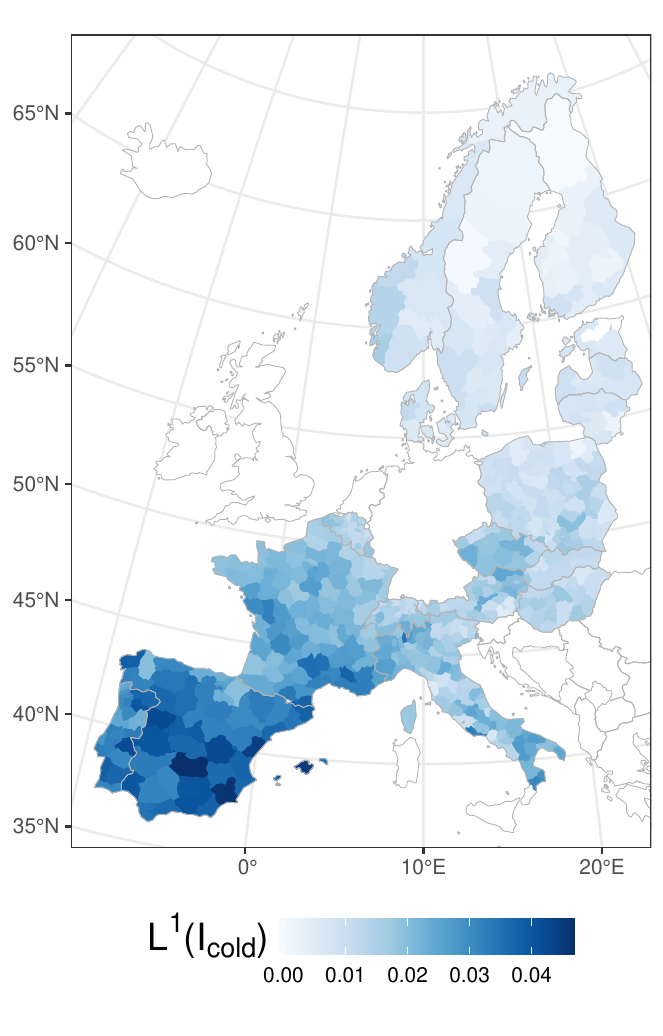}
\begin{minipage}{1.1\textwidth}
\vspace{-0.5cm}
\nsubcap{\label{fig:alereg2}}
\end{minipage}
\end{subfigure}
\hspace{0.1cm}
\begin{subfigure}{0.3\textwidth}
\centering
\includegraphics[width = 0.85\textwidth]{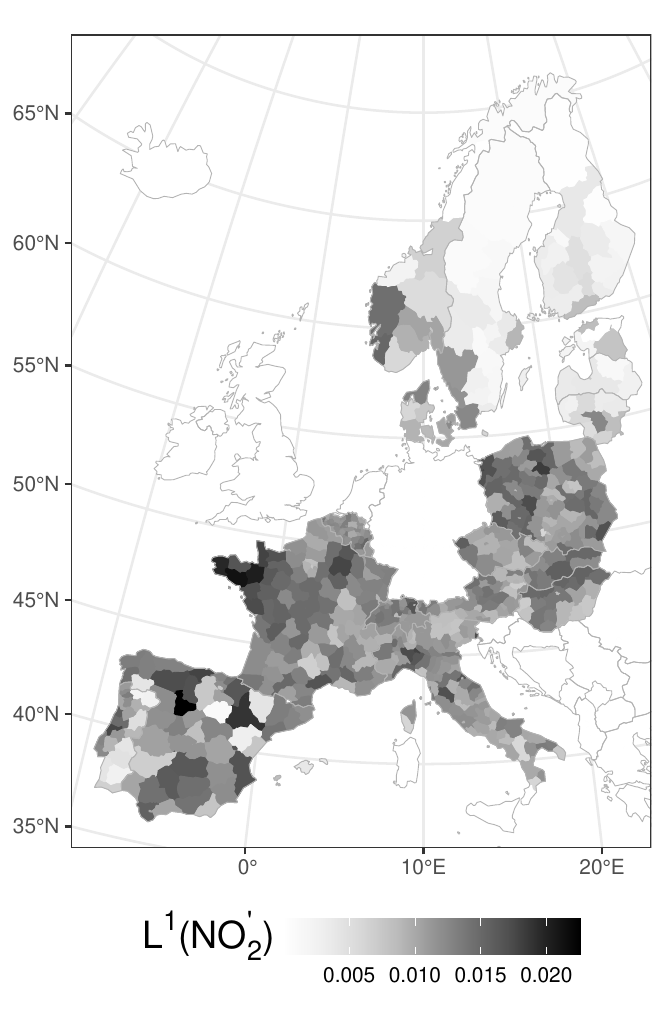}
\begin{minipage}{1.1\textwidth}
\vspace{-0.5cm}
\nsubcap{\label{fig:alereg3}}
\end{minipage}
\end{subfigure}
\caption{ALE regional effect for the hot-week index $I_{\text{hot}}$ (a), the lagged cold-week index $\smash{L^1(I_{\text{cold}})}$ (b) and the lagged weekly average of the daily nitrogen dioxide anomalies $\smash{L^1(\text{NO}_{2}')}$ (c). The ALE effects are fitted on the training data restricted to each region and visualized at the values 1.5, 1.5, and 20 $\mu g/m^3$ respectively.\label{fig:alereg}}
\end{figure}

\paragraph{ALE regional effects.} We examine regional differences in how environmental factors impact deviations from the mortality baseline. To achieve this, we re-compute the ALE main effect of a specific feature by restricting the training data to region $r$. Figures~\ref{fig:alereg1} and~\ref{fig:alereg2} visualize the results for the hot and lagged cold-week index respectively, at a value of 1.5, see~(\ref{eq:Tind}) in the Supplementary Material. Figure~\ref{fig:alereg3} shows the ALE regional effect of the feature related to the lagged weekly average of the daily NO$_2$ anomalies ($\smash{L^1(\text{NO}_2')}$), at a value of 20 $\mu g/m^3$. This corresponds to the situation wherein the daily NO$_2$ baseline concentrations are surpassed by 20 $\mu g/m^3$ on average throughout the previous week.  

We observe that the hot-week index exhibits the most pronounced effect in the southern NUTS 3 regions of Spain, Portugal, Italy, and certain Eastern European countries. Conversely, northern regions appear less impacted by high temperatures. The cold-week index follows a somewhat comparable pattern. In the right panel, no clear geographical pattern emerges for the weekly average of last week's daily nitrogen dioxide anomalies, except for nearly zero ALE values in northern NUTS 3 regions.

\begin{figure}[ht!]
\centering
\begin{subfigure}{0.3\textwidth}
\centering
\includegraphics[width = \textwidth]{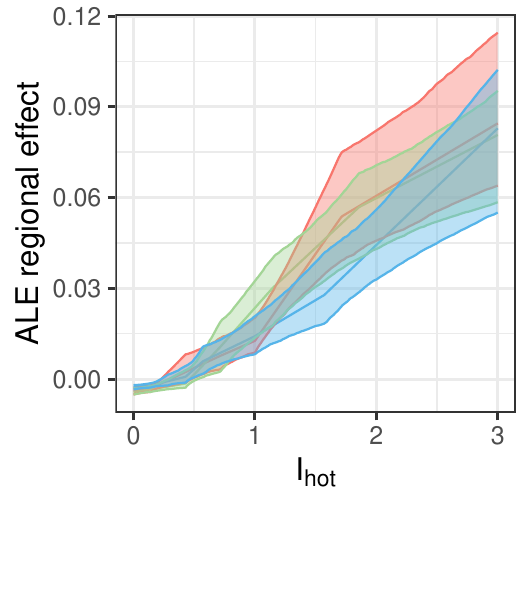}
\begin{minipage}{1.15\textwidth}
\vspace{-0.5cm}
\nsubcap{\label{fig:aleregboot1}}
\end{minipage}
\end{subfigure}
\hspace{0.1cm}
\begin{subfigure}{0.3\textwidth}
\centering
\includegraphics[width = \textwidth]{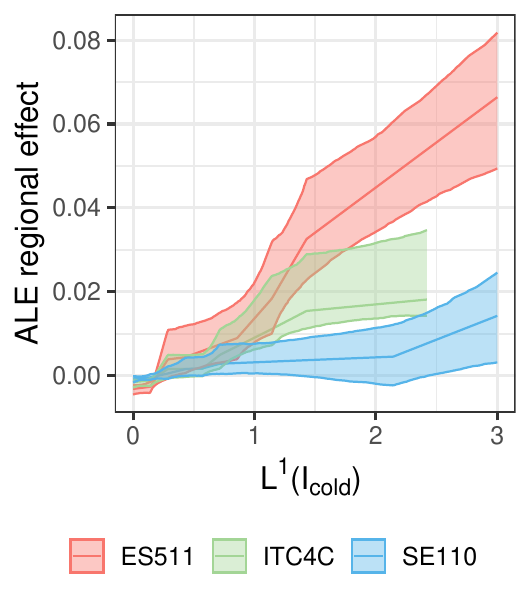}
\begin{minipage}{1.15\textwidth}
\vspace{-0.5cm}
\nsubcap{\label{fig:aleregboot2}}
\end{minipage}
\end{subfigure}
\hspace{0.1cm}
\begin{subfigure}{0.3\textwidth}
\centering
\includegraphics[width = \textwidth]{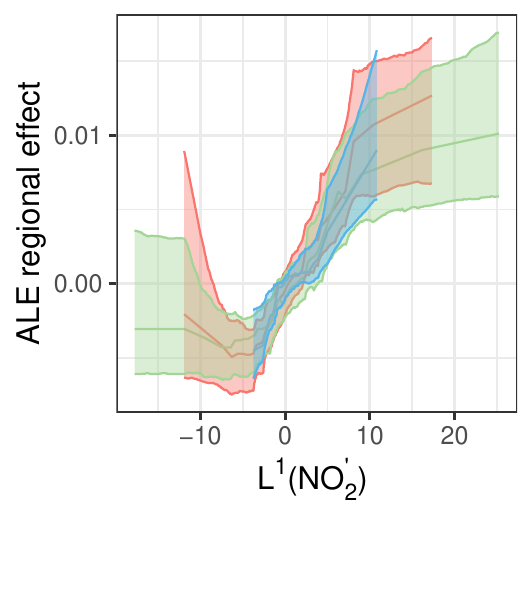}
\begin{minipage}{1.15\textwidth}
\vspace{-0.5cm}
\nsubcap{\label{fig:aleregboot3}}
\end{minipage}
\end{subfigure}
\caption{ALE regional effect in the NUTS 3 regions of Barcelona (red), Milano (green), and Stockholm (blue) for the hot-week index $I_{\text{hot}}$ (a), lagged cold-week index $\smash{L^1(I_{\text{cold}})}$ (b) and lagged weekly average of the daily nitrogen dioxide anomalies $\smash{L^1(\text{NO}_{2}')}$. The ALE effects are estimated on the training data restricted to region $r$. We show a point-wise confidence interval based on 1\ 000 bootstrap samples of the training data.\label{fig:aleregboot}}
\end{figure}

Figure~\ref{fig:aleregboot} illustrates the regional ALE effects for the NUTS 3 regions of Barcelona, Milano, and Stockholm for the same features shown in Figure~\ref{fig:alereg}. Figure~\ref{fig:aleregboot1} shows a similar marginal impact of the hot-week index on deviations from the mortality baseline across the three considered regions, with a slightly lower observed impact in Stockholm at values for the hot-week index between 1 and 2. This could explain the behaviour in Figure~\ref{fig:alereg1}. Figure~\ref{fig:aleregboot2} indicates that the marginal impact of the lagged cold-week index on mortality deviations from the baseline is clearly lower in Stockholm, and clearly higher in Barcelona. This implies that individuals in Stockholm may tolerate extreme cold temperatures better. For the lagged nitrogen dioxide anomalies in Figure~\ref{fig:aleregboot3}, we see a similar marginal impact across the three considered regions.

\begin{figure}[ht!]
\centering
\begin{subfigure}{0.3\textwidth}
\centering
\includegraphics[width = \textwidth]{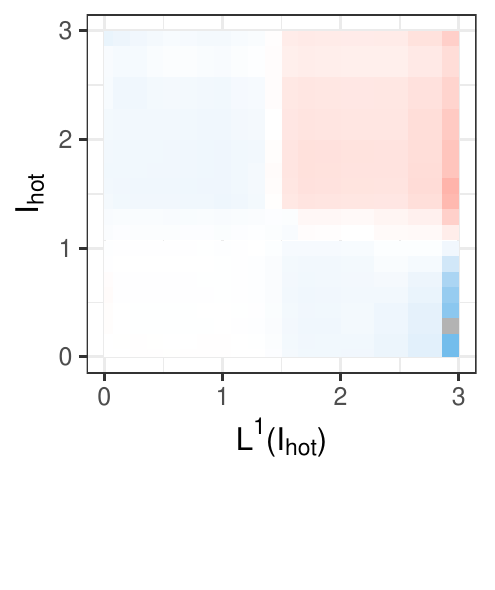}
\begin{minipage}{1.15\textwidth}
\vspace{-0.5cm}
\nsubcap{\label{fig:aleint1}}
\end{minipage}
\end{subfigure}
\begin{subfigure}{0.3\textwidth}
\centering
\includegraphics[width = \textwidth]{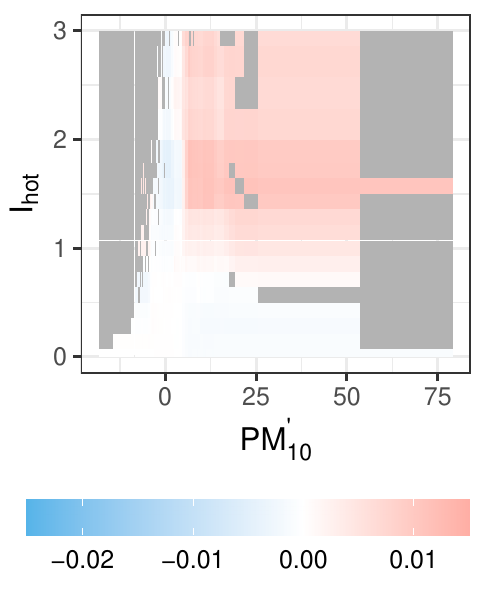}
\begin{minipage}{1.15\textwidth}
\vspace{-0.5cm}
\nsubcap{\label{fig:aleint2}}
\end{minipage}
\end{subfigure}
\begin{subfigure}{0.3\textwidth}
\centering
\includegraphics[width = \textwidth]{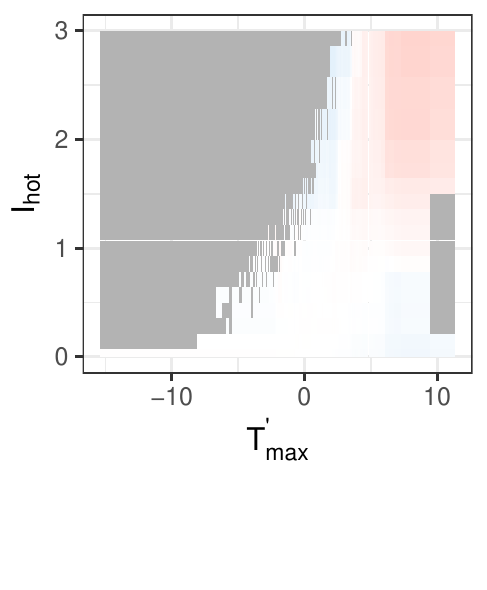}
\begin{minipage}{1.15\textwidth}
\vspace{-0.5cm}
\nsubcap{\label{fig:aleint3}}
\end{minipage}
\end{subfigure}
\caption{ALE interaction effects between the hot-week index $I_{\text{hot}}$ with three other features: the lagged hot-week index $\smash{L^1(I_{\text{hot}})}$ (a), the weekly average of the daily PM$_{10}$ anomalies $\text{PM}_{10}'$ (b), and the weekly average of the daily maximum temperature anomalies $T_{\max}'$ (c). The gray squares indicate that there is no data available. \label{fig:aleint}}
\end{figure}

\paragraph{ALE interaction effects.} Using the method outlined in Suppl.~Mat.~\ref{app:intertools}, we calculate the ALE interaction effect of the feature related to the hot-week index ($I_{\text{hot}}$) with three other features: the one-week lagged hot-week index ($\smash{L^1(I_{\text{hot}})}$), the weekly average of the daily PM$_{10}$ anomalies ($\text{PM}_{10}'$), and the weekly average of the daily maximum temperature anomalies ($T_{\max}'$). Figure~\ref{fig:aleint1} illustrates that elevated levels of the hot-week index for both the current and preceding week have a larger impact on excess mortality. However, when the previous week is hot while the current week is not, we observe the opposite effect. This is an indication of harvesting. Suppl.~Mat.~\ref{subsec:harvesting} further examines harvesting effects across various environmental factors, evaluating the XGBoost model's ability to capture these interactions. Moreover, Figure~\ref{fig:aleint2} shows that heightened PM$_{10}$ values in combination with a hot week lead to higher excess mortality. Lastly, Figure~\ref{fig:aleint3} indicates that when the hot-week index is high and maximum temperatures significantly exceed their baseline, a higher impact is observed. 

\subsection{Back-testing} \label{subsec:backtesting}
We back-test the proposed weekly mortality model by projecting the death counts in the European NUTS 3 regions for the year 2019. Hereto, we recalibrate the model on weekly mortality and environmental data for the years 2013 to 2018.

We first calibrate the weekly mortality baseline model, as presented in~\eqref{eq:baselinemodelspec} and~\eqref{eq:baselinedeathsstructure}, using the weekly death counts for the years 2013-2018. This calibration follows the strategy outlined in Section~\ref{subsec:calibratingblm}. For computational purposes, we adopt the same penalty parameter values as obtained during the calibration process of Section~\ref{subsec:calibration.casestudy} using death counts from 2013 to 2019. Nevertheless, we anticipate that this choice will only have a minor impact on the calibrated baseline death counts. We denote the resulting fitted baseline structure as $\smash{\hat{\mu}_{t,w}^{(r)}}$, for week $w$ in year $t$ and region $r$. Subsequently, we use this calibrated baseline model to forecast the death counts in the year 2019 as follows:
\begin{align*}
\hat{b}_{2019,w}^{(r)} = E_{2019,w} \cdot \text{exp}\left( \left( \hat{\boldsymbol{\beta}}^{(r)}\right)^T z_{2019,w}\right),
\end{align*}
with $\hat{\boldsymbol{\beta}}$ the estimated parameter vector from the 2013-2018 calibration and $z_{2019,w}$ the vector consisting of the covariates in the baseline model evaluated in the year 2019, see Section~\ref{subsec:calibratingblm}.

Next, we continue with the environmental data spanning the years 2013 to 2018, as outlined in Section~\ref{subsec:data}. We create environmental anomalies and extreme environmental indices following the feature engineering process detailed in Section~\ref{subsec:featureengineering.casestudy}. These engineered features serve as input in the machine learning model, enabling us to estimate the excess or deficit deaths relative to the calibrated baseline for the aforementioned years. During the calibration phase of the machine learning model, we solely focus on the tuning parameters related to the maximum tree depth and the number of trees and maintain the remaining parameters at the values determined in Section~\ref{subsec:calibration.casestudy}. Using the calibrated environmental baseline models established from data spanning 2013 to 2018, we then proceed with the feature engineering process using the registered observations for the year 2019. The resulting 2019 environmental features are then used as inputs for the calibrated machine learning model to forecast death counts across the European NUTS 3 regions in the year 2019.

\begin{figure}[htb!]
\centering
\begin{subfigure}{0.32\textwidth}
\centering
\includegraphics[width=\textwidth]{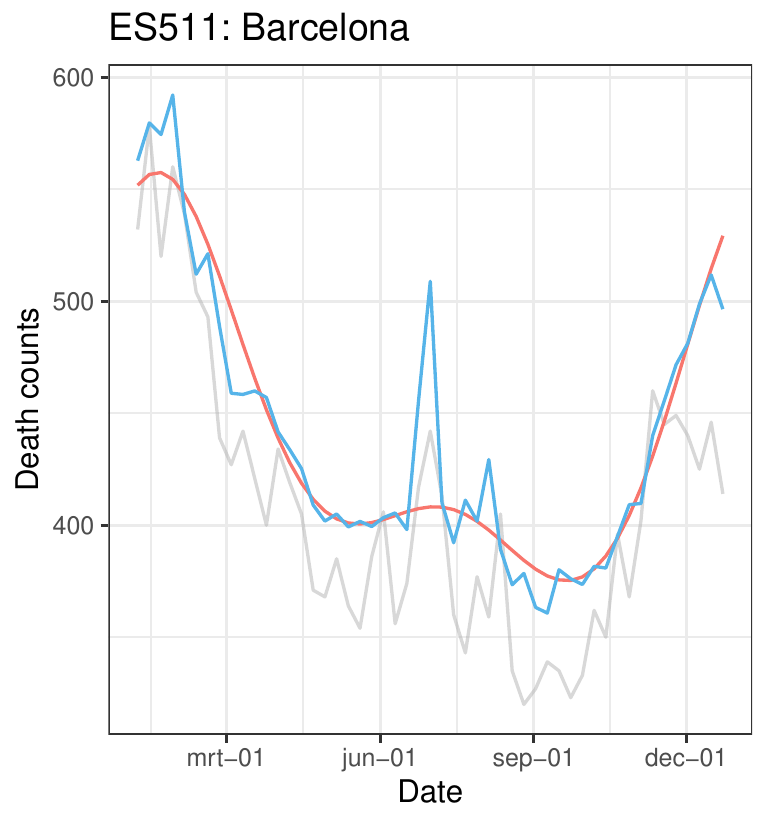}
\begin{minipage}{1.15\textwidth}
\vspace{-0.5cm}
\nsubcap{\label{fig:projectdt1}}
\end{minipage}
\end{subfigure}
\begin{subfigure}{0.32\textwidth}
\centering
\includegraphics[width=\textwidth]{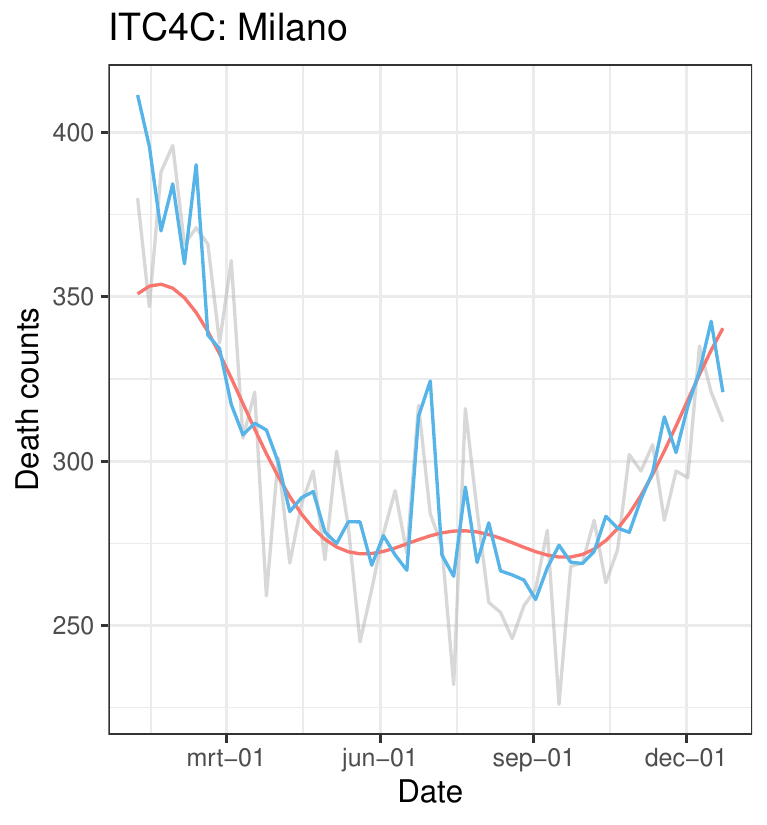}
\begin{minipage}{1.15\textwidth}
\vspace{-0.5cm}
\nsubcap{\label{fig:projectdt2}}
\end{minipage}
\end{subfigure}
\begin{subfigure}{0.32\textwidth}
\centering
\includegraphics[width=\textwidth]{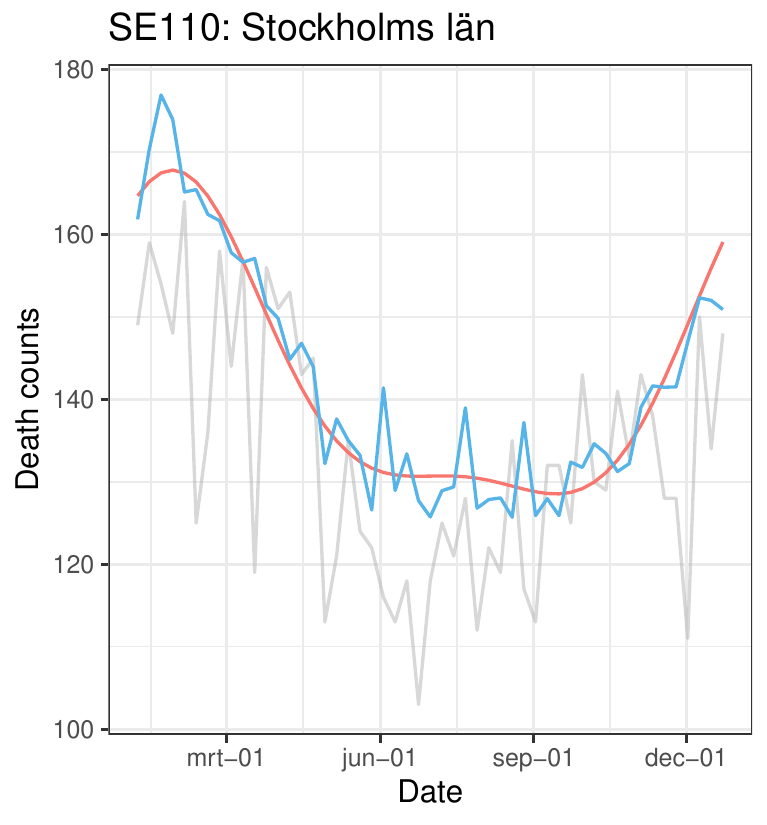}
\begin{minipage}{1.15\textwidth}
\vspace{-0.5cm}
\nsubcap{\label{fig:projectdt3}}
\end{minipage}
\end{subfigure}
\caption{The observed and projected death counts for the year 2019 in the NUTS 3 regions of Barcelona (a), Milano (b), and Stockholm (c). The observed death counts are visualized in gray and the death counts as obtained from the weekly baseline mortality model and the XGBoost model in red and blue, respectively.  \label{fig:projectdt}}
\end{figure}

Figure~\ref{fig:projectdt} illustrates the outcomes for the regions of Barcelona, Milano, and Stockholm. The results indicate a clear advantage of the machine learning model (blue line) over the baseline model (red line) in Milano, see Figure~\ref{fig:projectdt2}. Specifically, the XGBoost model correctly predicts higher-than-baseline deaths in the first weeks of 2019 and during the summer of 2019 in Milano. However, in Barcelona and Stockholm, in Figures~\ref{fig:projectdt1}, \ref{fig:projectdt3}, we find that the calibrated baseline model consistently overestimates the observed deaths in 2019 and henceforth fails to accurately capture the region-specific seasonal mortality trend. Nevertheless, the machine learning model fulfils its task as the pattern and spikes, present in the observed and estimated death counts from the XGBoost model, largely coincide. This is particularly true for Barcelona, while in the region around Stockholm, it is less pronounced. This aligns with our earlier observation that the northern regions appear to be less impacted by extreme environmental levels, see Figure~\ref{fig:alereg} for example. Since the back-test relies on data from only a single year, we refrain from drawing further conclusions. 

To compare the performance of our weekly mortality model, integrating environmental features, across all considered European NUTS 3 regions, Suppl.~Mat.~\ref{appsubsec:back-test} displays the relative change in the Poisson deviance of the XGBoost model relative to the one of the weekly baseline mortality model. We find that, based on the one year-prediction results, adding environmental features is beneficial in about 55$\%$ of the NUTS 3 regions. Northern regions seem to experience little benefit from the integration of environmental features, while southern regions generally benefit more. 

\section{Conclusions} \label{sec:conclusions}
This paper proposes a comprehensive framework for integrating environmental features, related to weather and air pollution factors, into a weekly mortality model using fine-grained, open source data. Building upon the established weekly Serfling-type mortality model, we enhance this so-called weekly baseline model by incorporating environmental anomalies and extreme environmental indices through a machine learning approach to explain short-term deviations from the mortality baseline. 

Through a case study spanning over 550 NUTS 3 regions in 20 European countries among individuals aged 65 and older, we find clear short-term associations between temperature anomalies and extreme indices and deviations from the mortality baseline. While some air pollution features also contribute significantly, other weather-related variables such as humidity, rainfall, and wind speed are only of minor importance. To address the correlation among the environmental features, we employ Accumulated Local Effect plots to investigate the short-term marginal impact of the environmental features as well as their interactions on deviations from the mortality baseline. We hereby find that the hot-week index, indicating the frequency of days surpassing the 95$\%$ temperature threshold within a week, exhibits the most pronounced effect on excess deaths. Moreover, regional disparities emerge, with southern regions estimated to have higher excess deaths associated with both hot and cold week indices compared to northern regions. Our analysis further unveils a harvesting effect for the hot-week index, suggesting a mortality deficit following hot preceding weeks. Additionally, through back-testing, we validate the model's ability to explain excess deaths relative to the baseline in an out-of-sample year.

The proposed modeling framework can provide a valuable tool for evaluating the impact of specific environmental scenarios on mortality outcomes. By generating region-specific environmental scenarios for temperature, weather factors, and air pollution concentrations, we can assess the associated risks of excess deaths across different regions. Through feature engineering and utilization of machine learning techniques like XGBoost, we will then be able to estimate risk ratios, representing multipliers that augment or diminish the baseline number of deaths. Such findings can inform policymakers about the potential health implications of environmental changes, guiding proactive measures to mitigate risks and protect vulnerable populations. However, when estimating mortality risk using environmental scenario projections, incorporating the benefits and costs of climate change adaptation becomes an additional challenge, as demonstrated by \cite{carleton2022valuing}.

In future research, it will be valuable to conduct a similar study on cause-of-death mortality data to investigate how environmental factors influence excess deaths across various causes. Additionally, we can explore the robustness of our proposed weekly mortality model by exploring alternative weekly mortality models as baselines, as suggested by \cite{Scholey2021robustness} and by using weather-station data instead of fine-grained gridded data from the Copernicus Climate Data Store and the Copernicus Atmospheric Monitoring Service.

\section*{Data and code availability statement} 
The datasets used in this paper are publicly available. We consulted the deaths by week, sex, 5-year age group and NUTS 3 region on Eurostat (\url{https://ec.europa.eu/eurostat/databrowser/view/demo_r_mweek3/}, downloaded on January 20, 2024). We retrieved the weather-related variables from the Copernicus Climate Data Store (\url{https://cds.climate.copernicus.eu/}, downloaded on January 22, 2024) and the air pollution variables from the Copernicus atmospheric monitoring service (\url{https://ads.atmosphere.copernicus.eu/}, downloaded on January 22, 2024). The code used for the implementation and analysis of the case study presented in this paper is available on the following GitHub repository (\url{https://github.com/jensrobben/EnvVar-Mortality-Europe}). The code is written in \texttt{R} and can be accessed and downloaded for further reference and replication of the obtained results.

\section*{Funding statement} 
Katrien Antonio acknowledges funding from the FWO and F.R.S.-FNRS under the Excellence of Science (EOS) programme, project ASTeRISK (40007517). The FWO WOG network W001021N is acknowledged as well. Katrien Antonio acknowledges support from the Chair of Excellence on Digital Insurance And Long‑term risk (DIALog) by CNP Assurances. This study is furthermore part of the research programme at the Research Centre for Longevity Risk. RCLR is a joint initiative of NN Group and the University of Amsterdam, with additional funding from the Dutch government’s Public-Private Partnership (PPP) programme.

\section*{Conflict of interest disclosure} 
The authors declare no conflict of interest.

\newpage

\spacingset{1.5}

\appendix

\begin{center}
\title{\bf \LARGE Supplementary material for “The short-term association between environmental variables and mortality: evidence from Europe”}
\maketitle
\end{center}
\spacingset{1}

\renewcommand\thefigure{\thesection.\arabic{figure}}
\renewcommand\thetable{\thesection.\arabic{table}}  
\setcounter{figure}{0}
\setcounter{table}{0}
\section[Methodological approaches in epidemiological and medical literature]{Methodological approaches in the epidemiological and medical literature: a literature review} \label{appendix:literaryreview}
In exploring the relationship between environmental factors and mortality statistics, various methodologies have been proposed in the epidemiological and medical literature. 

\paragraph{Minimum mortality temperature bands.} \cite{keatinge2000heat} conduct an observational study to investigate heat-related mortalities across Europe. Hereto, they compute the daily death rates $m_{65{\text -}74,t}^{(r)}$ of individuals aged 65-74 at time $t$ per million population for several regions $r$ including North Finland, South Finland, Baden-Württemberg, Netherlands, London, North Italy, and Athens from 1988 to 1992. They then retrieve region-specific, daily mean temperatures from the Royal Meteorological Office and calculate the average death rate per million population across successive temperature bands of $3^{\circ}$C, each separated by $0.1^{\circ}$C. Hereto, let $T^{\circ}$ be any temperature in degrees Celsius and define the set of time points at which the temperature $T^{\circ}$ is exceeded in region $r$:
\begin{align*}
\mathcal{S}_r(T^{\circ}) = \left\{ t \mid T ^{\circ}\leq \texttt{Tavg}_{t}^{(r)} < T^{\circ} + 3^{\circ}\text{C} \right\},
\end{align*}
with $\texttt{Tavg}_{t}^{(r)}$ the region-specific daily mean temperature at time $t$ (in daily date format). The average region-specific death rate per million population in temperature band $[T^{\circ},T^{\circ}+3^{\circ}\text{C}]$ is then defined as:
\begin{align*}
\bar{m}^{(r,T^{\circ})}_{65{\text -}74} =  \dfrac{1}{|\mathcal{S}_r(T^{\circ})|} \displaystyle \sum_{t \in \mathcal{S}(T^{\circ})} m_{65{\text -}74,t}^{(r)}.
\end{align*}
Subsequently, they identify the temperature band with the lowest death rate in each region, i.e., $[T^{(\star,r)}, T^{(\star,r)} + 3^{\circ}\text{C}]$, also called the region-specific minimum mortality temperature (band), and find that these bands are significantly higher in regions with hot summer temperatures, e.g., Athens with $[22.7^{\circ}\text{C},25.7^{\circ}\text{C}]$, compared to regions with cold summer temperatures, e.g., north Finland with $[14.3^{\circ}\text{C},17.3^{\circ}\text{C}]$. They then calculate the annual heat-related mortality in each region $r$ as the average number of days the minimum mortality temperature band was exceeded in a year, multiplied by the mean difference of the daily death rates per million population at temperatures above the minimum mortality temperature band with the daily death rate per million population within this band. These differences are calculated as:
$$m_{65{\text -}74,t}^{(r)} -  \bar{m}^{(r,T^{\circ})}_{65{\text -}74},$$
for time points $t$ at which the average temperature $\texttt{Tavg}_t^{(r)}$ exceeds $T^{(\star,r)} + 3^{\circ}\text{C}$. They find that regions with hot summers did not exhibit significantly higher annual heat-related mortality compared to colder regions. They conclude that European populations have effectively acclimated to average summer temperatures, and can be expected to adapt to the predicted global warming, see \cite{hulme1998climate}, with only a small increase in heat-related mortality. However, such an approach typically overlooks the non-linear relationship between temperature and the (daily) mortality statistics.
  
\paragraph{Time series models.} Yet another strand in the literature relies on time series regression models. Here, researchers typically assume an overdispersed Poisson distribution for the daily death counts, with as an exploratory variable the observed daily temperature values or daily measurements of a particular air pollutant, alongside other time-varying confounding factors \citep{armstrong2006models}. As an example, the structure of such a time series regression model, with the daily average temperature $\texttt{Tavg}$ as the explanatory variable of interest, is typically represented as:
\begin{equation} \label{eq:tsreg}
\begin{aligned}
\log \mathbb{E}\left[D_{t}^{(r)}\right] = \alpha^{(r)} + f\left(\texttt{Tavg}_{t}^{(r)} \mid \boldsymbol{\beta}^{(r)} \right)& + (\texttt{confounding covariates})_{t}^{(r)} + \\ &\hspace{-0.5cm}(\texttt{smooth function of time})_{t}^{(r)}, 
\end{aligned}
\end{equation}
where $D_{t}^{(r)}$ is the death count random variable from a specific region $r$ at time $t$ and $\alpha$ denotes the intercept term.  Furthermore $\smash{f^{(r)}\left(\texttt{Tavg}_{t}^{(r)} \mid \boldsymbol{\beta} \right)}$ denotes the function capturing the relationship between the region-specific daily average temperature and the daily death counts at time $t$ and region $r$, also known as the temperature-mortality association, parametrized through a vector $\boldsymbol{\beta}$. Examples of confounding factors include daily measurements of air pollution or other weather-related covariates, such as humidity, rainfall, and wind speed, that may influence the daily death counts. The smooth function of time is incorporated into the model to accommodate for seasonal effects and demographic shifts \citep{armstrong2006models}. While earlier studies such as \cite{group1997cold} and \cite{lovett1986analysing} favoured Poisson regression models with a linear functional form for modeling the temperature-mortality association, later research by \cite{braga2002effect} suggests a more nuanced U-, V-, or J-shaped association between daily death counts and daily temperature values, necessitating a non-linear functional form. Hereto, \cite{armstrong2006models} mentions that natural cubic splines are typically used for modeling the temperature-mortality association.

Studies by \cite{braga2001time} and \cite{pattenden2003mortality} reveal that the impact of a cold spell on mortality may persist for a week or longer, while the effects of heat are more immediate. Additionally, \cite{verhoeff1996air} applies time series Poisson regression models to find a positive correlation between ozone levels and daily mortality in Amsterdam, with effects persisting for up to two days. Ozone is a significant contributor to smog at ground level and can harm the respiratory system \citep{mudway2000ozone}. \cite{sunyer1996air} show similar findings in Barcelona for other air pollutants such as sulphur and nitrogen dioxide. Sulphur dioxide mainly originates from fossil fuel burnings and irritates the respiratory tract and eyes, posing heightened risks to asthmatic patients, while nitrogen dioxide is a toxic gas that induces respiratory issues and can lead to a reduced lung function \citep{devalia1994effect}. These studies indicate that the effects of a particular event, such as a cold spell, heat wave, or elevated air pollution levels, may extend beyond the period in which it occurs, and possibly exhibit delayed impacts over time. 

\paragraph{Distributed lag models.} To adequately address such delayed effects, researchers have proposed distributed lag models (DLMs). A DLM is a linear, additive regression model for time series data that predicts the current values of a dependent variable using both current and past (lagged) values of an explanatory variable. \cite{almon1965distributed} initially proposes the DLM to model quarterly capital expenditures in manufacturing industries, but the framework has been applied in other fields such as epidemiology, as highlighted by \cite{pope1996time}. \cite{schwartz2000distributed} further refines the DLM and applies it to a case study to estimate a biologically plausible lag structure between daily air pollution levels and daily deaths across ten cities in the United States (US). Additionally, \cite{braga2001time} examine the delayed impact of temperature and humidity on overall daily mortality in 12 US cities. In scenarios where the daily average temperature serves as the explanatory variable of interest, the DLM's model specification is similar to the time series regression model in~\eqref{eq:tsreg}, but with the following lagged functional form to model the temperature-mortality association:
\begin{equation} \label{eq:dlm}
\begin{aligned}
f\left(\texttt{Tavg}_{t}^{(r)} \mid \boldsymbol{\beta}^{(r)} \right) = \displaystyle \sum_{l=0}^{L} \beta_l^{(r)} \cdot \texttt{Tavg}^{(r)}_{t-l},
\end{aligned}
\end{equation}
where $L$ is the maximum number of lags considered. To avoid collinearity issues, \cite{almon1965distributed} and \cite{schwartz2000distributed} suggest to constrain the parameters corresponding to the different lags. More specifically, they assume that the $\smash{\beta_l^{(r)}}$'s, for $l=0,1,..,L$, follow a polynomial function with parameters $\smash{\eta_k^{(r)}}$, for $k=0,1,..,v_L$:
\begin{align} \label{eq:constrdlm}
\beta_l^{(r)} = \displaystyle \sum_{k=0}^{v_L} \eta_k^{(r)} l^k.
\end{align}
Hence, the $\smash{\beta_l^{(r)}}$'s are modelled as smooth functions using polynomial basis functions of the lag dimension. 

To enhance modeling flexibility, \cite{zanobetti2000generalized} propose a comprehensive framework by combining DLMs with generalized additive models. Expanding on our example where the daily average temperature is used as explanatory variable of interest, they generalize the time series regression model in~\eqref{eq:tsreg} using the DLM structure in~\eqref{eq:dlm} to:
\begin{equation} \label{eq:gadlm}
\begin{aligned}
\log \mathbb{E}\left[D_{t}^{(r)}\right] = \alpha^{(r)} + \left(\boldsymbol{\gamma}^{(r)}\right)^T \boldsymbol{z}_t^{(r)} + \displaystyle \sum_{i=1} ^d g_i\left(s_{i,t}^{(r)}\right) + \displaystyle \sum_{l=0}^L \beta_l^{(r)} \cdot \texttt{Tavg}^{(r)}_{t-l},
\end{aligned}
\end{equation}
where $\boldsymbol{z}_t^{(r)}$ denotes a set of region-specific confounding covariates modelled linearly through the region-specific parameter vector $\smash{\boldsymbol{\gamma}^{(r)}}$, and $\smash{g_i(s_{i,t}^{(r)})}$'s denote smooth functions of time or of confounding covariates in region $r$. The parameter vector $\smash{\boldsymbol{\beta}^{(r)}}$ follows the polynomial function in~\eqref{eq:constrdlm} to ensure smoothness along the lag dimension. \cite{zanobetti2000generalized} demonstrate the utility of their approach by modeling the association between lagged daily air pollution levels and daily death counts in Milano. Due to the incorporation of lagged effects, DLMs are valuable in identifying harvesting effects, where particular events, such as extreme temperatures, disproportionately impact vulnerable individuals. This precipitates events like deaths, leading to short-term higher mortality rates relative to the baseline \citep{schwartz2001there}. Afterwards, the mortality rates are lower than expected. This phenomenon is known as harvesting \citep{zanobetti2000generalized}. In conclusion, the main advantage of DLMs lies in their ability to incorporate a detailed representation of the time-course of the temperature-mortality or air pollution-mortality relationship and provide information about the overall effect even in the presence of delayed contributions or harvesting. 

\paragraph{Distributed lag non-linear models.} While the extension of \cite{zanobetti2000generalized} provides a more flexible approach, DLMs can only be used to model the lag structure of linear effects, see~\eqref{eq:gadlm}. As such, their proposed model structure exhibits limitations in representing the lag structure of non-linear relationships. Therefore, \cite{gasparrini2010distributed} introduce Distributed Lag Non-linear Models (DLNMs), a versatile family of models capable of describing non-linear effects along the predictor space and the lag dimension. Here, we select a set of $v_E+1$ basis functions for the explanatory variable of interest, denoted as $b_j(\cdot)$, for $j = 0,1,...,v_E$. Examples include polynomials or spline functions. The DLNM specification is similar to the generalized additive DLM shown in~\eqref{eq:gadlm}. However, in this model, the lagged functional form of the explanatory variable is expressed as \citep{gasparrini2010distributed}:
\begin{align} \label{eq:dnlm}
f\left(\texttt{Tavg}_{t}^{(r)} \mid \boldsymbol{\beta}^{(r)} \right) = \displaystyle \sum_{l=0}^L \sum_{j=0}^{v_E} \beta_{j,l}^{(r)} \cdot b_j(\texttt{Tavg}_{t-l}^{(r)} ).
\end{align} 
To address collinearity issues, similar constraints as in~\eqref{eq:constrdlm} can be imposed:
\begin{align} \label{eq:constrdnlm}
\beta_{j,l}^{(r)} = \displaystyle \sum_{k=0}^{v_L} \eta_{j,k}^{(r)} l^k,
\end{align}
for all $j=0,1,..,v_E$ and $l=0,1,..,L$. By utilizing the $v_E+1$ basis functions for the explanatory variable, they ensure a smooth, non-linear relationship between temperature and daily death counts. Meanwhile, the constraints on $\smash{\beta_{j,l}^{(r)}}$ using $v_L+1$ polynomial basis functions guarantee a smooth delayed impact across consecutive lags. For interpretation purposes, we presented here a somewhat simplified representation of the DLNM. \cite{gasparrini2010distributed} extend the DLNM structure from~\eqref{eq:dnlm} by introducing a general set of basis functions for the lag dimension, rather than the polynomial basis functions used in~\eqref{eq:constrdnlm}.

The application of DLNMs is widespread in the epidemiological literature, particularly to study the health impacts of air pollution and weather factors across multiple locations. In this setting, researchers typically adopt a two-stage analytical design. In the first stage, location-specific temperature-mortality or air pollution-mortality associations are estimated using time series regression models such as DLMs or DLNMs. These estimated location-specific relationships are entirely described by the basis functions $b_{j,l}(\cdot)$, the lag dimension, and the estimated parameters $\hat{\boldsymbol{\eta}}^{(r)} \in \mathbb{R}^{(v_E+1)\times (v_L+1)}$ in the DL(N)M structure, see e.g., \eqref{eq:dnlm} and~\eqref{eq:constrdnlm}. The estimated parameters then form the response for a (multivariate) meta-analytical modeling technique in the second stage \citep{gasparrini2012multivariate, gasparrini2013reducing}. As such, meta-regression, a particular meta-analytical technique, groups and synthesizes results from different locations obtained from the first stage location-specific regression models while adjusting for region-specific covariate effects. More specifically, the estimated parameters from the DLNM of location $r$ are grouped into a column vector of size $(v_E+1)\times (v_L+1)$, denoted as $\hat{\boldsymbol{\eta}}^{(r)}$. Such a multivariate meta-regression model is defined as:
\begin{align*}
\hat{\boldsymbol{\eta}}^{(r)}  \sim \mathcal{N}\left(\boldsymbol{U}^{(r)} \boldsymbol{\zeta} , \boldsymbol{S}^{(r)} + \boldsymbol{\Psi}\right),
\end{align*}
where $\boldsymbol{U}^{(r)}$ encompasses region-specific covariate effects, referred to as meta-variables. The matrices $\boldsymbol{S}^{(r)}$ and $\boldsymbol{\Psi}$ represent the within and between-location covariance matrices, respectively. The parameter vector $\boldsymbol{\zeta}$ describes the association between the meta-variables and the estimated parameters from the DLNM structure. Employing this two-stage approach, \cite{gasparrini2015mortality} conduct a systematic assessment of the impact of temperature on mortality across various countries in the world, revealing that on average $7.71\%$ of the total daily death counts can be attributed to both heat and cold, see \cite{gasparrini2014attributable} for the concept of attributable risk in the context of distributed lag models. Moreover, they find that cold temperatures exert a more significant impact on mortality than heat. Additionally, \cite{gasparrini2011impact} use a DLNM combined with meta-regression to examine the impact of heat waves on daily death counts. They find that the excess risk associated with heat waves in the United States could be largely attributed to the single, independent effects of daily high temperatures, with a slight additional effect observed in heat waves lasting more than four days. However, a multi-country, multi-community investigation using DLNMs and meta-analysis by \cite{guo2017heat} reveals significant cumulative associations between heat waves and mortality across all countries, although the significance varies by community. Interestingly, they find that heat waves exhibit stronger associations with mortality in areas experiencing moderate cold and moderate hot temperatures compared to cold and hot areas.

\setcounter{figure}{0}
\setcounter{table}{0}
\section{Construction of a weekly exposure measure} \label{app:weekexpo}
We retrieve the population count of people aged 65+ on January 1 of each year $t$ by sex and NUTS 3 region from Eurostat.\footnote{The database can be accessed on the web page: \url{https://ec.europa.eu/eurostat/databrowser/view/demo_r_pjanaggr3/default/table?lang=en}.} We construct an estimate for the weekly exposure  as \citep{stmfnote}:
\begin{align} \label{eq:expo}
E_{t,w}^{(r)} = \dfrac{P_{t}^{(r)} + P_{t+1}^{(r)}}{2\cdot 52.18},
\end{align}
for $t\in\{2013,..,2019\}$, with $P_{t}^{(r)}$ the population count of people aged 65+ at January, $1$ of year $t$ in region $r\in\mathcal{R}$. The number $52.18$ represents the average number of weeks per year. While the current approach assumes a constant weekly exposure within a year, a more advanced method could employ linear interpolation for a smoother transition of weekly exposures between consecutive years.

\setcounter{figure}{0}
\setcounter{table}{0}
\section{Weekly mortality baseline model: specification of the penalty matrix} \label{app:choicepenaltymatrix}
Consider the following specification of the penalty matrix $\boldsymbol{S} = (s_{ij})_{i,j\in \mathcal{R}}$:
\begin{equation}\label{eqA:penmat}
s_{ij} = \begin{cases}
|\mathcal{N}_i| & \:\:\:\text{if} \:\: i=j \\
-1 & \:\:\: \text{if $i\neq j$ are neighboring regions} \\
0 & \:\:\: \text{elsewhere},
\end{cases}
\end{equation}
with $i,j\in\mathcal{R}$ and $\mathcal{N}_i$ the set of neighbors of region $i \in \mathcal{R}$, not including the region itself.\footnote{We define neighboring regions as regions that have a common border.} The matrix $\boldsymbol{S}$, as defined in~\eqref{eqA:penmat}, is symmetric and the rows and columns sum to zero.

We now calculate the product $\boldsymbol{\beta}_j^T \boldsymbol{S} \boldsymbol{\beta}_j$. Below we omit the subscript $j$ for notational purposes. We obtain:
\begin{align*}
\boldsymbol{\beta}^T \boldsymbol{S} \boldsymbol{\beta} &= \left(\displaystyle \sum_{j= 1}^R \beta_j s_{j1}\right)\beta_1 + 
\left(\displaystyle \sum_{j= 1}^R \beta_j s_{j2}\right)\beta_2 + \ldots + \left(\displaystyle \sum_{j=1}^R \beta_j s_{jR}\right)\beta_R \\
&= \left(|\mathcal{N}_1| \beta_1 - \displaystyle \sum_{j \in \mathcal{N}_1} \beta_j \right) \beta_1 + \left(|\mathcal{N}_2| \beta_2 - \displaystyle \sum_{j \in \mathcal{N}_2} \beta_j \right) \beta_2 + \ldots + \left(|\mathcal{N}_R| \beta_R - \displaystyle \sum_{j \in \mathcal{N}_R} \beta_j \right) \beta_R.
\end{align*}
If we rewrite the terms in brackets, we obtain:
\begin{align*}
\boldsymbol{\beta}^T \boldsymbol{S} \boldsymbol{\beta} &= \displaystyle \sum_{j\in\mathcal{N}_1} \left(\beta_1 - \beta_j\right)\beta_1 + \displaystyle \sum_{j\in\mathcal{N}_1} \left(\beta_2 - \beta_j\right)\beta_2 + \ldots + \displaystyle \sum_{j\in\mathcal{N}_R} \left(\beta_R - \beta_j\right)\beta_R.
\end{align*}
Next we group the terms containing $\beta_1$, $\beta_2$, etc., and we rewrite this to:
$$\begin{aligned}
\boldsymbol{\beta}^T \boldsymbol{S} \boldsymbol{\beta} = \displaystyle &\sum_{j\in\mathcal{N}_1} \left(\beta_1 - \beta_j\right)\beta_1 + \displaystyle \sum_{j\in\mathcal{N}_1} \left(\beta_j - \beta_1\right) \beta_j \:+ 
\sum_{\mathclap{j\in\mathcal{N}_2\backslash \{1\} }} \left(\beta_2 - \beta_j\right)\beta_2 + \displaystyle \sum_{\mathclap{j\in\mathcal{N}_2\backslash \{1\}}} \left(\beta_j - \beta_2\right) \beta_j \: + \ldots +\\
&\hspace{0.5cm}\sum_{\mathclap{j\in\mathcal{N}_{R-1}\backslash \{1,..,R-2\} }} \left(\beta_{R-1} - \beta_j\right)\beta_{R-1} + \displaystyle \sum_{\mathclap{j\in\mathcal{N}_{R-1}\backslash \{1,..,R-2\} }} \left(\beta_j - \beta_{R-1}\right) \beta_j,
\end{aligned}$$
where, e.g., $\mathcal{N}_k\backslash \{1,..,k-1\}$ represents the set of neighbors of region $k$, excluding the first $k-1$ regions, for $k = 1,2,...,R-1$. After simplifying, we obtain:
$$\begin{aligned}
\boldsymbol{\beta}^T \boldsymbol{S} \boldsymbol{\beta} = \displaystyle &\sum_{j\in\mathcal{N}_1} \left(\beta_1 - \beta_j\right)^2 \: + \: \sum_{\mathclap{j\in\mathcal{N}_2\backslash \{1\}}} \left(\beta_2 - \beta_j\right)^2 \: + \: \ldots \:  + \: \sum_{\mathclap{j\in\mathcal{N}_{R-1}\backslash \{1,..,R-2\} }} \left(\beta_{R-1} - \beta_j\right)^2.
\end{aligned}$$
We conclude that the imposed penalty matrix penalizes the sum of the squared differences between the parameters of neighboring regions. 

\begin{figure}[ht!]
\begin{floatrow}
\ffigbox{%
	\includegraphics[width = 0.475\textwidth]{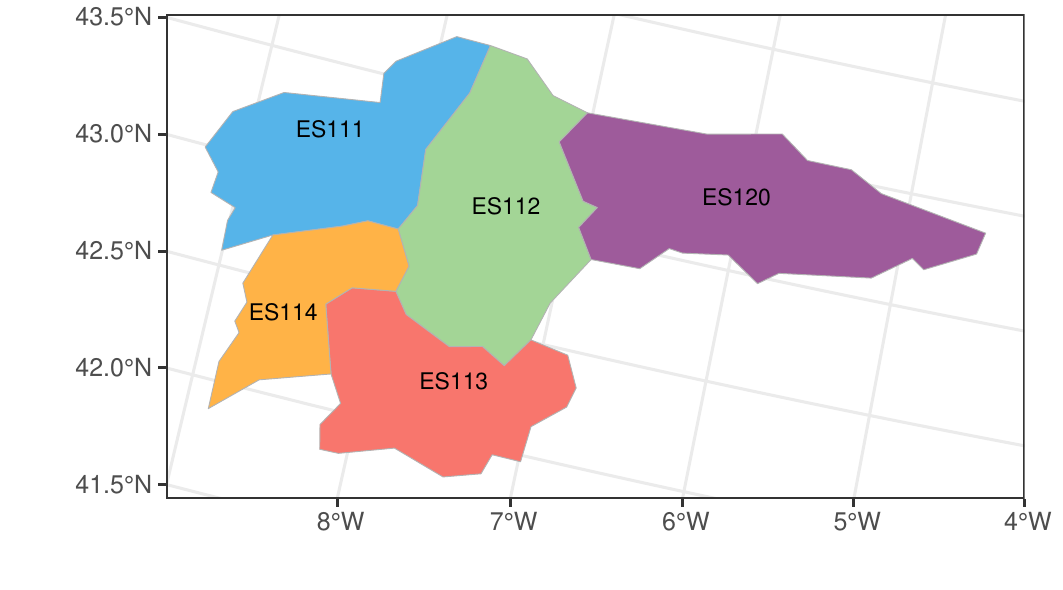}
}{%
  \caption{Five NUTS 3 regions in Spain.\label{fig:exreg}}%
}
\capbtabbox{%
\vspace{-2cm}
$\begin{footnotesize}
\bordermatrix{
  & ES111 & ES112 & ES113 & ES114 & ES120 \cr
ES111 &  2 & -1 &  0 & -1 &  0 \cr
ES112 & -1 &  4 & -1 & -1 & -1 \cr
ES113 &  0 & -1 &  2 & -1 &  0 \cr
ES114 & -1 & -1 & -1 &  3 &  0 \cr
ES120 &  0 & -1 &  0 &  0 &  1}
\end{footnotesize} \vspace{1.25cm}$
}{%
  \caption{Quadratic penalty matrix $\boldsymbol{S}$.\label{tab:penmat}}%
}
\end{floatrow}
\end{figure}

A toy example is shown in Figure~\ref{fig:exreg} and Table~\ref{tab:penmat}, where we focus on five NUTS 3 regions in Spain and specify the corresponding quadratic penalty matrix $\boldsymbol{S}$.

\setcounter{figure}{0}
\setcounter{table}{0}
\section{XGBoost: algorithm, parameter tuning, and interpretation tools} \label{appendix:xgboost}

\subsection{Algorithm} \label{appendix:xgboostalgo}
For notational purposes, we denote $\boldsymbol{x}_{t,w}^{(r)}$ as the input vector for the XGBoost model. This vector consists of the (one-week lagged) environmental anomalies, the (one-week lagged) extreme environmental indices, and the region and season-specific covariates. Given that we assume a Poisson distribution for the random variable $D_{t,w}^{(r)}$, representing the number of deaths in week $w \in \mathcal{W}_t$ of year $t\in \mathcal{T}$ and in $r \in \mathcal{R}$, we construct the XGBoost model on a logarithmic scale to ensure non-negativity of the outcome values. Hereto, we introduce the function:
\begin{align*}
\mathfrak{f}: \mathcal{R}^q \rightarrow \mathcal{R} : \boldsymbol{x} \mapsto \log f(\boldsymbol{x}),
\end{align*}
where $q$ is the dimension of the input vector $\boldsymbol{x}_{t,w}^{(r)}$. Algorithm~\ref{algo:xgboost} sketches the XGBoost algorithm in pseudo-code \citep{chen2016xgboost}. Since the predictions are constructed on logarithmic scale, the final outcomes of the XGBoost algorithm are:
\begin{align*}
\hat{f}_{\text{XGBoost}}\left(\boldsymbol{x}_{t,w}^{(r)}\right) =  \exp \left(
\hat{\mathfrak{f}}_{\text{XGBoost}}\left(\boldsymbol{x}_{t,w}^{(r)}\right)\right),
\end{align*} 
for all $t\in\mathcal{T}$, $w \in \mathcal{W}_t$, and $r \in \mathcal{R}$. In Algorithm~\ref{algo:xgboost}, we consider the negative Poisson log-likelihood as loss function:
\begin{align} \label{eq:poisloss}
\mathcal{L}\left(d_{t,w}^{(r)},  \hat{d}_{t,w}^{(r)}\right) = \hat{b}_{t,w}^{(r)} \cdot f\left(\boldsymbol{x}_{t,w}^{(r)}\right) - d_{t,w}^{(r)} \cdot \left( \log f\left(\boldsymbol{x}_{t,w}^{(r)}\right) + \log \hat{b}_{t,w}^{(r)}\right) + \log d_{t,w}^{(r)} !.
\end{align}
Here, $\smash{\hat{d}_{t,w}^{(r)} = \hat{b}_{t,w}^{(r)} \cdot f(\boldsymbol{x}_{t,w}^{(r)})}$ is the estimated death count in week $w$ of year $t$ for region $r$.

\begin{table}[htb!]
\centering
\begin{algorithm}[H]
\hspace{-0.5cm}	
    \SetAlgoLined
    \KwIn{the training data $\left\{\left(\boldsymbol{x}_{t,w}^{(r)},\: \hat{b}^{(r)}_{t,w}, \:d_{t,w}^{(r)}\right)\right\}_{r\in\mathcal{R},t\in\mathcal{T},w\in\mathcal{W}_t}$}
\hspace{-0.5cm}    Set initial model predictions: $\hat{\mathfrak{f}}_{(0)}\left(\boldsymbol{x}_{t,w}^{(r)}\right) = 0$ for all $r\in \mathcal{R}$, $t\in \mathcal{T}$ and $w\in \mathcal{W}_t$.\\
\hspace{-0.5cm}    	\For{$n$ \KwTo{} $1:\texttt{nrounds}$}{
    	\vspace{0.1cm}
\hspace{-0.25cm}    	Calculate the gradient and hessian of the loss function $\mathcal{L}$:\\
\begin{small}
    	\begin{equation*}
\begin{aligned}
& \hat{g}_{(n)}\left(\boldsymbol{x}_{t,w}^{(r)}\right)=\left[\frac{\partial \mathcal{L}\left(d_{t,w}^{(r)}, \mathfrak{f}\left(\boldsymbol{x}_{t,w}^{(r)}\right)\right)}{\partial \mathfrak{f}\left(\boldsymbol{x}_{t,w}^{(r)}\right)}\right]_{\mathfrak{f}\left(\boldsymbol{x}_{t,w}^{(r)}\right)=\hat{\mathfrak{f}}_{(n-1)}\left(\boldsymbol{x}_{t,w}^{(r)}\right)} = \hat{b}_{t,w}^{(r)} \cdot e^{\hat{\mathfrak{f}}_{(n-1)}\left(\boldsymbol{x}_{t,w}^{(r)}\right)} - d_{t,w}^{(r)} . \\
& \hat{h}_{(n)}\left(\boldsymbol{x}_{t,w}^{(r)}\right)=\left[\frac{\partial^2 \mathcal{L}\left(d_{t,w}^{(r)}, \mathfrak{f}\left(\boldsymbol{x}_{t,w}^{(r)}\right)\right)}{\partial \mathfrak{f}\left(\boldsymbol{x}_{t,w}^{(r)}\right)^2}\right]_{\mathfrak{f}\left(\boldsymbol{x}_{t,w}^{(r)}\right)=\hat{\mathfrak{f}}_{(n-1)}\left(\boldsymbol{x}_{t,w}^{(r)}\right)} = \hat{b}_{t,w}^{(r)} \cdot e^{\hat{\mathfrak{f}}_{(n-1)}\left(\boldsymbol{x}_{t,w}^{(r)}\right)}.
\end{aligned}
\end{equation*}
\end{small}
\vspace{0.1cm}
\hspace{-0.25cm}	Fit a single regression tree $\delta_{(n)}$ of maximum depth \texttt{max\_depth} and of minimum child weight \texttt{min\_child\_weight} on the training data $\smash{ \footnotesize \left(\boldsymbol{x}_{t,w}^{(r)},\: \hat{b}^{(r)}_{t,w}, \: -\nicefrac{\hat{g}_{(n)}\left(\boldsymbol{x}_{t,w}^{(r)}\right)}{\hat{h}_{(n)}\left(\boldsymbol{x}_{t,w}^{(r)}\right)}\right)}$:\footnotemark
\begin{small}
\begin{align*}
\hat{\delta}_{(n)}=\underset{\delta\in \Delta}{\arg \min } \displaystyle \sum_{r \in \mathcal{R}} \sum_{t\in \mathcal{T}} \sum_{w\in \mathcal{W}_t} \frac{1}{2} \hat{h}_{(n)}\left(\boldsymbol{x}_{t,w}^{(r)}\right) \left[\delta\left(\boldsymbol{x}_{t,w}^{(r)}\right)-\frac{\hat{g}_{(n)}\left(\boldsymbol{x}_{t,w}^{(r)}\right)}{\hat{h}_{(n)}\left(\boldsymbol{x}_{t,w}^{(r)}\right)}\right]^2 .
\end{align*}
\end{small}
\vspace{0.1cm}
Update the model predictions with learning rate \texttt{eta}:
\begin{small}
\begin{align*}
\hat{\mathfrak{f}}_{(n)}\left(\boldsymbol{x}_{t,w}^{(r)}\right) = \hat{\mathfrak{f}}_{(n-1)}\left(\boldsymbol{x}_{t,w}^{(r)}\right) + \texttt{eta} \cdot \hat{\delta}_n\left(\boldsymbol{x}_{t,w}^{(r)}\right).
\end{align*}
\end{small}
    }    
   
   \KwOut{$\hat{\mathfrak{f}}_{\text{XGBoost}}\left(\boldsymbol{x}_{t,w}^{(r)}\right) = \hat{\mathfrak{f}}_{(\texttt{nrounds})} \left(\boldsymbol{x}_{t,w}^{(r)}\right)$ for all $r\in \mathcal{R}$, $t \in \mathcal{T}$ and $w \in \mathcal{W}_t$.}
    \caption{XGBoost algorithm for Poisson distributed outcomes. \label{algo:xgboost}}
\end{algorithm}
\end{table}

The XGBoost algorithm relies on six tuning parameters, detailed in Table~\ref{tab:tuningparamgbm}. It is important to note that the XGBoost algorithm encompasses a broader set of parameters beyond those explicitly listed in Table~\ref{tab:tuningparamgbm}. The additional parameters, which we classify as hyper-parameters, extend the configurational possibilities of the algorithm. \cite{chen2019package} provide a comprehensive overview of all parameters associated with the XGBoost algorithm.

\begin{table}[ht!]
  \centering
  \begin{tabular}{p{0.25\linewidth}p{0.7\linewidth}}
    \toprule
    \textbf{Tuning parameter} & \textbf{Explanation} \\
    \midrule
        \texttt{nrounds} & The maximum number of boosting iterations. \\
    \texttt{eta} & Learning rate: adjusts the impact of each tree by multiplying it with a factor \texttt{eta} when incorporating it into the current approximation.  \\
    \texttt{max$\_$depth} & The maximum depth of a tree. \\
    \texttt{subsample} & Subsample ratio of the training data. A value of, e.g., 0.75, means that the algorithm randomly selects 75$\%$ of the training observations each time it constructs a tree.   \\
    \texttt{colsample$\_$bytree} & Subsample ratio of the columns or features when constructing trees. \\
    \texttt{min$\_$child$\_$weight} & Minimum sum of observation weights, i.e. hessian values, required in each node of a tree. \\
    \bottomrule
  \end{tabular}
  \caption{Tuning parameters in the XGBoost algorithm.   \label{tab:tuningparamgbm}
}
\end{table}

\subsection{Parameter tuning with cross-validation} \label{appendix:xgboosttuning}
For simplicity, let $\mathcal{T} = \{0,1,2,...,T-1\}$ represent the considered time range consisting of $T$ years. We choose the number of folds ($K$) to be equal to the number of years in our training data, i.e., $K = T$. Each fold uses a specific year $t \in \mathcal{T}$ as validation data and the remaining years as training data. Consequently, for each fold, the validation data consists of either $53 \cdot |\mathcal{R}|$ observations for a leap year or $52 \cdot |\mathcal{R}|$ observations otherwise.

The $T$-fold cross-validation process, visualized in Figure~\ref{tikz:cv}, involves training the XGBoost model on $T-1$ folds and predicting the observations in the remaining hold-out fold, considering various parameter combinations from a predefined tuning grid. This cycle is repeated $T$ times, as illustrated in Figure~\ref{tikz:cv}.\footnotetext{We select a random subset of $\texttt{subsample}\times 100\%$ of the training instances and $\texttt{colsample\_bytree}\times100 \%$ of the features.} Subsequently, for each parameter combination in the tuning grid, the Poisson loss, see~\eqref{eq:poisloss}, is computed on the predictions made for the hold-out fold. The average of these loss values across the $T$ different hold-out folds is calculated for each parameter combination. The optimal values for the tuning parameters correspond to the parameter combination that yields the smallest average loss value.

\begin{figure}[!ht]
\centering
\begin{adjustbox}{width=0.55\textwidth}
\begin{tikzpicture}

  \draw[->] (0,0) -- (11.5,0) node[below,yshift=-0.2cm] {Time};
  \foreach \x[evaluate=\x as \evalx using int(\x/2)] in {0,2,...,6}
    \draw (\x,0.2) -- (\x,-0.2) node[below] {$\evalx$};
  \draw (6.9,-0.3) node[below] {$\ldots$};
  \draw (8,0.2) -- (8,-0.2) node[below] {$T-1$};  
  \draw (10,0.2) -- (10,-0.2) node[below] {$T$};  
    
  \foreach \year in {0,1,2,3} {
    \fill[newblue] (\year*2,0.5) rectangle (\year*2+2,2.5);
  }
  \fill[newred] (8,0.5) rectangle (10,2.5);
  
  \foreach \year in {0,1,2,4} {
    \fill[newblue] (\year*2,2.5) rectangle (\year*2+2,4.5);
  }
  \fill[newred] (6,2.5) rectangle (8,4.5);
  \draw (7,3.5) node {$\ddots$};
  
  \foreach \year in {0,1,3,4} {
    \fill[newblue] (\year*2,4.5) rectangle (\year*2+2,6.5);
  }
  \fill[newred] (4,4.5) rectangle (6,6.5);
  
  \foreach \year in {0,2,3,4} {
    \fill[newblue] (\year*2,6.5) rectangle (\year*2+2,8.5);
  }
  \fill[newred] (2,6.5) rectangle (4,8.5);
  
  \foreach \year in {1,2,3,4} {
    \fill[newblue] (\year*2,8.5) rectangle (\year*2+2,10.5);
  }
  \fill[newred] (0,8.5) rectangle (2,10.5);
  
  \foreach \year in {0,1,2,3,4} {
    \draw (\year*2,10.5) rectangle (\year*2+2,11.8);
  }
  
  \draw (11,1.5) node {Fold $T$};
  \draw (11,3.5) node {$\vdots$};
  \draw (11,5.5) node {Fold $3$};
  \draw (11,7.5) node {Fold $2$};
  \draw (11,9.5) node {Fold $1$}; 
  
  \draw (1,11.15) node {Year $1$}; 
  \draw (3,11.15) node {Year $2$};  
  \draw (5,11.15) node {Year $3$}; 
  \draw (7,11.15) node {$\cdots$};   
  \draw (9,11.15) node {Year $T$};  
  
  \draw[color=black] (0,10.5) -- (10,10.5);
  \draw[color=black] (0,8.5) -- (10,8.5);
  \draw[color=black] (0,6.5) -- (10,6.5);
  \draw[color=black] (0,4.5) -- (10,4.5);
  \draw[color=black] (0,2.5)    -- (10,2.5);
  \draw[color=black] (0,0.5)  -- (10,0.5); 
  
  \draw[color=black] (0,0.5)  -- (0,10.5);
  \draw[color=black] (10,0.5) -- (10,10.5);
  
  
  \foreach \year in {0.5,2.5,4.5,...,8.5} {
    \draw[<->] (10.3,\year + 0.1) -- (10.3,\year + 1.9);
  }
  
 \foreach \year in {0.5,2.5,4.5,...,10.5} {
    \draw (10.1,\year) -- (10.5,\year);
  }
  
  \draw[<->] (0.1,12.2) -- (9.9,12.2) node[midway,above] {{\Large Full dataset}};
  \draw (0,12) -- (0,12.4);
  \draw (10,12) -- (10,12.4);  
\end{tikzpicture}
\caption{Visualisation of the proposed $T$-fold cross-validation for tuning the parameters in the XGBoost model. The blue boxes refer to the training folds and the red boxes to the validation folds in each split.\label{tikz:cv} }
\end{adjustbox}
\end{figure}

\subsection{Interpretation tools} \label{app:intertools}
We assume that the training data, i.e.,
\begin{align}\label{eq:trdata}
\left\{\left(\boldsymbol{x}_{t,w}^{(r)}, \hat{b}_{t,w}^{(r)}, d_{t,w}^{(r)}\right)\right\}_{r\in\mathcal{R},t\in\mathcal{T}, w\in\mathcal{W}_t},
\end{align}
consists of feature vectors with $q$ variables:
\begin{align*}
\boldsymbol{x}_{t,w}^{(r)} = \left(x_{1,t,w}^{(r)}, x_{2,t,w}^{(r)}, \ldots, x_{q,t,w}^{(r)}\right) \in \mathbb{R}^q.
\end{align*}
For simplicity, we refer in the below explanation to $X_l$ as the $l$-th feature, for $l=1,..,q$, as random variable and to $f(\cdot)$ as the fitted
XGBoost model $\hat{f}_{\text{XGBoost}}(\cdot)$.

\paragraph{Feature importance} \label{app:varimp}
Following the notation from Section~\ref{subsec:calibratingmlm} in the paper, we let $\Delta \mathcal{L}_n(X_l)$ be the total reduction in the Poisson loss function, see~\eqref{eq:poisloss}, caused by splits associated to feature $X_l$ in the tree built during iteration $n$ of the XGBoost algorithm. We then calculate the (unscaled) feature importance of the feature $X_l$ as:
\begin{align*}
\mathcal{V}_{\text{imp}}(X_l) = \frac{1}{\texttt{nrounds}}\displaystyle \sum_{n=1}^{\footnotesize \texttt{nrounds}} \Delta \mathcal{L}_n(X_l),
\end{align*}
where we normalize the obtained importance measure to ensure interpretability and comparability. As a result, the sum of the feature importances for all features adds up to one.

\paragraph{Accumulated Local Effects} \label{app:ale}
The ALE effect of a feature $X_l$ at value $x$ is defined as:
\begin{align} \label{eq:ale1d}
f_{l,\text{ALE}}(x) = \displaystyle \int_{z_{0,l}}^{x} \mathbb{E}\left[ \frac{\partial f(X_1, X_2, ..., X_p)}{\partial X_l} \Big| X_l = z_l \right] dz_l - c_l
\end{align}
where $z_{0,l}$ is a lower bound on the range of $X_l$, typically chosen as the minimum value observed in the data. We interpret the integrand in~\eqref{eq:ale1d} as the expected change in the prediction function $f(\cdot)$ when slightly changing the feature $X_l$ around the value $z_l$, also known as the local effect. Afterwards we accumulate these local effects by integrating them from $z_{0,l}$ to $x$ to unravel the global effect of the feature $X_l$ on the response. Taking the conditional expectation in~\eqref{eq:ale1d} isolates the effect of the feature $X_l$ on the prediction function $f(\cdot)$ from the effects of all other, (possibly) correlated features. Further, \cite{apley2020visualizing} introduce the constant $c_l$ such that the expectation of the random variable $f_{l,\text{ALE}}(X_l)$ equals zero. 

To investigate interaction effects between two environmental features, say $X_k$ and $X_l$, we calculate the ALE interaction effect \citep{apley2020visualizing}. Let $z_{0,k}$ and $z_{0,l}$ be the lower bounds on the ranges of $X_k$ and $X_l$ respectively, then:
\begin{align} \label{eq:ale2d}
f_{k,l,\text{ALE}}(x,y) = \displaystyle \int_{z_{0,k}}^x \int_{z_{0,l}}^{y} \mathbb{E}\left[ \frac{\partial^2 f(X_1, X_2, ..., X_p)}{\partial X_l \partial X_k} \Big| X_k = z_k, X_l = z_l \right] dz_l dz_k - c_{k,l},
\end{align}
where now the conditional expectation is taken with respect to the joint distribution of the random variables $X_j$ for $j\neq k,l$. Equation~\eqref{eq:ale2d} considers the second order partial effects and accumulates these effects in two dimensions. As such, the main effects of the features are not considered in the calculation and the ALE interaction effect should therefore be interpreted as an additional interaction effect on top of the main effects. Further, the constant $c_{k,l}$ is determined such that the expectation of $f_{k,l,\text{ALE}}(X_k,X_l)$ equals zero. Appendix~\ref{app:ale} outlines the estimation procedure for calculating the ALE (interaction) effect on the training data.

Since we do not have a closed-form expression for the prediction function $f$ of the fitted XGBoost model, the expressions of the ALE (interaction) effects in~\eqref{eq:ale1d} and \eqref{eq:ale2d} can not be calculated analytically. Therefore, we estimate these ALE effects using the training data represented in~\eqref{eq:trdata}. We explain the procedure for estimating the ALE effect of a single feature $X_l$. 

We consider the feature space of the random variable $X_l$ in the training data and divide it into $K$ non-overlapping bins, i.e., $z_{0,l} < z_{1,l} < \ldots < z_{K,l}$, where $z_{0,l}$ and $z_{K,l}$ are the covariate's observed minimum and maximum value respectively. Further, let $\mathfrak{X}_l(k)$ represent the set of training instances for which the $l$-th feature value falls into the $k$-th bin, i.e.,
\begin{align*}
\mathfrak{X}_l(k) = \left\{ \boldsymbol{x}_{t,w}^{(r)} \mid z_{k-1} < x_{l,t,w}^{(r)} < z_k\right\},
\end{align*}
with $k\in \{1,2,...,K\}$. We denote $n_l(k) = |\mathfrak{X}_l(k)|$ for the number of training instances that fall into the $k$-th bin. Now consider any $x$ in the feature space of $X_l$ and let $k_l(x)$ be the number of the bin the value $x$ belongs to. Next, we approximate the partial order derivative in~\eqref{eq:ale1d} with a first-order finite difference and obtain \citep{molnar2020interpretable}:
\begin{align*}
\tilde{f}_{l,\text{ALE}}(x) = \displaystyle \sum_{k=1}^{k_l(x)} \frac{1}{n_l(k)} \displaystyle \sum_{\boldsymbol{x}_{t,w}^{(r)} \in \mathfrak{X}_l(k)} \left[ f\left(z_{k,l},\boldsymbol{x}_{-l,t,w}^{(r)}\right) - f\left(z_{k-1,l},\boldsymbol{x}_{-l,t,w}^{(r)}\right)\right] 
\end{align*}
where $\smash{\boldsymbol{x}_{-l,t,w}^{(r)}}$ is the feature vector where the $l$-th feature value has been removed from. The centred ALE effect then equals:
\begin{align} \label{eq:ale1dest}
\hat{f}_{l,\text{ALE}}(x) = \tilde{f}_{l,\text{ALE}}(x) - \frac{1}{n} \displaystyle \sum_{r\in\mathcal{R}} \sum_{t\in \mathcal{T}} \sum_{w\in\mathcal{T}} \tilde{f}_{l,\text{ALE}}\left(x_{l,t,w}^{(r)}\right),
\end{align} 
with $n$ the number of training instances. We estimate the ALE interaction effects in~\eqref{eq:ale2d} in a similar way. However, instead of considering one-dimensional bins, we now need to consider rectangular grid cells. Furthermore, we approximate the second-order partial derivative by a second-order finite difference, see \cite{apley2020visualizing} and \cite{molnar2020interpretable} for further details.

\setcounter{figure}{0}
\setcounter{table}{0}

\section{Feature engineering} \label{appendix:feature-engineering}

\subsection{Daily aggregation of hourly air pollutant levels} \label{subsubsec:hourlytodaily}
We consider an air pollution factor from Table~\ref{tab:envvar} in the paper and denote it as $x_{t,w,d,h}^{(\text{long,lat})}$. This feature represents the concentration of a specific air pollutant in year $t$, week $w$, day $d$, and hour $h$, located at longitude-latitude coordinates (long,lat). We then compute the daily minimum, average, and maximum concentrations of the air pollutant, measured at the coordinates (long,lat) as follows:
\begin{align*}
\overset{\wedge}{x}_{t,w,d}^{(\text{long,lat})} &= \min\left\{ x_{t,w,d,h}^{(\text{long,lat})} \: \big| \:h= 0,1,...,23 \right\} \\
\overline{x}_{t,w,d}^{(\text{long,lat})} &= \text{avg}\left\{ x_{t,w,d,h}^{(\text{long,lat})} \:\big|\: h= 0,1,...,23 \right\}\\
\overset{\vee}{x}_{t,w,d}^{(\text{long,lat})} &= \max\left\{ x_{t,w,d,h}^{(\text{long,lat})} \:\big|\: h= 0,1,...,23 \right\}.
\end{align*}
For the weather factors listed in Table~\ref{tab:climatevar} in the paper, we do not need to make the conversion from an hourly to a daily time scale, since these factors are already defined on a daily time scale at specific longitude-latitude coordinates.

\subsection{Population-weighted spatial aggregation to NUTS 3 level} \label{subsubsec:populationweighted}
Let $\tilde{x}_{t,w,d}^{(\text{long,lat})}$ denote the measurement of a particular environmental factor at coordinates (long,lat), year $t$, week $w$, and day $d$. This feature can encompass minimum, maximum, or average air pollution measurements, as defined in Section~\ref{subsubsec:hourlytodaily}, or any weather-related factor from Table~\ref{tab:climatevar} in the paper.

We consider the grid $\mathcal{G}$ of longitude-latitude coordinates (long,lat) on which the feature $\tilde{x}_{t,w,d}^{(\text{long,lat})}$ is registered. We then introduce the function:
\begin{align*}
m_\mathcal{G}: \mathcal{G} \rightarrow \mathcal{R}: \: \text{(long,lat)} \mapsto r,
\end{align*}
which maps longitude-latitude coordinates from the grid $\mathcal{G}$ to the NUTS 3 region $r$ it belongs. Now, define the set $\mathcal{I}(r)$ that consists of all (long,lat) coordinates of the grid $\mathcal{G}$ contained in the NUTS 3 region $r$:
\begin{align} \label{eq:grid1}
\mathcal{I}(r) = \left\{\text{(long,lat)} \in \mathcal{G} \mid  m_{\mathcal{G} }(\text{long},\text{lat}) = r\right\}.
\end{align}

We create a weighted aggregated feature at the NUTS 3 geographical level as follows:
\begin{align} \label{eq:popweightavg}
\tilde{x}_{t,w,d}^{(r)} = \displaystyle \smashoperator[r]{\sum_{\substack{\text{(long,lat)} \in \mathcal{I}(r)}}} \omega_{\text{(long,lat)}} \cdot \tilde{x}_{t,w,d}^{(\text{long},\text{lat})},
\end{align}
where $\omega_{\text{(long,lat)}}$ are weights that sum up to one within each grid $\mathcal{I}(r)$. These weights aim to compute a population-weighted version of the daily environmental factors, a method that has also been explored in the existing literature \citep{balakrishnan2019impact,de2021comparative}. This weighting strategy is necessary due to the uneven distribution of a population within a NUTS 3 region, which tends to concentrate around specific locations. Therefore, we put more weight on the daily measurements of the environmental factors at longitude-latitude coordinates where more people are located, relative to the population in that NUTS 3 region. We assume these weights $\omega_{\text{(long,lat)}}$ to remain constant over time. This assumption is consistent with the idea that the proportion of the population within region $r$ associated with grid point (long,lat) $\in \mathcal{I}(r)$ remains relatively stable throughout the time span under consideration.\footnote{Our current method assigns each grid point (long,lat) $\in \mathcal{G}$ exclusively to one NUTS 3 region $r$. Alternatively, an interpolation method could be explored, enabling grid points near the borders of NUTS 3 regions to also contribute to the calculation of the feature at NUTS 3 level in~\eqref{eq:popweightavg}.} 

We compute the weights using the Socioeconomic Data and Applications Center (SEDAC) in NASA's Earth Observing System Data and Information System (EOSDIS).\footnote{This data center is available on \url{https://sedac.ciesin.columbia.edu/}.} Particularly, we use the high-dimensional population count dataset, defined on a grid with a spatial resolution of 2.5 arc-minutes ($\approx 5 \text{ km}$) and a temporal dimension of five years \citepalias{nasadataset}. We extract the population gridded dataset at the year $2015$ to construct the population weights in~\eqref{eq:popweightavg}.\footnote{We select the year $2015$ since it falls within our considered time range.} 

\subsubsection{Construction of population weights} \label{appendix:population-weights}
Similarly to~\eqref{eq:grid1}, we define the set of (long,lat) coordinates of the grid $\mathcal{G}^{(2)}$ that fall within the boundaries of a NUTS 3 region $r\in\mathcal{R}$ as:
\begin{align*}
\mathcal{I}_2(r) = \left\{\text{(long,lat)} \mid \text{(long,lat)} \in \mathcal{G}^{(2)} \:\: \text{and}\:\: m_{\mathcal{G}^{(2)} }(\text{long},\text{lat}) = r\right\}.
\end{align*}

\begin{figure}[htb!]
\centering
\includegraphics[width=0.75\textwidth]{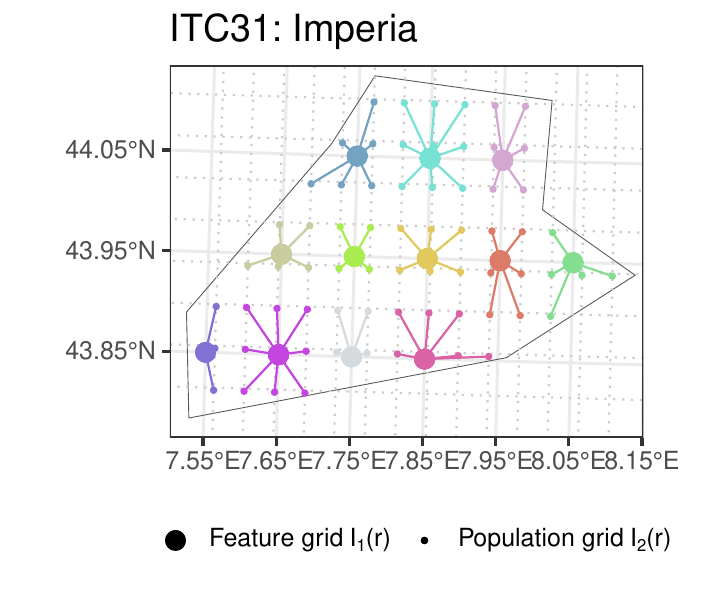}
\caption{Visualisation of the calculation of the population weights at each longitude-latitude coordinate of the feature grid (big dots) in the NUTS 3 region Imperia (Northern Italy). We attribute the population count at each population grid point (small dots) to the closest feature grid point (big dots). The solid, light-grey lines represent the feature grid $\mathcal{G}^{(1)}$, while the dotted, light-grey lines represent the population grid $\mathcal{G}^{(2)}$. \label{fig:vispopweight}}
\end{figure}

Next, we appoint each longitude-latitude coordinate in the population grid restricted to region $r$, i.e., $\mathcal{I}_2(r)$, to its closest longitude-latitude coordinate in the feature grid restricted to region $r$, i.e., $\mathcal{I}_1(r)$. Hereto, we define, for any (long,lat) $\in\mathcal{I}_1(r)$, the set:
\begin{align*}
\mathcal{P}^{(\text{long},\text{lat})}(r) = \left\{ (a,b) \in \mathcal{I}_2(r) \: \big| \: d_{2}((a,b),(\text{long},\text{lat})) \leq d_{2}((a,b),(l_1,l_2)) \:\:\:\: \forall \:(l_1,l_2) \in \mathcal{I}_1^{(r)}\right\}, 
\end{align*}
where $d_2$ refers to the 2-norm or Euclidean distance. We then calculate the population weights at grid points (long,lat) $\in \mathcal{I}_1(r)$ as:
\begin{align} \label{eq:popweights}
\omega_{(\text{long},\text{lat})} = \dfrac{\displaystyle \smashoperator[r]{\sum_{\substack{(a,b) \in \mathcal{P}^{(\text{long},\text{lat})}(r)}}} P^{(a,b)} }{\displaystyle \smashoperator[r]{\sum_{\substack{(a,b) \in \mathcal{I}_2(r)}}} P^{(a,b)}},
\end{align}
where $P^{(a,b)}$ refers to the population count at any grid point $(a,b$) in the population grid $\mathcal{G}^{(2)}$. The nominator in~\eqref{eq:popweights} equals the population count attributed to the grid point (long,lat) of the feature grid and the denominator equals the total population count in the entire NUTS 3 region $r$. By construction, the sum of the weights in each region $r$ equals one. Figure~\ref{fig:vispopweight} illustrates the steps to calculate the population weights in~\eqref{eq:popweightavg}.

In the scenario where a specific, very small NUTS 3 region $r \in\mathcal{R}$ contains no grid points from the feature grid $\smash{\mathcal{G}^{(1)}}$, i.e., when $\mathcal{I}_1(r) = \emptyset$, we determine the feature value $\smash{\tilde{x}_{t,m,d}^{(r)}}$ as the value $\smash{\tilde{x}_{t,m,d}^{(\text{long},\text{lat})}}$ at the closest longitude-latitude coordinates from the feature grid to the boundaries of that region.

\subsection{Daily extreme environmental indices} \label{subsubsec:extremeenvironmentalindices}
To capture the effects of extreme environmental conditions on mortality, we construct daily, region-specific features that indicate whether the daily, environmental measurement in that region exceeds a certain high quantile or falls below a certain low quantile. This quantile-based approach has been previously used in the literature to analyze the effects of heat waves \citep{gasparrini2011impact, hajat2006impact, anderson2009weather}. Hereto, we propose to calculate the region-specific $5\%$ and $95\%$ quantile of the entire set of historical observations on the daily minimum temperature \texttt{Tmin}, the average temperature \texttt{Tavg}, and the maximum temperature \texttt{Tmax} throughout the considered time period 2013-2019. We define the extreme high temperature index for region $r$ on day $d$ in week $w$ of year $t$ as:
\begin{align} \label{eq:Tind}
\begin{small}
\text{T.ind}_{t,w,d}^{(r,95\%)} = \mathlarger{\mathlarger{\mathbbm{1}}}\left\{\texttt{Tmax}_{t,w,d}^{(r)} \geq q_{\texttt{Tmax}}^{(r,95\%)}\right\} + \mathlarger{\mathlarger{\mathbbm{1}}}\left\{\texttt{Tavg}_{t,w,d}^{(r)} \geq q_{\texttt{Tavg}}^{(r,95\%)}\right\} + 
\mathlarger{\mathlarger{\mathbbm{1}}}\left\{\texttt{Tmin}_{t,w,d}^{(r)} \geq q_{\texttt{Tmin}}^{(r,95\%)}\right\},
\end{small}
\end{align}
with, e.g., $\mathbbm{1}\{\texttt{Tmax}_{t,w,d}^{(r)} \geq q_{\texttt{Tmax}}^{(r,95\%)}\}$ an indicator function which equals one when the maximum temperature in region $r$ on day $d$ in week $w$ of year $t$ exceeds its historical 95$\%$ quantile, and zero otherwise. We refer to this feature as the hot-day index. The index can take values 0, 1, 2, or 3, where a value of three corresponds to a very hot day. We define a similar cold-day index using the $5\%$ quantile, where a value of three corresponds to a very cold day. Similarly, we create daily indices for surpassing high thresholds of the other environmental factors and refer to them as extreme environmental indices.

\subsection{Daily environmental anomalies} \label{subsubsec:baselinemodels}
Alongside the extreme environmental indices, constructed in Section~\ref{subsubsec:extremeenvironmentalindices}, we engineer environmental anomalies that measure deviations from daily baseline conditions, see, e.g., \citep{ballester2023heat} for an application of temperature anomalies to estimate the heat-related mortality during the summer of 2022 in European regions. Some of the daily environmental features exhibit a clear seasonal pattern. For these features, we first fit a region-specific baseline model using robust linear regression with a single set of sine and cosine Fourier terms as covariates \citep{gervini2002class}.\footnote{By adopting robust linear regression, the fitted baseline model becomes less susceptible to the impact of extreme levels within the observed environmental features. We use the \texttt{lmRob} function in the \texttt{R}-package \texttt{robust}.} In this way, we capture the region-specific baseline environmental conditions on each day within a year. We have:
\begin{align} \label{eq:baselinemodelfeatures}
\tilde{x}_{t,w,d}^{(r)} = \alpha_0^{(r)} + \alpha_1^{(r)} \sin\left(\frac{2\pi \text{day}_{t,w,d}}{365.25}\right) + \alpha_2^{(r)} \cos\left(\frac{2\pi \text{day}_{t,w,d}}{365.25}\right) +  \epsilon_{t,w,d}^{(r)},
\end{align}
where $\smash{\epsilon_{t,w,d}^{(r)}}$ is a normally distributed error term, $\text{day}_{t,w,d}$ denotes the number of the day in year $t$ for ISO-date $(t,w,d)$, and $365.25$ represents the average number of days per year.\footnote{Alternatively, the exact number of days in a year can be used, with 366 days in a leap year and 365 days otherwise.} In case a feature does not exhibit a clear annual, seasonal pattern, we do not consider a baseline model for that feature.

\begin{figure}[ht!]
\centering
\begin{subfigure}[b]{0.3\textwidth}
\includegraphics[width=\textwidth]{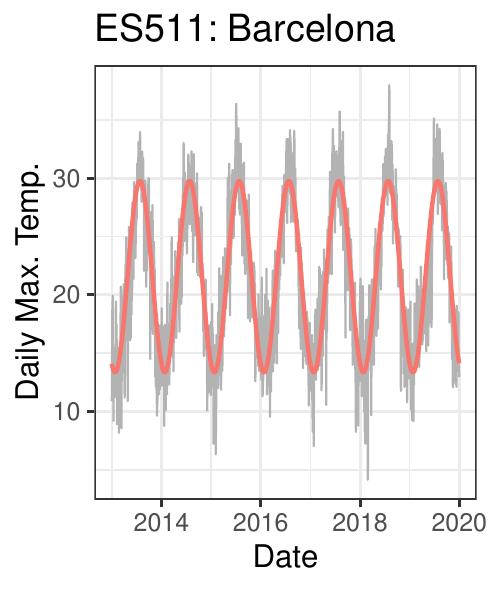}
\begin{minipage}{1.175\textwidth}
\vspace{-0.5cm}
\nsubcap{\label{fig:visbaselinefeat1}}
\end{minipage}
\end{subfigure}
\hspace{0.1cm}
\begin{subfigure}[b]{0.3\textwidth}
\includegraphics[width=\textwidth]{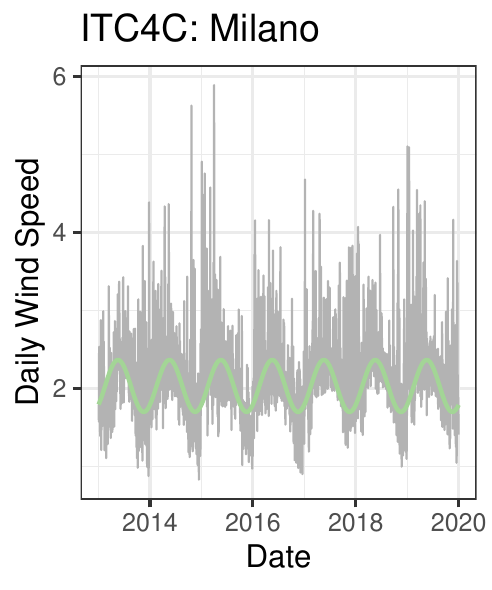}
\begin{minipage}{1.15\textwidth}
\vspace{-0.5cm}
\nsubcap{\label{fig:visbaselinefeat2}}
\end{minipage}\end{subfigure}
\hspace{0.1cm}
\begin{subfigure}[b]{0.3\textwidth}
\includegraphics[width=\textwidth]{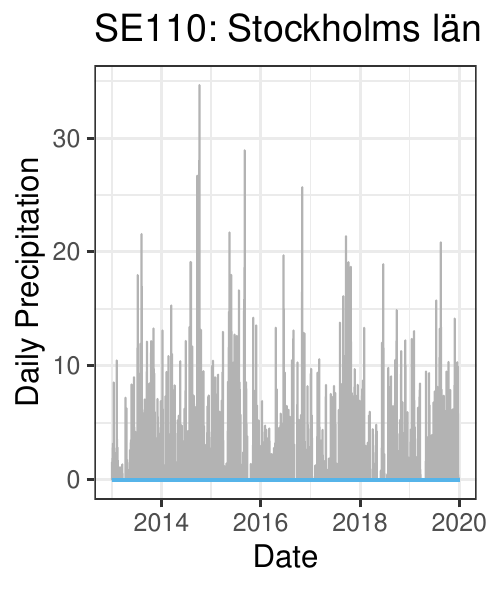}
\begin{minipage}{1.175\textwidth}
\vspace{-0.5cm}
\nsubcap{\label{fig:visbaselinefeat3}}
\end{minipage}\end{subfigure}
\caption{We visualize the baseline model across the years 2013-2019, fitted by means of robust linear regression with one sine-cosine Fourier pair for the daily maximum temperatures observed in Barcelona (a) and for the daily wind speed levels in Milano (b). In panel (c), we show the daily precipitation amounts in Stockholm with a zero baseline trend. \label{fig:visbaselinefeat}}
\end{figure}

Figure~\ref{fig:visbaselinefeat} shows an example of the daily baseline trend for three weather factors as registered in three different NUTS 3 regions. Figure~\ref{fig:visbaselinefeat1} visualizes the seasonal baseline trend for the daily maximum temperature levels in Barcelona (ES511), showing higher maximum temperatures during summer days compared to winter days. Figure~\ref{fig:visbaselinefeat2} shows the seasonal baseline trend for the daily wind speed levels. Furthermore, we assume a zero baseline trend for the variable \texttt{Rain}, visualized by the blue line in Figure~\ref{fig:visbaselinefeat3}. In the remainder of this section, we will work with the excesses or deviations from the baseline levels, also referred to as anomalies. These anomalies are the residuals from the region-specific baseline in~\eqref{eq:baselinemodelfeatures} for the different environmental features. The anomalies of an environmental feature $\tilde{x}_{t,w,d}^{(r)}$ with a non-zero baseline equal:
\begin{align*}
\Delta \tilde{x}_{t,w,d}^{(r)}  = \tilde{x}_{t,w,d}^{(r)}  - \hat{\tilde{x}}_{t,w,d}^{(r)},
\end{align*}
where $\hat{\tilde{x}}_{t,w,d}^{(r)}$ represents the estimated daily baseline environmental condition in region $r$ at day $d$ of week $w$ in year $t$. In case of a zero baseline trend, i.e., for wind speed, we take $\Delta \tilde{x}_{t,w,d}^{(r)}  = \tilde{x}_{t,w,d}^{(r)}$.

\subsection{Weekly averages of daily environmental anomalies and extreme indices} \label{subsubsec:weeklyaggregation}
We align the daily time scale of the environmental anomalies (see Section~\ref{subsubsec:baselinemodels}) and the extreme environmental indices (see Section~\ref{subsubsec:extremeenvironmentalindices}), with the weekly time scale of the death counts. Hereto, let $\smash{\tilde{a}_{t,w,d}^{(r)}}$ be any environmental anomaly or extreme environmental index, registered in region $r$ at day $d$ of week $w$ in year $t$. We then compute the weekly averages of this environmental feature as:
\begin{align*}
\overline{a}_{t,w}^{(r)} = \text{avg} \left\{\tilde{a}_{t,w,d}^{(r)} \mid d = 1,2,...,7\ \right\},
\end{align*}
for all $t\in \mathcal{T}$, $w \in \mathcal{W}_t$, and $r\in \mathcal{R}$. This weekly averaging process captures effects such as the heightened impact of prolonged periods of environmental stress on mortality rates, e.g., multiple hot days within a week. Table~\ref{tabapp:features} lists the final features which we use as inputs in the machine learning model.

\begin{table}[!htb]
  \centering
\adjustbox{max width=0.48\textwidth}{  \begin{tabular}{ll}
    \toprule
    \textbf{Feature} & \pbox{5.5cm}{\textbf{Weekly average of the daily}}\\
    \midrule
        $T_{\max}'$ & maximum temperature anomalies \\
        $T_{\min}'$ & minimum temperature anomalies \\
        $I_{\text{hot}}$ & hot-day indicator, see~(\ref{eq:Tind}) \\
        $I_{\text{cold}}$ & cold-day indicator \\
        $H'$  & average relative humidity anomalies \\
        $I_{\text{highH}}$  & high humidity indicator \\
        $I_{\text{lowH}}$  & low humidity indicator \\
        $P'$ & total precipitation levels \\
        $I_{\text{highP}}$  & high precipitation indicator \\
        $I_{\text{lowP}}$  & low precipitation indicator \\
        $W'$ & average wind speed anomalies \\ 
        $I_{\text{highW}}$  & high wind speed indicator \\
        $I_{\text{lowW}}$  & low wind speed indicator \\       
    \bottomrule
  \end{tabular}}
  \quad
\adjustbox{max width=0.48\textwidth}{\begin{tabular}{ll}
    \toprule
    \textbf{Feature} & \pbox{5.5cm}{\textbf{Weekly average of the daily}}\\
    \midrule
        $\text{O}_3'$  & average ozone anomalies \\
        $I_{\text{high}\text{O}_3}$  & high ozone indicator \\
        $I_{\text{low}\text{O}_3}$  & low ozone indicator \\
        $\text{NO}_{2}'$  & average nitrogen dioxide anomalies \\
        $I_{\text{high}\text{NO}_2}$  & high nitrogen dioxide indicator \\
        $I_{\text{low}\text{NO}_2}$  & low nitrogen dioxide indicator \\
        $\text{PM}_{10}'$  & average PM$_{10}$ anomalies \\
        $I_{\text{high}\text{PM}_{10}}$  & high PM$_{10}$ indicator \\
        $I_{\text{low}\text{PM}_{10}}$  & low PM$_{10}$ indicator \\
        $\text{PM}_{2.5}'$  & average PM$_{2.5}$ anomalies \\
        $I_{\text{high}\text{PM}_{2.5}}$  & high PM$_{2.5}$ indicator \\
        $I_{\text{low}\text{PM}_{2.5}}$  & low PM$_{2.5}$ indicator \\      
        \: & \: \vspace{-0.25cm}\\ 
    \bottomrule
  \end{tabular}}
  \caption{Weekly region-specific, environmental features at a NUTS 3 geographical level. The prime ' refers to anomalies, and $I$ refers to the indicator features.  \label{tabapp:features}
}
\end{table}

\begin{figure}[ht!]
\centering
\includegraphics[width=0.85\textwidth]{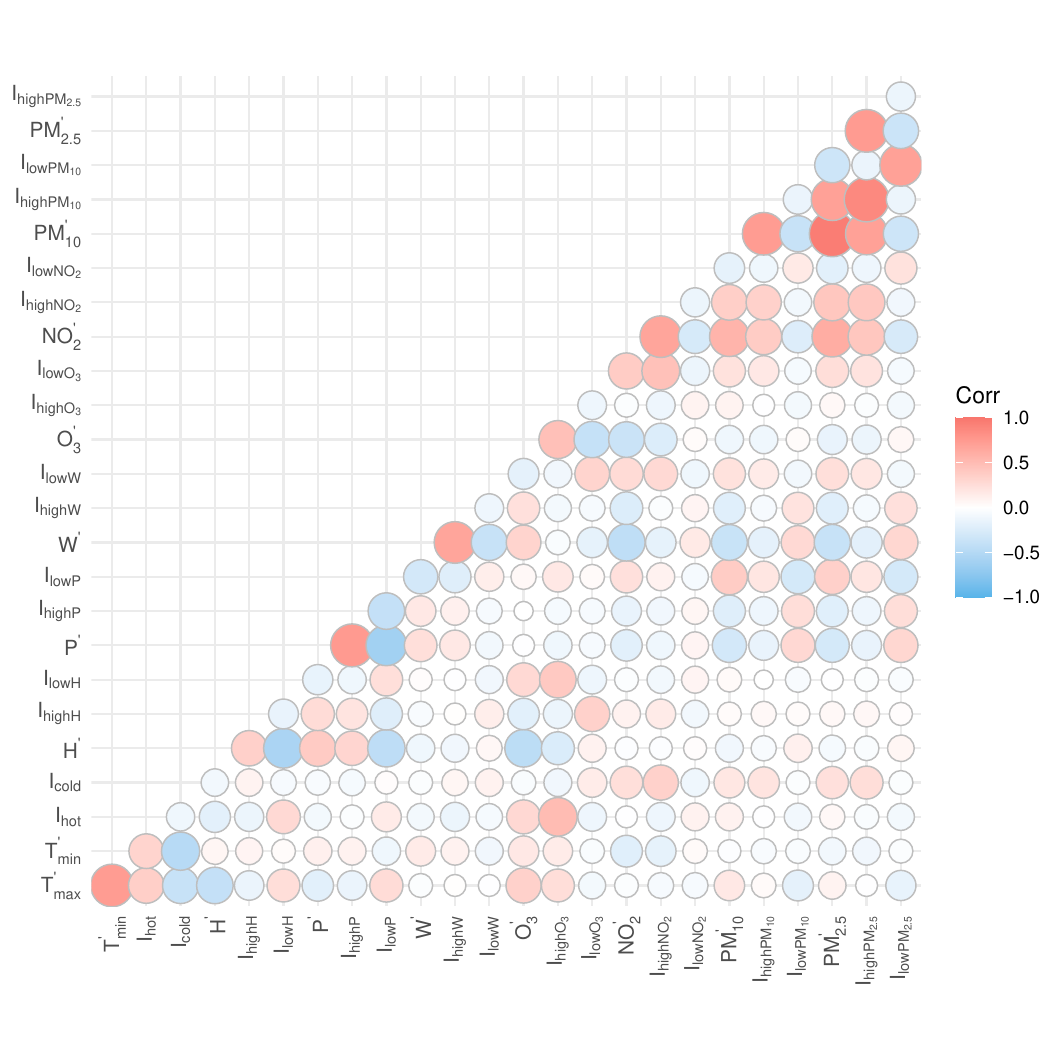}
\caption{Correlations between the weekly environmental features, constructed through the outlined
 feature engineering process, and presented in Table~\ref{tabapp:features}.\label{fig:corrplot}}
\end{figure}

Figure~\ref{fig:corrplot} illustrates the correlations among the weekly environmental features, as listed in Table~\ref{tabapp:features}. The most notable positive correlations appear between the weekly averages of daily PM$_{2.5}$ and PM$_{10}$ anomalies (0.95), and between the weekly averages of daily maximum and minimum temperature anomalies (0.73). Additionally, unsurprisingly, strong positive correlations are observed between the weekly averages of various environmental anomalies and their respective high quantile indices. Conversely, notable negative correlations emerge between the weekly averages of daily environmental anomalies and their corresponding low quantile indices.

\section{Case study: calibration}
\subsection{Calibrating the baseline model for weekly death counts} \label{subsubsec:baselinemodel.casestudy}
\begin{figure}[htb!]
\centering
\includegraphics[width=0.85\textwidth]{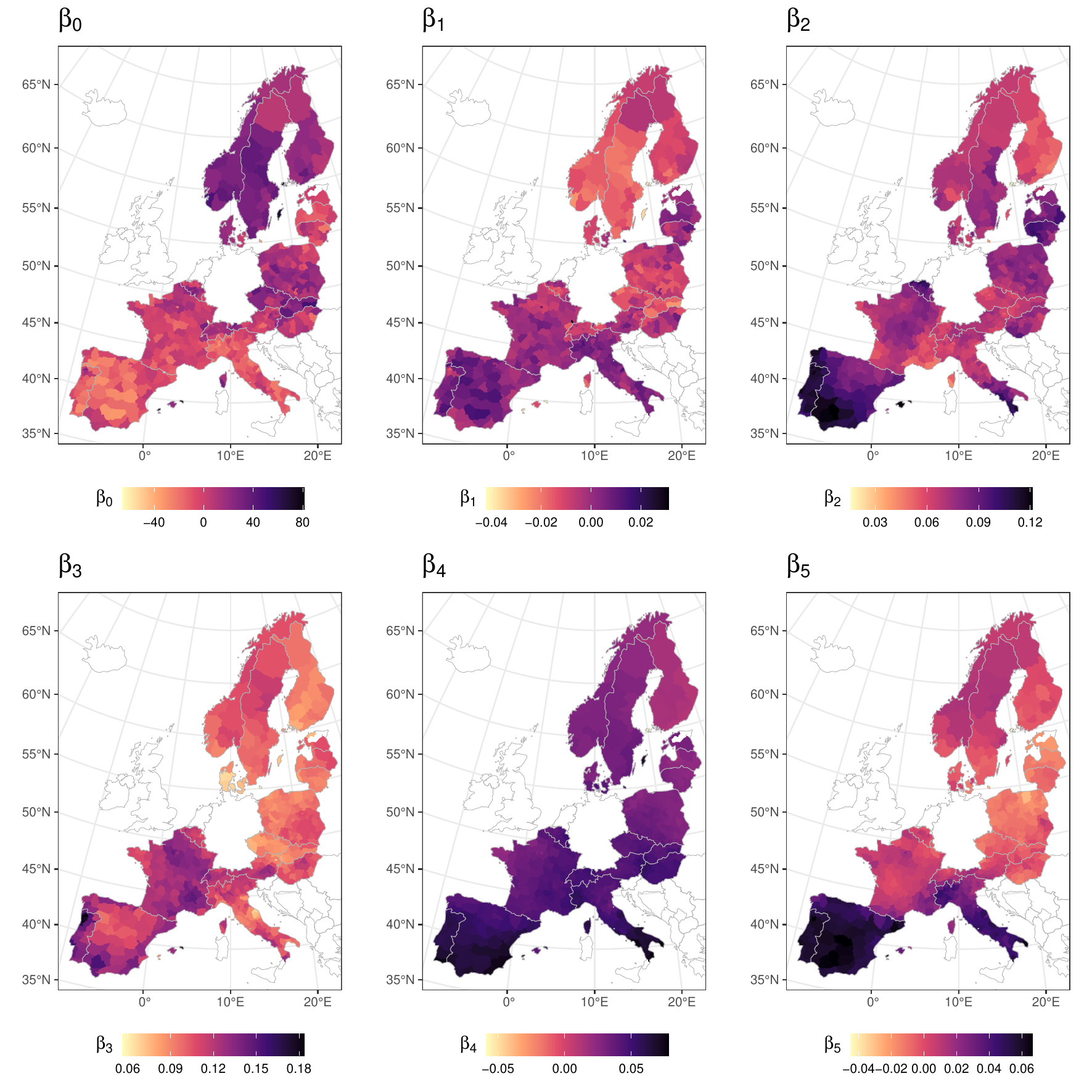}
\caption{The fitted parameters $\hat{\beta}_0^{(r)}$, $\hat{\beta}_1^{(r)}$, and $\hat{\beta}_2^{(r)}$ (top panels) and $\hat{\beta}_3^{(r)}$, $\hat{\beta}_4^{(r)}$, and $\hat{\beta}_5^{(r)}$ (bottom panels) in the spatially smoothed Poisson GLM belonging to the intercept, ISO year $t$ and the sine and cosine Fourier terms with a period of 52 or 26 weeks, see~(\ref{eq:baselinedeathsstructure}) in the paper.\label{fig:fitparam}}
\end{figure}

Figure~\ref{fig:fitparam} illustrates the fitted parameters in the spatially smoothed baseline model, showing smooth variations in the parameters between adjacent regions. The estimated parameter $\smash{\hat{\beta}_1^{(r)}}$ for the ISO year effect exhibits a negative sign in northern European regions and a slightly positive sign in southern regions. This suggests that, from 2013 to 2019, mortality rates slightly increased in the south while declining in the north. Moreover, the larger magnitude of the estimated parameters related to the sine and cosine Fourier terms in southern regions suggests that the associated seasonal patterns (with 52-week or 26-week periods) have a more pronounced effect on weekly death counts in southern regions compared to northern regions.

Figure~\ref{fig:insampledtr} in the paper illustrates the estimated baseline mortality rates $\hat{q}_{t,w}^{(r)}$ in red, as obtained in~(\ref{eq:calcmortrates}) when using the estimated force of mortality specified in~(\ref{eq:baselinedeathsstructure}). In line with the findings from Figure~\ref{fig:fitparam}, we observe, for example, that the seasonal variation in Barcelona is more pronounced than in Milano and Stockholm. Furthermore, the mortality rates in the southern regions show a relatively stable (Barcelona) or slightly increasing (Milano) trend, whereas the mortality rates in the region of Stockholm clearly exhibit a decreasing trend.

\subsection{Tuning and calibrating the machine learning model} \label{subsubsec:xgboostmodel.casestudy}
We consider the following tuning grid for the six XGBoost parameters of interest:
\begin{align*}
&\texttt{nrounds} \in \{10,20,...,5\ 000\}, \hspace{0.3cm}
\texttt{eta} \in \{0.01,0.05,0.1\}, \hspace{0.3cm}
\texttt{min\_child\_weight} \in \{10,100,1000\} \\
&\texttt{subsample} \in \{0.50,0.75\},\hspace{0.3cm}
\texttt{colsample\_bytree} \in \{0.50,0.75\},\hspace{0.3cm}
\texttt{max.depth} \in \{1,3,5,7,9\}.
\end{align*}
We tune these six parameters by means of 7-fold cross-validation, following the approach outlined in Section~\ref{subsec:calibratingmlm}.\footnote{We consider a 7-fold cross-validation as the calibration period $\mathcal{T}$ consists of seven years, i.e., the years 2013-2019.} Throughout this cross-validation process, the offset, representing the estimated baseline deaths, is considered to be known. While, in principle, we should recalibrate this offset for each set of training fold combinations, we refrain from doing so due to computational constraints. Moreover, the robust construction of the baseline model ensures that excluding one year from the training set will only have a negligible impact on the baseline fit and consequently on the offset.

The optimal set of tuning parameters, see Table~\ref{tab:tuningparamgbm}, consists of $490$ boosting iterations (\texttt{nrounds}), a learning rate of $0.01$ (\texttt{eta}), a minimum child weight of $1\ 000$ (\texttt{min\_child\_weight}), a maximum depth of $7$ for each tree (\texttt{max.depth}), and a subsample ratio of $75\%$ and $50\%$ for the training data (\texttt{subsample}) and the columns (\texttt{colsample\_bytree}), respectively. Subsequently, we retrain the machine learning model using this optimal parameter set over the entire calibration period spanning the years 2013-2019.

\subsection{Model fits, findings, and discussions}
\subsubsection{In-sample fit and model performance} \label{app:residualsfit}
\begin{figure}[htb!]
\centering
\includegraphics[width=\textwidth]{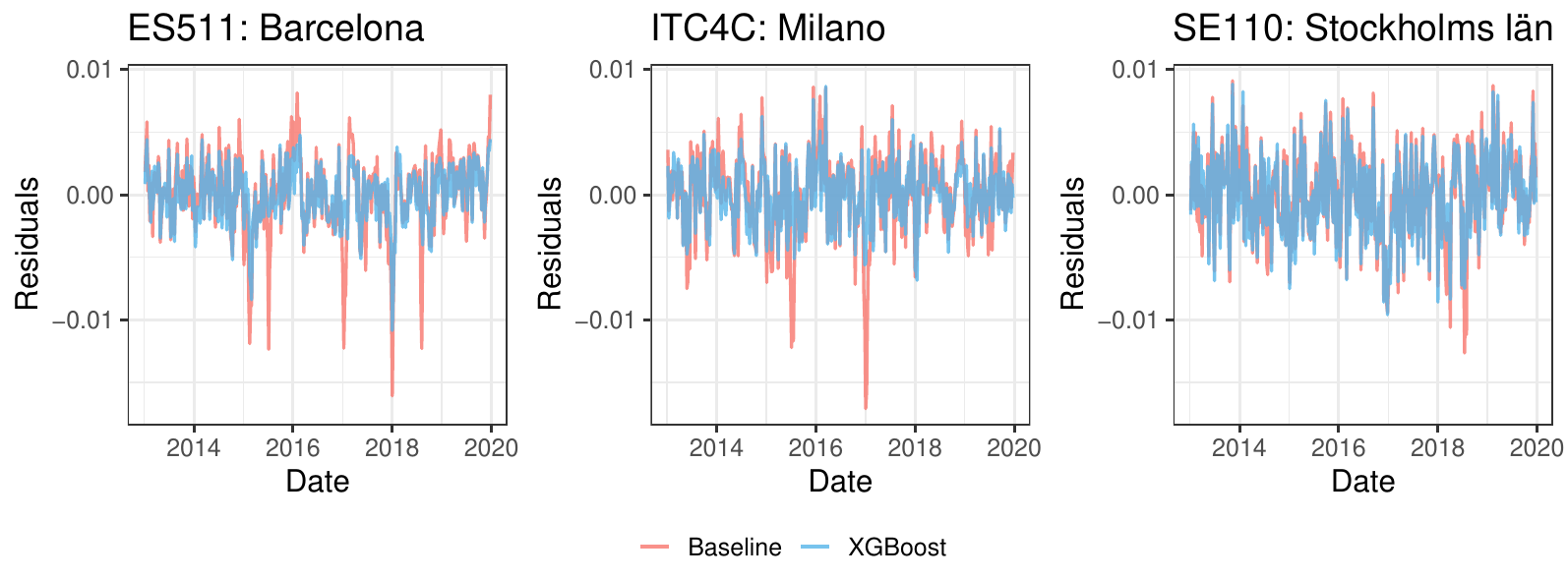}
\caption{The residuals of the estimated weekly mortality rates derived from the baseline model (red) and the XGBoost model (blue) in Barcelona (left), Milano (middle), and Stockholm (right).   \label{fig:insampledtrApp}}
\end{figure}

The findings in Section~\ref{subsubsec:insamplefit} of the paper are confirmed by Figure~\ref{fig:insampledtrApp}, which visualizes the residuals corresponding to the estimated mortality rates, calculated as:
\begin{align*}
\text{resid}_{t,w}^{(r)} = \hat{q}_{t,w}^{(r)} - q_{t,w}^{(r)},
\end{align*}
where $\smash{\hat{q}_{t,w}^{(r)}}$ and $\smash{q_{t,w}^{(r)}}$ represent the estimated and observed mortality rate in week $w$ of year $t$ in region $r$, respectively. The estimated mortality rate is derived from either the baseline or the machine learning model (XGBoost). We observe less peaks in the residuals of the machine learning model compared to the residuals of the baseline model, particularly during the summer and winter weeks. 

\subsubsection{Harvesting effects} \label{subsec:harvesting}
Harvesting effects occur when, for example, environmental related excess mortality in the previous week leads to a mortality deficit in the current week \citep{schwartz2001there}. We first conduct an exploratory analysis to empirically identify potential harvesting effects within our training data spanning from 2013 to 2019, across all considered NUTS 3 regions. We then assess if we observe similar effects using the predictions obtained from the machine learning model, calibrated in Section~\ref{subsubsec:xgboostmodel.casestudy}. 

To quantify the difference in observed or estimated deaths relative to the mortality baseline for a specific week $w$ in year $t$ and region $r$, we calculate the excess death proportion or the so-called `P-score' \citep{Scholey2021robustness,msemburi2023estimates}, defined as:
\begin{align*}
\text{EDP}_{t,w}^{(r)} = \dfrac{d_{t,w}^{(r)} - \hat{b}_{t,w}^{(r)}}{\hat{b}_{t,w}^{(r)}}, \hspace{0.5cm} \widehat{\text{EDP}}_{t,w}^{(r)} = \dfrac{\hat{d}_{t,w}^{(r)} - \hat{b}_{t,w}^{(r)}}{\hat{b}_{t,w}^{(r)}},
\end{align*}
for $t\in \mathcal{T}$, $w \in \mathcal{W}_t$, and $r \in \mathcal{R}$. Furthermore, $d_{t,w}^{(r)}$, $\hat{b}_{t,w}^{(r)}$, and $\hat{d}_{t,w}^{(r)}$ represent, respectively, the observed deaths, the deaths estimated by the baseline mortality model (see~(\ref{eq:btwr}) in the paper), and the deaths estimated by the machine learning model (see~(\ref{eq:machinemodelspec}) in the paper). Next, we consider a specific feature $x_J$ from Table~\ref{tabapp:features} for any $J \in \{1,2,...,25\}$, and we denote its value in week $w$ of year $t$ in region $r$ as $\smash{x_{J,t,w}^{(r)}}$. Furthermore, we denote its lagged version as $\smash{L^1(x_{J,t,w}^{(r)}) := x_{J,t,w-1}}$.\footnote{Note that $x_{J,t,0}$ represents $x_{J,t-1,53}$ (leap year) or $x_{j,t-1,52}$ (non-leap year)} We then partition the domain of this feature into $B$ equal intervals denoted as $[a_{J,0}, a_{J,1})$, $[a_{J,1},a_{J,2})$, ..., $[a_{J,B-1}, a_{J,B}]$, where $a_{J,0}$ and $a_{J,B}$ denote the minimum and maximum values of $x_J$ observed in the training data. This creates a two-dimensional grid for the interaction $\smash{L^1(x_J)}\times x_J$. We calculate the average relative change in observed deaths with respect to the mortality baseline in each grid cell $[a_{J,i-1},a_{J,i}) \times [a_{J,j-1},a_{J,j}]$, for $i,j \in \{1,2,...,B\}$, as: 
\begin{align} \label{eq:harv}
\mathcal{D}_{J}^{(i,j)}  = \dfrac{1}{n_{J,i,j}} \displaystyle \sum_{r\in\mathcal{R}}\sum_{t\in\mathcal{T}}\sum_{w\in\mathcal{W}_t} \text{EDP}_{t,w}^{(r)}  \cdot \mathbbm{1}\left\{x_{J,t,w-1}^{(r)} \in [a_{J,i-1},a_{J,i}) \: \wedge \: x_{J,t,w}^{(r)} \in [a_{J,j-1},a_{J,j}) \right\},
\end{align}
where $\mathbbm{1}\{\cdot\}$ is an indicator function, equaling one when the argument holds true and zero otherwise, and $n_{J,i,j}$ denotes the number of training observations falling inside the grid cell $[a_{J,i-1},a_{J,i}) \times [a_{J,j-1},a_{J,j})$. Similarly, we calculate the average relative change in the XGBoost estimated deaths with respect to the mortality baseline by using $\smash{\widehat{\text{EDP}}_{t,w}^{(r)}}$ in~\eqref{eq:harv}. Comparing the observed and estimated relative differences provides insights into how effectively the XGBoost model captures interaction effects present in the training data.

\begin{figure}[ht!]
\centering
\begin{subfigure}{0.49\textwidth}
\centering
\includegraphics[width = \textwidth]{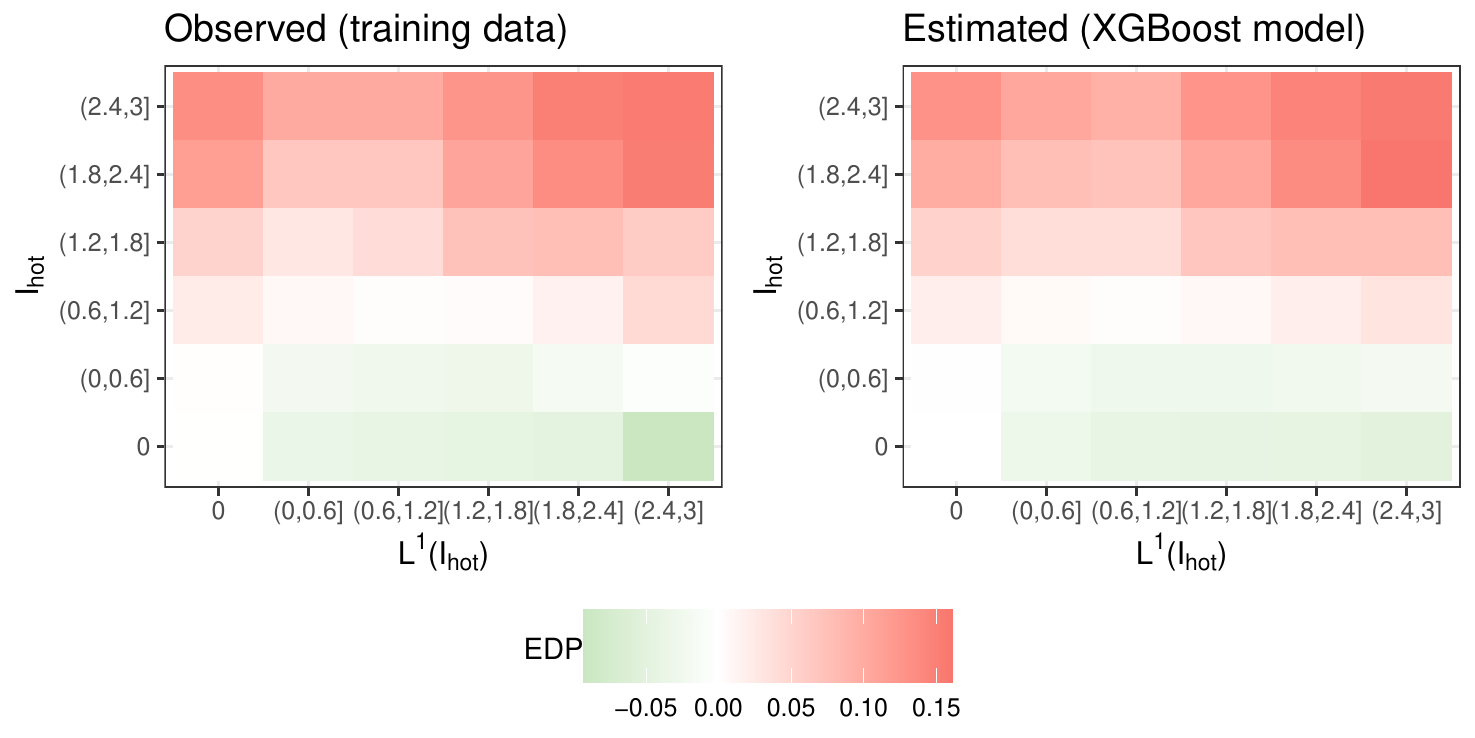}
\begin{minipage}{1.05\textwidth}
\vspace{-0.5cm}
\nsubcap{\label{fig:EPD1}}
\end{minipage}
\end{subfigure}
\begin{subfigure}{0.49\textwidth}
\centering
\includegraphics[width = \textwidth]{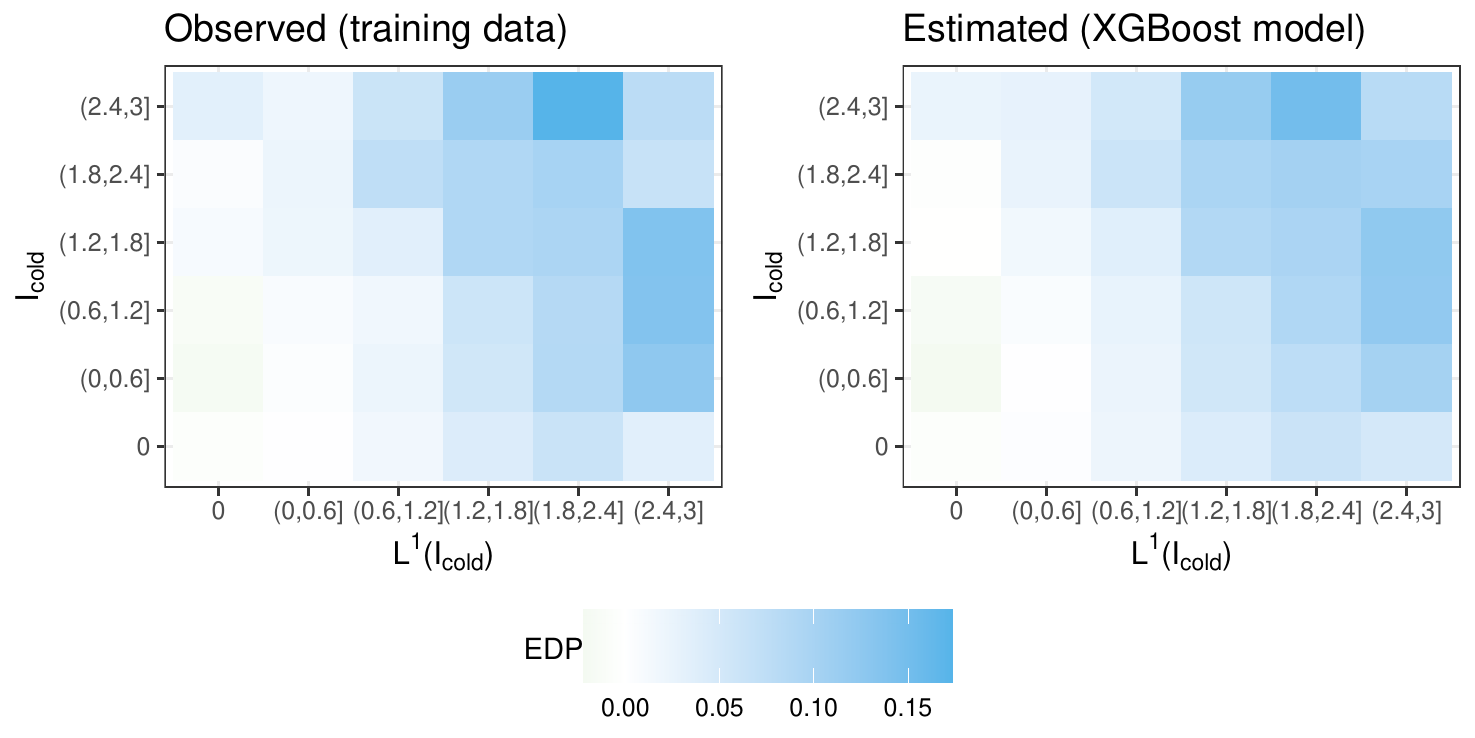}
\begin{minipage}{1.05\textwidth}
\vspace{-0.5cm}
\nsubcap{\label{fig:EPD2}}
\end{minipage}
\end{subfigure}
\begin{subfigure}{0.49\textwidth}
\centering
\includegraphics[width = \textwidth]{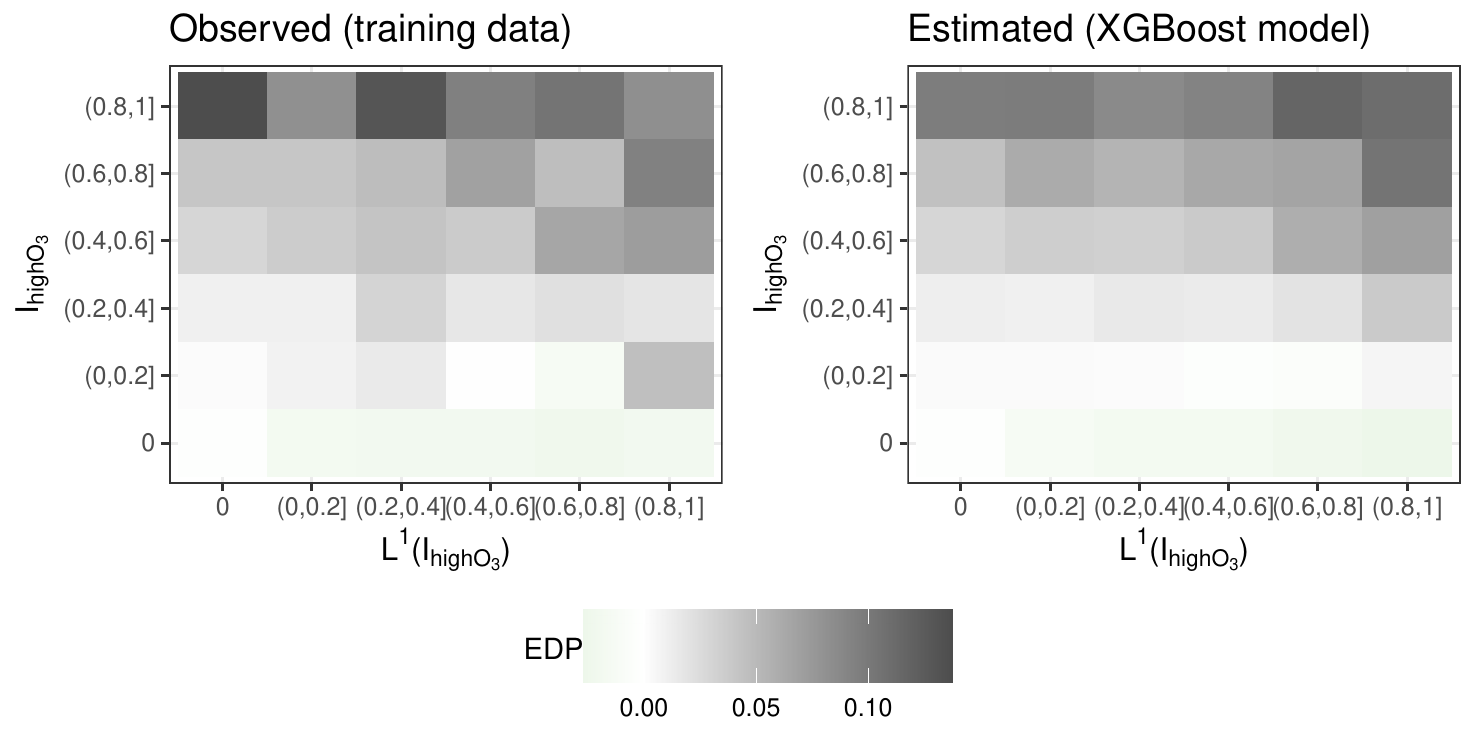}
\begin{minipage}{1.05\textwidth}
\vspace{-0.5cm}
\nsubcap{\label{fig:EPD3}}
\end{minipage}
\end{subfigure}
\begin{subfigure}{0.49\textwidth}
\centering
\includegraphics[width = \textwidth]{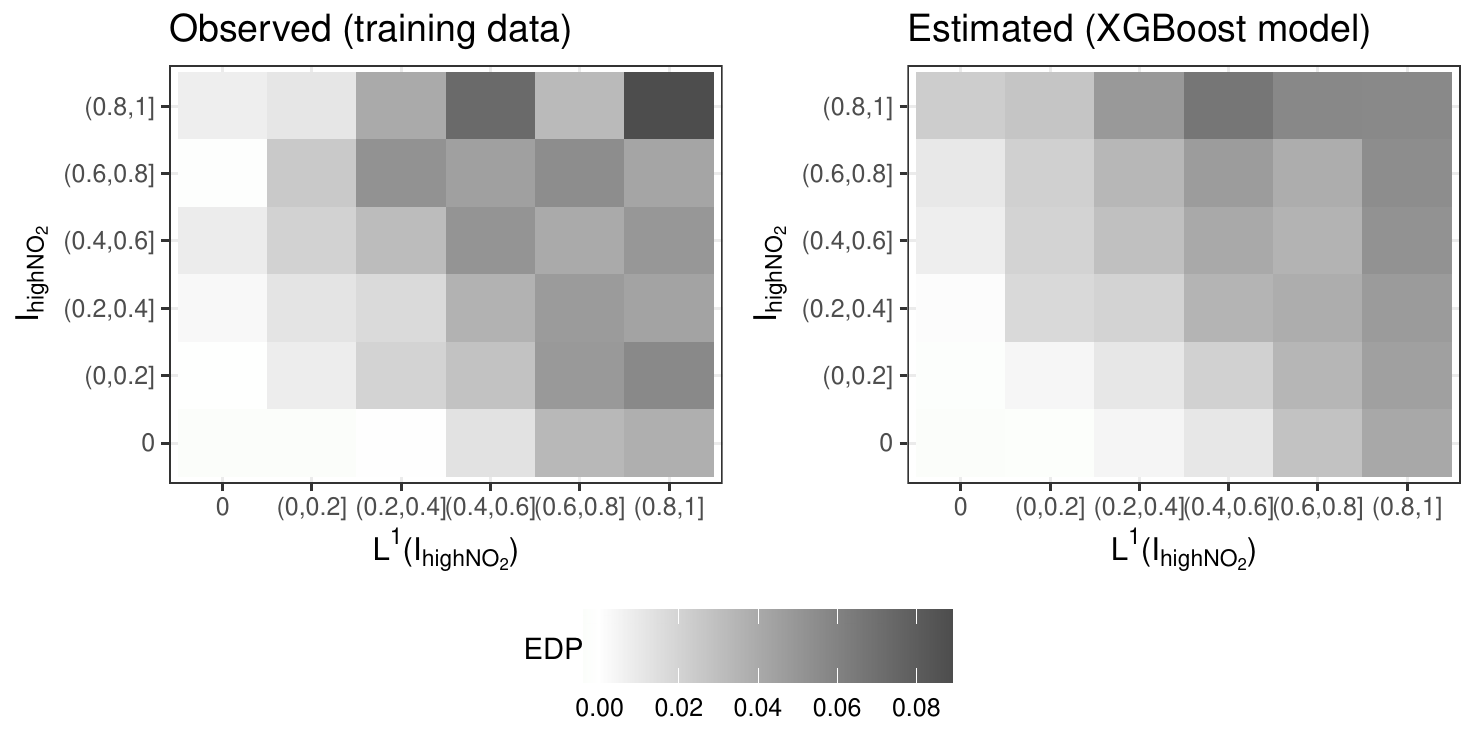}
\begin{minipage}{1.05\textwidth}
\vspace{-0.5cm}
\nsubcap{\label{fig:EPD4}}
\end{minipage}
\end{subfigure}
\caption{The average relative difference in observed and estimated deaths relative to the weekly mortality baseline model, defined in grid cells of the feature and its lagged version for the weekly average of the hot-day indicator $I_{\text{hot}}$ (a), the weekly average of the cold-day indicator $I_{\text{cold}}$ (b), the weekly average of the daily high ozone indicator $I_{\text{highO}_3}$ (c), and the weekly average of the daily high nitrogen dioxide indicator $I_{\text{highNO}_2}$ (d).\label{fig:EPD}}
\end{figure}

Figure~\ref{fig:EPD} illustrates the results of the aforementioned method for four extreme environmental indices: the weekly average of the hot-day indicator $I_{\text{hot}}$, cold-day indicator $I_{\text{cold}}$, daily high ozone indicator $I_{\text{highO}_3}$, and the daily high nitrogen dioxide indicator $I_{\text{highNO}_2}$. The term ``harvesting effect" is employed when, e.g., high temperatures in the current week lead to excess deaths, yet the subsequent week with normal temperatures results in a mortality deficit compared to the baseline. This concept applies to any weather or air pollution factor. Based on Figure~\ref{fig:EPD}, the harvesting effect appears most pronounced for the hot-week index, see the green color in the bottom right corner of Figure~\ref{fig:EPD1}. This suggests that following a very hot week succeeded by a week of normal temperatures, observed death counts are lower on average by approximately 5$\%$ relative to the baseline (more green colors). Conversely, Figure~\ref{fig:EPD2} reveals that the cold-week index does not exhibit a harvesting effect in our observations, since we even observe additional excess deaths one week after a cold week, regardless of the temperature. Moreover, no harvesting effect is detected for the weekly extreme NO$_2$ index (see Figure~\ref{fig:EPD3}), while Figure~\ref{fig:EPD4} shows a slight harvesting effect for the weekly extreme ozone index. We also observe that two consecutive weeks with high air pollution levels of NO$_2$ or O$_3$ result in higher excess deaths, indicating the presence of interaction effects in the data. Lastly, we observe a close correspondence between the observed and estimated excess death proportions, indicating that the XGBoost model effectively captures the interaction effects present in the data.

\subsection{Statistical in-sample tests} \label{appendix:in-sample-test}
\paragraph{Weekly, region-specific observations} We perform a statistical in-sample test to investigate the benefit of incorporating environmental factors into the weekly mortality model. Hereto, we compare the difference in the in-sample Poisson deviance between the XGBoost model and the baseline model. We calculate the Poisson deviance of the XGBoost model as:
\begin{align} \label{eqA:poisdev}
\text{POI\_Dev} = 2 \displaystyle \sum_{r\in\mathcal{R}} \sum_{t \in \mathcal{T}} \sum_{w \in \mathcal{W}_t} \left(d_{t,w}^{(r)} \cdot \log \frac{d_{t,w}^{(r)}}{\hat{d}_{t,w}^{(r)}} - \left( d_{t,w}^{(r)} - \hat{d}_{t,w}^{(r)}\right) \right),
\end{align}
where $\smash{\hat{d}_{t,w}^{(r)}}$ represents the estimated number of deaths by the XGBoost model. We compute the Poisson deviance of the baseline model in a similar way, but replace the estimated number of deaths $\smash{\hat{d}_{t,w}^{(r)}}$ by the estimated baseline number of deaths $\smash{\hat{b}_{t,w}^{(r)}}$. In a global analysis across all NUTS 3 regions, the Poisson deviance on the entire training dataset is 252\ 912.3 when calculated with the baseline model and 220\ 766.0 when calculated with the XGBoost model, representing a reduction of 12.71$\%$.

\begin{figure}[ht!]
\centering
\includegraphics[width=0.4\textwidth]{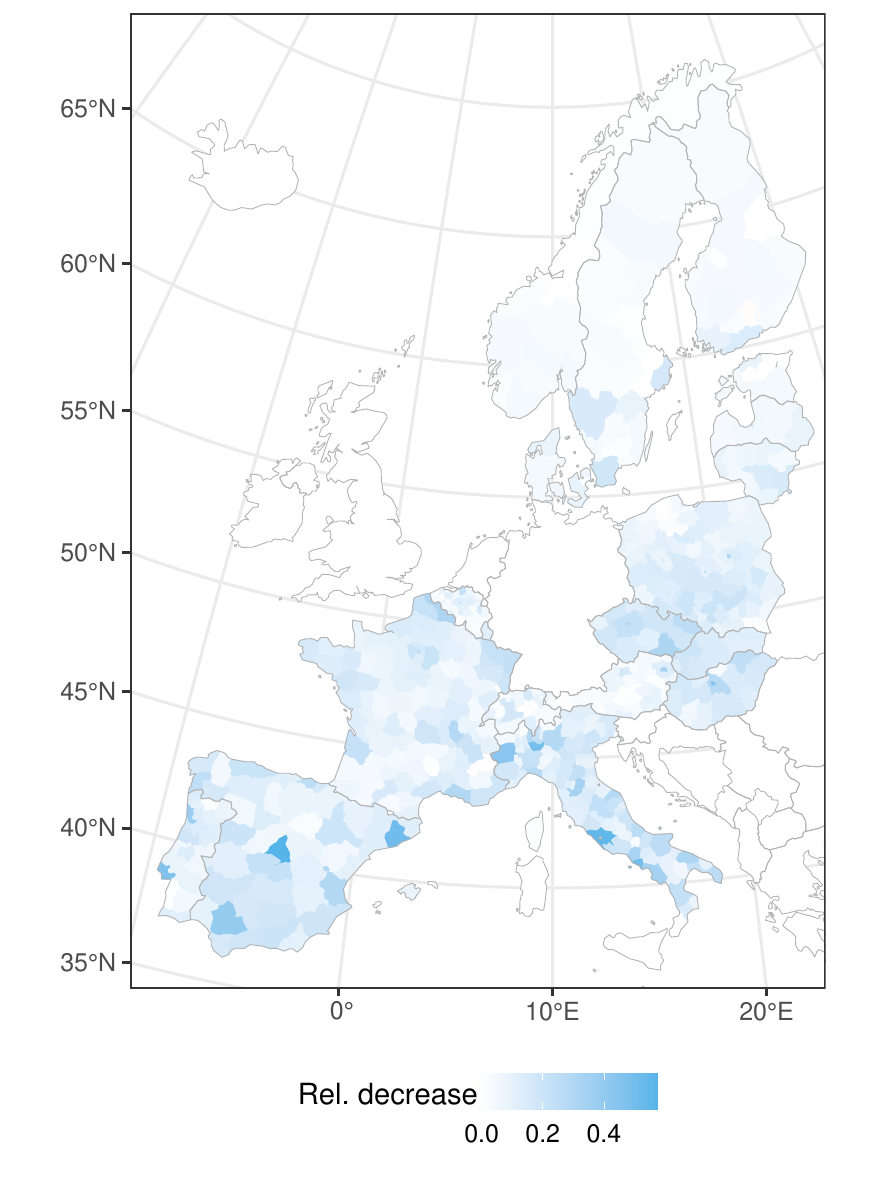}
\caption{The relative change in the Poisson deviance of the XGBoost model, incorporating weather and air pollution factors, and the baseline model in each NUTS 3 region.\label{fig:percdecr}}
\end{figure}

Additionally, we conduct a similar analysis on a regional basis by comparing the in-sample Poisson deviance on the training datasets restricted to each region, calculated with both the baseline and XGBoost model. Figure~\ref{fig:percdecr} shows the relative change in these Poisson deviances in each region with respect to the baseline model. We observe that the most significant change in the Poisson deviance is observed in the southern regions, indicating a substantial improvement when incorporating weather and air pollution factors into the weekly baseline mortality model. Conversely, in the northern regions, the added value of such incorporation appears to be limited.

\subsection{Statistical out-of-sample test} \label{appsubsec:back-test}
\begin{figure}[htb!]
\centering
\includegraphics[width=0.375\textwidth]{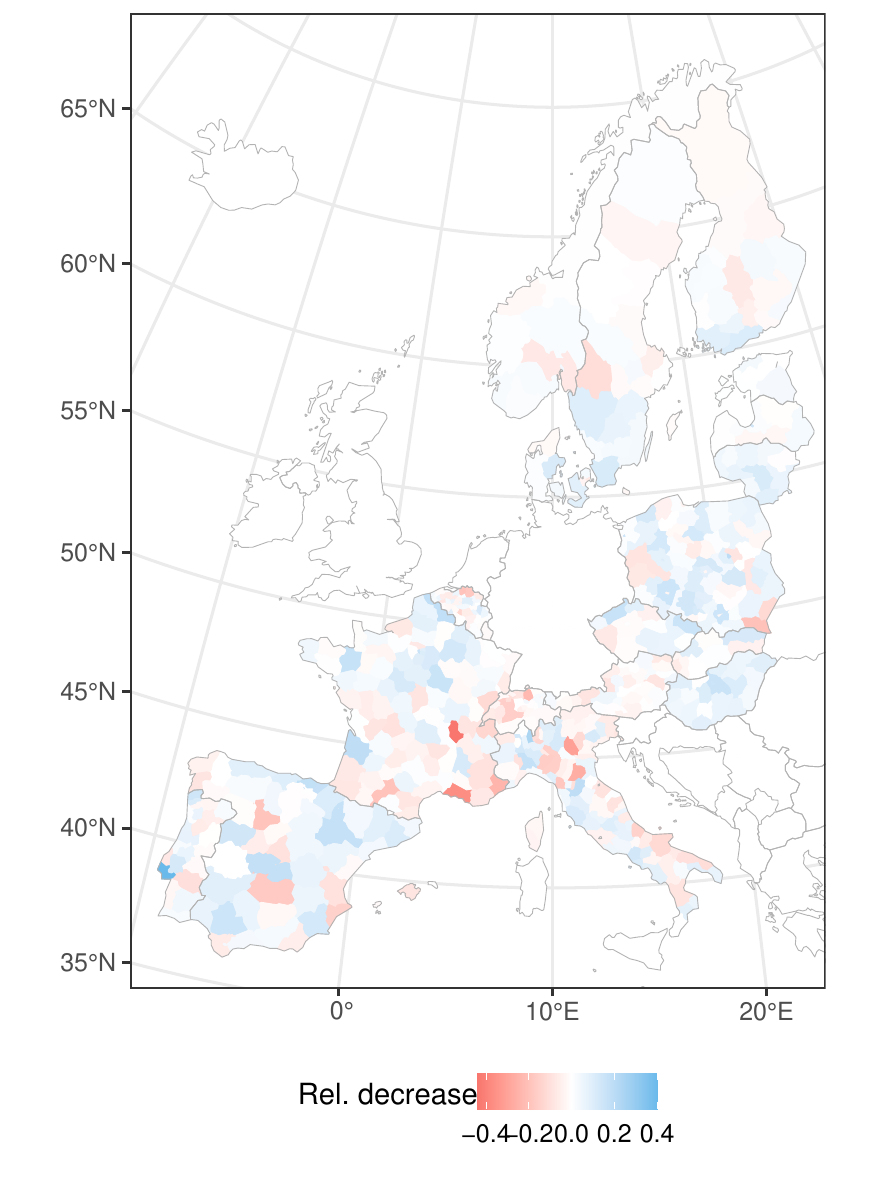}
\caption{The relative change in the Poisson deviance of the death counts estimated by the XGBoost model, integrating environmental features, relative to the Poisson deviance in the weekly baseline model, in each NUTS 3 region for the year 2019.\label{fig:percdecr2019}}
\end{figure}

We evaluate the out-of-sample Poisson deviance in each region for the year 2019 using both the weekly mortality baseline model and the XGBoost model incorporating environmental factors, see Section~\ref{subsec:backtesting} of the paper. Figure~\ref{fig:percdecr2019} illustrates the relative change in the Poisson deviance of the deaths estimated by the XGBoost model relative to the deaths estimated by the weekly baseline model, defined as:
\begin{align*}
\dfrac{\text{POI\_Dev}_{b} - \text{POI\_Dev}_{\text{XGB}}}{\text{POI\_Dev}_{b}},
\end{align*}
where $\text{POI\_Dev}_b$ and $\text{POI\_Dev}_{\text{XGB}}$ are the Poisson deviances calculated using the deaths obtained from the estimated weekly baseline and XGBoost model, respectively, see~\eqref{eqA:poisdev}. A positive value for the above defined criterion indicates an improvement of the XGBoost model with respect to the baseline model.

{
\bibliography{References}}
\end{document}